\documentclass[preprint,12pt]{elsarticle}

\usepackage{graphicx,import}
\usepackage{caption}
\usepackage{subcaption}
\usepackage{setspace}
\usepackage{xcolor}
\usepackage{framed}
\usepackage{ulem}
\usepackage[makeroom]{cancel}
\usepackage{float}

\usepackage{amssymb}
\usepackage{amsmath}
\usepackage[symbol]{footmisc}
\usepackage{lineno}
\usepackage{nomencl}
\usepackage{multicol}	% for multi-colume
\makenomenclature
\usepackage{etoolbox}

\renewcommand\nomgroup[1]{%
  \item[\bfseries
  \ifstrequal{#1}{O}{Others}{%
  \ifstrequal{#1}{C}{Constants}{%
  \ifstrequal{#1}{G}{Greek symbols}{%
  \ifstrequal{#1}{S}{Subscripts/superscripts}{%
  \ifstrequal{#1}{A}{Abbreviations}{%
  }}}}}%
]}

\journal{arXiv} 

\begin{document}

\begin{spacing}{1.25}

\begin{frontmatter}

\title{An improved Coupled Level Set and Volume of Fluid (i-CLSVoF) framework for sessile droplet evaporation}


%\def\correspondingauthor{\footnote{Corresponding author.
%\ \ \ E-mail address: huihuang.xia@kit.edu (H.Xia)}}

\author{Huihuang Xia\corref{cor1}}
\ead{huihuang.xia@kit.edu}
\cortext[cor1]{Corresponding author}

%\author[add1]{Huihuang Xia\correspondingauthor\textsuperscript{,}}
\author{Marc Kamlah}

\address{Institute for Applied Materials (IAM), Karlsruhe Institute of Technology (KIT), Hermann-von Helmholtz-Platz 1 76344 Eggenstein-Leopoldshafen, Germany}

\begin{abstract}
Surface-tension-dominant droplet evaporation is ubiquitous and of importance to many applications. We present an improved Coupled Level Set and Volume of Fluid (i-CLSVoF) framework without explicit interface reconstruction for modelling micro-sized droplets with and without evaporation. In the i-CLSVoF framework, an improved surface tension force model with additional filtering steps to filter un-physical spurious velocities is developed and implemented. A simple, yet efficient, velocity-potential based approach is proposed to reconstruct a divergence-free velocity field for the advection of the free surface during droplet evaporation. This approach fixes the numerical issues resulting from the evaporation-induced velocity jump at the interface. The smeared mass source term approach incorporated in this work guarantees greater numerical stability than the non-smeared approach. Three different evaporation models (constant mass flux, thermally driven evaporation and droplet evaporation at room temperature) are implemented in the i-CLSVoF. Corresponding numerical benchmark cases (dam break, droplet relaxation and droplet evaporation subjected to different evaporation models) are conducted to validate the surface tension and the evaporation models. Good agreement between the numerical and corresponding analytical solutions is found. The model developed in this work shows convincing performance in modelling surface-tension-dominant flow with and without evaporation.
\end{abstract}

% keyword \sep keyword

\end{frontmatter}

\begin{framed}
%\begin{multicols}{2}
\nomenclature[A]{VoF}{Volume of Fluid}
\nomenclature[A]{LS}{Level Set}
\nomenclature[A]{CLSVoF}{Coupled Level Set and Volume of Fluid}
\nomenclature[A]{s-CLSVoF}{simple Coupled Level Set and Volume of Fluid}
\nomenclature[A]{i-CLSVoF}{improved Coupled Level Set and Volume of Fluid}
\nomenclature[A]{CSF}{Continuum Surface-tension Force}
\nomenclature[A]{AMR}{Adaptive Mesh Refinement}
\nomenclature[A]{FVM}{Finite Volume Method}
\nomenclature[A]{RHS}{Right Hand Side}
\nomenclature[A]{LHS}{Left Hand Side}

\nomenclature[C]{$\boldsymbol{g}$}{gravitational acceleration constant [$m/s^2$]}
\nomenclature[C]{$c_p$}{specific heat capacity [$J/(kg \cdot K)$]}
\nomenclature[C]{$k$}{thermal conductivity [$W/(m \cdot K)$]}
\nomenclature[C]{$h_{ev}$}{enthalpy/latent heat of evaporation [$J/kg$]}
\nomenclature[C]{$T_{sat}$}{saturation temperature [$K$]}

\nomenclature[G]{$\rho$}{density [$kg/m^3$]}
\nomenclature[G]{$\alpha$}{volume fraction [$-$]}
\nomenclature[G]{$\phi$}{velocity potential [$m^2/s$]}
\nomenclature[G]{$\psi$}{signed level-set function [$-$]}
\nomenclature[G]{$\sigma$}{surface tension coefficient [$N/m$]}
\nomenclature[G]{$\lambda$}{density ratio [$-$]}

\nomenclature[O]{$I$}{indicator function [$-$]} 
\nomenclature[O]{$\mathbf{U}$}{velocity [$m/s$]}
\nomenclature[O]{$p$}{pressure [$Pa$]}
\nomenclature[O]{$\boldsymbol{n}_f$}{normal vector at face centre}
\nomenclature[O]{$\boldsymbol{n}_s$}{normal vector at cell centre}
\nomenclature[O]{$\dot m$}{mass source per unit volume [$kg/(m^3 \cdot s)$]}
\nomenclature[O]{$J$}{mass source per unit area [$kg/(m^2 \cdot s))$]}
\nomenclature[O]{$T$}{temperature [$K$]}
\nomenclature[O]{$K$}{interface curvature [$1/m$]}
\nomenclature[O]{$\chi_e$}{evaporation coefficient [$-$]}
\nomenclature[O]{$\Delta$}{Laplace operator}
\nomenclature[O]{$\nabla$}{gradient operator}
\nomenclature[O]{$D_v$}{vapour diffusivity [$m^2/s$]}
\nomenclature[O]{$M$}{molar mass [$kg/mol$]}
\nomenclature[O]{$X_v$}{vapour mole fraction [$-$]}
\nomenclature[O]{$Y$}{vapour mass fraction [$-$]}

\nomenclature[S]{$l$}{liquid}
\nomenclature[S]{$g$}{gas}
\nomenclature[S]{$st$}{surface tension}
\nomenclature[S]{$filt$}{filtering}
\nomenclature[S]{$\infty$}{infinity}
\printnomenclature

%\end{multicols}
\end{framed}

%\modulolinenumbers[2]	% every n line to show line number
%\linenumbers

\section{Introduction}
\label{S:1}
Modelling droplet evaporation is of great importance for many applications, such as inkjet printing \cite{wijshoff2018drop}, spray coating \cite{christodoulou2020model}, and combustion of fuel droplets \cite{saufi2021interface}. The key issues in the computational modelling of droplet evaporation are three-fold: free-surface tracking or capturing, the phase change from liquid to vapour, and accurate calculations of the surface tension force \cite{zang2019evaporation}. We begin this paper with a review of previous work to address each issue.
\subsection{Free-surface tracking/capturing}
For interface tracking, the front tracking method is widely used; here, the basic idea is to use so-called marker points to identify the interface's location \cite{irfan2017front, inguva2022front}. Alternatively, the phase-field method identifies the interface through the value of an order parameter: the order parameter is $1$ in the liquid phase, $-1$ in the gas phase, with a value in between representing the diffuse interface \cite{jamshidi2019suitability}. The Volume of Fluid (VoF) method is another popular interface capturing approach in which the volume fraction field is computed and the interface inferred. The Level Set (LS) method captures the free surface by the signed LS function, with the zero LS at the interface, positive in the liquid and negative in the gas phase. Furthermore, the LS method can guarantee a very sharp interface without interface diffusion \cite{sussman1994level}.
\subsection{The surface tension force}
The surface tension force plays a significant role in droplet wetting and evaporation \cite{popinet2018numerical, takada2016phase}. The water droplet on a leaf is a simple case to demonstrate the role of the surface tension force in forming a given contact angle between the droplet and the leaf surface. The surface tension force is also crucial to maintain the droplet shape under the influence of gravity and other external forces acting on the droplet \cite{ioannou2016droplet}. The Continuum Surface-tension Force (CSF) model was proposed to model the surface tension force as a volumetric body force \cite{brackbill1992continuum}. However, this conventional surface-tension model suffers from spurious currents or velocities which appear around the interface. Spurious currents destabilize the simulations and even influence the internal flow inside the droplets when studying droplets numerically \cite{mohammadrezaei2022surface}. Spurious currents partially result from numerical errors when calculating the interface curvature. Some numerical models have been developed to improve the calculation of the interface curvature and thus suppress the un-physical velocities. The geometric VoF represents the interface by a reconstructed thin interface inside each of the interface cells explicitly and is reported to have better performance in interface representations as well as reducing spurious velocities \cite{malan2021geometric}. Some open-source codes or libraries incorporate the geometric VoF, such as PARIS \cite{aniszewski2021parallel}, Basilisk \cite{popinet2015quadtree} , isoAdvector \cite{roenby2016computational}, interPlicFoam \cite{dai2019analytical} and VoFLibrary \cite{scheufler2019accurate}. 

In contrast to the geometric VoF approach, the algebraic VoF method is relatively simple and easy to implement as the interface is represented implicitly and without explicit interface reconstruction \cite{yin2019direct}. The algebraic VoF method is mass conserving but was reported to suffer from some interface diffusion \cite{jamshidi2019suitability}. Several methods have been proposed to address the interface diffusion problem, for instance,  incorporation of a surface compression term \cite{klostermann2013numerical}, adaptive interface compression \cite{aboukhedr2018simulation}, and coupling VoF to some other numerical methods, such as the LS method \cite{albadawi2013influence}. Concerning the advantages and shortcomings of both the VoF and LS method, the so-called Coupled Level Set and Volume of Fluid (CLSVoF) method was proposed to combine sharp interface representation and mass conservation \cite{sussman2000coupled}. The coupled approach improves the suppression of the spurious currents. However, according to our experience, relatively large spurious velocities still exist around the interface, especially for micro-sized droplets. 
\begin{table}[h]
\footnotesize %\normalsize
\centering
\caption{Summary of numerical methods to track/capture free surface.}
\label{methodsSummary}
\begin{tabular}{llll}
\hline
Authors (publication year)       & Method        & Code                & Applications         \\ \hline
Brackbill et al. (1992)\cite{brackbill1992continuum}  & VoF                & in-house code       & interfacial flows    \\
Sussman et al. (1998)\cite{sussman1998improved}    & LS                 & in-house code       & interfacial flows    \\
Popinet (2009)\cite{popinet2009accurate}           & VoF                & Basilisk            & interfacial flows    \\
Raeini et al. (2012)\cite{raeini2012modelling}     & VoF                & OpenFOAM            & porous media \\
Albadawi et al. (2013)\cite{albadawi2013influence}   & CLSVoF             & OpenFOAM            & bubble dynamics      \\
Yokoi 2014\cite{yokoi2014density}               & CLSVoF             & in-house code       & droplet splashing    \\ 
Roenby et al. (2016)\cite{roenby2016computational}     & VoF                & isoAvector/OpenFOAM & interfacial flows    \\
Irfan et al. (2017)\cite{irfan2017front}      & Front tracking     & in-house code       & phase change         \\
Dai et al. (2019)\cite{dai2019analytical}        & VoF                & OpenFOAM            & multiphase flows     \\
Scheufler et al. (2019)\cite{scheufler2019accurate}  & VoF                & VoFLibrary/OpenFOAM & interfacial flows    \\
Jamshidi et al. (2019)\cite{jamshidi2019suitability}   & Phase field        & OpenFOAM            & microfluids          \\
Aniszewski et al. (2021)\cite{aniszewski2021parallel} & VoF/Front tracking & PARIS               & multiphase flows     \\
Inguva et al. (2022)\cite{inguva2022front} & Front tracking       & in-house code		& two-phase flow \\ \hline
\end{tabular}
\end{table}
A short summary of methods used to track/capture the free surface and their applications are outlined in Table \ref{methodsSummary}.

\subsection{Phase change from liquid to vapour}
Several phase-change models have been developed to model the phase change from liquid to vapour.These include the constant mass flux model \cite{tanguy2007level, scapin2020volume}, the thermally driven model \cite{hardt2008evaporation, nabil2016interthermalphasechangefoam, sato2013sharp}, and the vapour mass fraction gradient model \cite{scapin2020volume, irfan2017front, palmore2019volume}. The challenging part in modelling phase change or evaporation is to address the velocity jump at the interface, which results in some numerical difficulties. Kunkelmann developed an approach that removes the source terms at the interface cells and defines positive and negative mass sources in the most adjacent liquid and gas cells, respectively. This approach was demonstrated to have good performance in modelling boiling \cite{kunkelmann2011numerical}. A similar method is also implemented into the open-source code Gerris for modelling droplet evaporation subject to a large mass transfer rate \cite{wang2019vaporization}. Both methods are highly dependent on the mesh resolution at the interface. Normally, Adaptive Mesh Refinement (AMR) is needed to cut the interface region into two different regions with negative and positive mass sources accurately. These numerical models are implemented in either in-house code or commercial codes. Thus, we propose to develop a comprehensive solver in the open-source C++ toolbox OpenFOAM to incorporate the simple yet efficient evaporation models to model the micro-sized sessile droplet evaporation with negligible influence of un-physical spurious currents.

This paper addresses the issues mentioned above related to suppressing un-physical spurious currents and the velocity jump due to phase change at the interface. It presents a simple yet efficient numerical framework to model the evaporation of micro-sized sessile droplets with sharp interface representation and suppressed spurious velocities. The main contribution of the work is coupling algebraic VoF to LS with an improved surface-tension force model incorporating filtering steps. (All these contributions are referred to here as the i-CLSVoF framework.) Furthermore, improved evaporation models are implemented into i-CLSVoF for studying sessile droplet evaporation with three different evaporation models. The model predicts the evaporation rates accurately with less influence of spurious velocities. This work consists of the following main sections: First, the mathematical description of the incompressible Newtonian flow is given in section $2$. The corresponding numerical method and the improved numerical framework i-CLSVoF are detailed in section $3$. Section $4$ presents the demonstrations of the numerical benchmark cases to demonstrate the performance of the numerical framework developed in this work.

\section{Mathematical formulation}
\label{S:2}
This section presents the equations governing the physics behind incompressible Newtonian flow based on the one-field formulation. The one-field formulation solves only one set of governing equations for both liquid and gas phases \cite{deshpande2012evaluating}. The indicator function $I(\boldsymbol{x},t)$
\begin{equation} \label{indicatorFunc}
I(\boldsymbol{x},t)=
\begin{cases}
1 & \text{$\boldsymbol{x}$ in liquid,} \\
0 & \text{otherwise,}
\end{cases}
\end{equation}
is used to identify the tracked phase (liquid phase in this work) at time $t$ and position $\boldsymbol{x}$. The fundamental fluid quantities such as fluid density $\rho$ and dynamic viscosity $\mu$ can be represented in the complete multiphase domain by the indicator function $I(\boldsymbol{x},t)$ using expressions like
\begin{equation} \label{one-FieldF}
\begin{aligned}
\rho = I(\boldsymbol{x},t)\rho_1 + [1-I(\boldsymbol{x},t)] \rho_2,\\
\mu = I(\boldsymbol{x},t)\mu_1 + [1-I(\boldsymbol{x},t)] \mu_2.\\
\end{aligned}
\end{equation}

\subsection{Governing equations for incompressible flow without phase change}
The physics behind the incompressible Newtonian fluid without phase change is governed by the Navier-Stokes equations
\begin{equation} \label{continuityEqn}
\nabla \cdot{\mathbf{U}}=0,
\end{equation}
\begin{equation} \label{momentumEqn}
\frac{\partial(\rho\mathbf{U})}{\partial t}+\nabla \cdot (\rho\mathbf{U}\mathbf{U})=-\nabla p+\nabla \cdot [\mu (\nabla \mathbf{U} + (\nabla \mathbf{U})^{T})]+\rho \boldsymbol{g}+ \mathbf{F}_{st},
\end{equation}
where $\mathbf{U}$ is the velocity field. Its divergence being equal to zero means that the flow is incompressible. In the momentum equation (Eqn. \ref{momentumEqn}), $p$ is the pressure field and $\rho \mathbf{g}$ the gravity force term. The surface tension force newly introduced in this work is incorporated into the momentum equation as the last term on the RHS (the surface-tension force model is detailed in Section 3). The VoF transport equation is solved to capture the free surface, by which the volume fraction field can be continuously updated. The material derivative of the volume fraction field $\alpha_l$ defines the VoF transport equation
\begin{equation} \label{VoFEqn}
\frac{D \alpha_l}{D t}=\frac{\partial \alpha_l}{\partial t}+\mathbf{U} \cdot \nabla \alpha_l=0.
\end{equation}
In OpenFOAM, the conservative VoF equation 
\begin{equation} \label{conVoFEqn}
\frac{\partial \alpha_l}{\partial t}+\nabla \cdot  (\alpha_l \mathbf{U})= 0,
\end{equation}
is solved where the divergence-free condition (Eqn. \ref{continuityEqn}) must be satisfied simultaneously to guarantee the incompressibility. The additional so-called interface compression term $\nabla \cdot [\alpha_l (1-\alpha_l)\mathbf{U}_r]$ with $\mathbf{U}_r$ being the artificial compression velocity is generally incorporated into the LHS of the VoF equation (Eqn. \ref{conVoFEqn}) to suppress the interface diffusion \cite{rusche2003computational}. As demonstrated in our benchmark study and the literature, however, the compression term enhances spurious currents \cite{hoang2013benchmark, nekouei2017volume}.

\subsection{Governing equations for incompressible flow with phase change}
The primary governing equations for the incompressible flow without phase change are modified to account for the phase change from the liquid phase to the vapour phase occurring at the liquid surface. The velocity field for the incompressible flow with phase change is not divergence-free anymore. We introduce a volumetric mass source term $\dot{m}$ (mass change rate per unit volume) on the RHS of the continuity equation giving
\begin{equation} \label{continuityEqnPC}
\nabla \cdot{\mathbf{U}}=\dot{m} (\frac {1}{\rho_g}-\frac {1}{\rho_l}).
\end{equation}
Here, $\dot{m}$ means the mass loss of the liquid phase which reappears at the vapour phase with the same amount. Correspondingly, $\rho_g$ and $\rho_l$ are the density of the gas and liquid phases, respectively. The momentum equation is the same as the incompressible flow without phase change. 

However, we need to modify the velocity field in the VoF transport equation (Eqn. \ref{conVoFEqn}) by the new interface velocity field $\mathbf{U}_{\Gamma}$ as
\begin{equation} \label{conVoFEqnPC}
\frac{\partial \alpha_l}{\partial t}+\nabla \cdot  (\alpha_l \mathbf{U}_{\Gamma})=\alpha_l \nabla \cdot \mathbf{U}_{\Gamma}.
\end{equation}
The reason behind that lies in our preliminary numerical simulations which confirmed that using the default one-field velocity $\mathbf{U}$ to solve the VoF transport equation tends to overestimate the evaporation rate, and we can find the same conclusion in the literature \cite{tanguy2007level}. 

The calculations of $\mathbf{U}_{\Gamma}$ can be derived from the interface mass flux balance. The interface mass flux per unit area $J \ [kg/(m^2 \cdot s)]$ when phase changes is derived from the mass flux balance across the interface,
\begin{equation} \label{massFlux}
J = \rho_l(\mathbf{U}_e-\mathbf{U}_{\Gamma}) \cdot \mathbf{n} = \rho_g(\mathbf{U}_g-\mathbf{U}_{\Gamma}) \cdot \mathbf{n},
\end{equation}
where $\mathbf{n}$ is the interface normal vector, and $\mathbf{U}_e$ and $\mathbf{U}_g$ are the fluid velocity in the liquid and gas phase, respectively. The interface velocity $\mathbf{U}_{\Gamma}$ can be accordingly derived as
\begin{equation} \label{UGamma}
\mathbf{U}_{\Gamma} = \mathbf{U}_{e} - \frac{J}{\rho_l} \mathbf{n},
\end{equation}
where the second term $J/\rho_l \cdot \mathbf{n}$ is the interface recession velocity and accounts for the interface shrinking during the evaporation process. The first term $\mathbf{U}_{e}$ is known as the extended divergence-free liquid velocity. 

A simple method based on the algebraic VoF is developed in this work to reconstruct the new divergence-free velocity field $\mathbf{U}_e$. The basic idea behind the reconstruction of $\mathbf{U}_{e}$ is to solve the velocity potential equation 
\begin{equation} \label{velPotEqn}
\begin{cases}
a \phi + \nabla ^2 \phi = \dot{m} (\frac {1}{\rho_g}-\frac {1}{\rho_l}), \\ 
\mathbf{U}_s = \nabla \phi \\
\end{cases}
\end{equation}
in the whole computational domain with the homogeneous Dirichlet boundary condition applied on the boundary to guarantee that the velocity potential at the domain boundary is zero. Here, $\phi$ is the velocity potential, and $\mathbf{U}_s$ denotes the evaporation-induced Stefan flow velocity (equal to the velocity potential gradient). 
\begin{figure}[h]
  \begin{center}
    \includegraphics[width=0.5\textwidth]{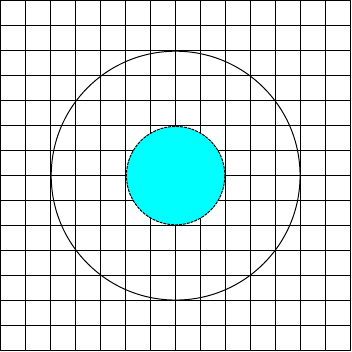}
  \end{center}
  \caption{The sub-domains for solving the velocity potential equation.}
  \label{velPotDomain}
\end{figure}
The critical parameter $a$ is used to divide the whole computation domain into two sub-domains (refer to Fig. \ref{velPotDomain}), where $a$ is zero in the liquid phase (blue circle) and within the three most adjacent cells around the interface (the circle with solid line). For the rest of the computational domain, $a$ can be any arbitrary non-zero value (the square of time-step size is used in the current study). As the RHS of the velocity potential equation (Eqn. \ref{velPotEqn}) has the same source term as the one of the continuity equation (Eqn. \ref{continuityEqnPC}) for the incompressible flow with phase change, the new divergence-free velocity field $\mathbf{U}_e$ is defined by subtracting the evaporation-induced Stefan flow velocity $\mathbf{U}_s$ from the one-field velocity field $\mathbf{U}$ as
\begin{equation} \label{Ue}
\mathbf{U}_e = \mathbf{U} - \mathbf{U}_s.
\end{equation}
Typically, the divergence of $\mathbf{U}_e$ should approximate $10^{-8}$ or even smaller values of $10^{-10}$, which can be regarded as zero numerically. As an alternative to solve and update the liquid volume fraction field $\alpha_l$ with the implicit source term (as shown in Eqn. \ref{conVoFEqnPC}), the divergence-free velocity field $\mathbf{U_e}$ can also be used to advect the free surface with either explicit 
\begin{equation} \label{VoFEqnExp}
\frac{\partial \alpha_l}{\partial t}+\nabla \cdot  (\alpha_l \mathbf{U}_e)= - \frac{\dot{m}}{\rho_l}
\end{equation}
or implicit
\begin{equation} \label{VoFEqnImp}
\frac{\partial \alpha_l}{\partial t}+\nabla \cdot  (\alpha_l \mathbf{U}_e)= \alpha_l \frac{-\dot{m}}{(\alpha_l + \delta_s)\rho_l}
\end{equation}
source term\footnote{When solving the VoF equation in OpenFOAM, the implicit source term is recommended.} accounting for the mass loss in the liquid phase due to the evaporation where $\delta_s$ in Eqn. \ref{VoFEqnImp} is a small number to guarantee a non-zero denominator. 

Tracking the temperature distribution around an evaporating droplet is crucial to model droplet evaporation subjected to the ambient temperature gradient. Conservation of thermal energy is given by the temperature equation
\begin{equation} \label{TEqn}
\frac{\partial (\rho c_pT)}{\partial t}+\nabla \cdot (\rho c_p \mathbf{U} T)=\nabla \cdot (k \nabla T)-\dot m h_{ev}+[\frac{\partial (\rho c_p)}{\partial t}+\nabla \cdot (\rho c_p \mathbf{U})]T,
\end{equation}
where $T$ is the temperature field, $c_p$ the specific heat capacity, $k$ the thermal conductivity and $h_{ev}$ the enthalpy or latent heat of evaporation. The second term on the RHS of the temperature equation is due to the evaporation-induced cooling, and the last term couples to the mass source term.

For droplet evaporation at room temperature, the vapour concentration gradient around the evaporating droplet drives the phase change from liquid to vapour. Accordingly, the vapour concentration is solved directly to model the phase change from liquid to vapour \cite{hu2002evaporation}. To simplify the characterization of the numerical model, in the current work, a dimensionless quantity called the vapour mass fraction field $Y$ is solved and updated by the convection-diffusion equation given as
\begin{equation} \label{YEqn}
\frac{\partial Y}{\partial t}+\nabla \cdot (Y \mathbf{U})=D_v \nabla^{2} Y,
\end{equation}
where $D_v$ is vapour diffusivity. The vapour convection-diffusion equation is solved on the whole computational domain with the prescribed so-called internal boundary conditions, i.e., all the liquid cells are assigned with the saturation mass fraction (detailed in Section 3). Accordingly, this internal boundary condition guarantees that vapour diffuses only from the liquid surface to the gas domain without un-physical diffusion going back to the droplet \cite{hassanvand2011direct}.

\section{Numerical method}
We solve the aforementioned equations within the Finite Volume Method (FVM) framework. The improved Coupled Level Set and Volume of Fluid (i-CLSVoF) framework is proposed in this work to suppress un-physical spurious velocities and improve numerical stabilities, especially when phase changes. The corresponding in-house solvers interDyMFoamX (without phase change) and interDyMEvapFoamX (with phase change) are accordingly developed.

\subsection{Interface capturing approaches}
In the VoF method, the volume fraction field $\alpha$ is defined as the volume-averaged volume integral of the phase indicator function $I(\boldsymbol{x},t)$
\begin{equation} \label{alpha}
\alpha = \frac{1}{V} \int_V I(\boldsymbol{x},t) dV.
\end{equation}
The basic idea behind the VoF interface capturing approach is to track the evolution of the volume fraction field for a given phase. Typically, the liquid is selected as the tracked phase, and accordingly, the liquid volume fraction field $\alpha_l$ is tracked.
% 2 figures side by side
\begin{figure}[h]
  \begin{subfigure}[h]{0.45\textwidth}
    \includegraphics[width=\textwidth]{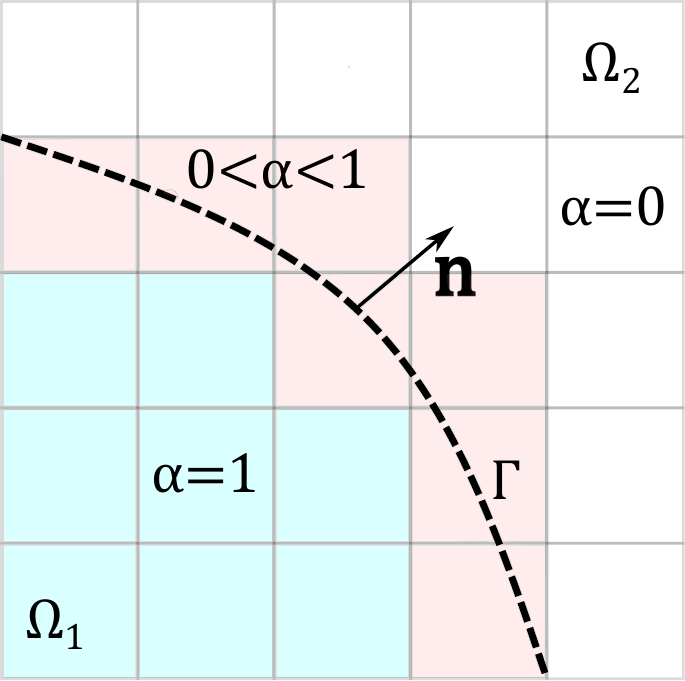}
    \caption{}
    \label{interfaceCaptureA}
  \end{subfigure}
  \hfill
  \begin{subfigure}[h]{0.45\textwidth}
    \includegraphics[width=\textwidth]{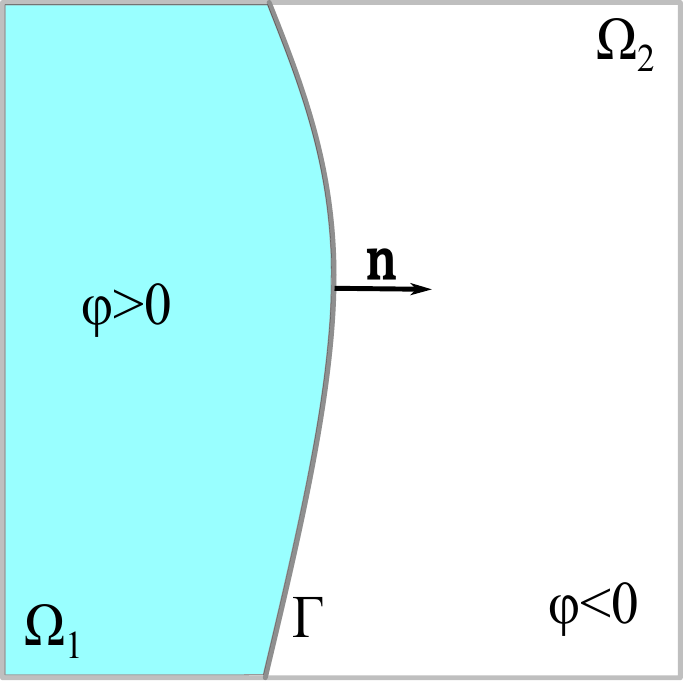}
    \caption{}
    \label{interfaceCaptureB}
  \end{subfigure}
  \caption{Interface capturing approaches: (a) the VoF method, (b) the LS method.}
  \label{interfaceCapture}
\end{figure}
As shown in Fig. \ref{interfaceCapture}, the computational domain has two sub-domains $\Omega_1$ (liquid) and $\Omega_2$ (gas) in the whole computational domain. When the liquid cells are full of liquid then the volume fraction of all liquid cells in $\Omega_1$ is $1$. Furthermore, the interface cells are partially filled with the liquid, so that the intermediate value between $0$ and $1$ are given there. The VoF method is mass-conserving, but as mentioned in the literature, the VoF method suffers from some interface diffusion, which can diffuse over several cells around the interface depending on the mesh resolution \cite{nguyen2017volume}. 

The LS method is another interface capturing method, and the quantity used in LS is the so-called signed LS function $\psi(\boldsymbol{x},t)$. The interface can be identified as zero level set, and in the liquid phase ($\Omega_1$), we have positive LS function and negative values in the gas phase ($\Omega_2$) \cite{sussman1994level}. The signed LS function is continuous and has a smooth transition from liquid to gas phase. It is obvious that the LS method can guarantee a more sharp interface compared to the VoF approach, but the LS method is reported to be not mass-conserving in the literature \cite{yamamoto2017validation}.

\begin{figure}[h]
  \begin{subfigure}[h]{0.5\textwidth}
    \includegraphics[width=\textwidth]{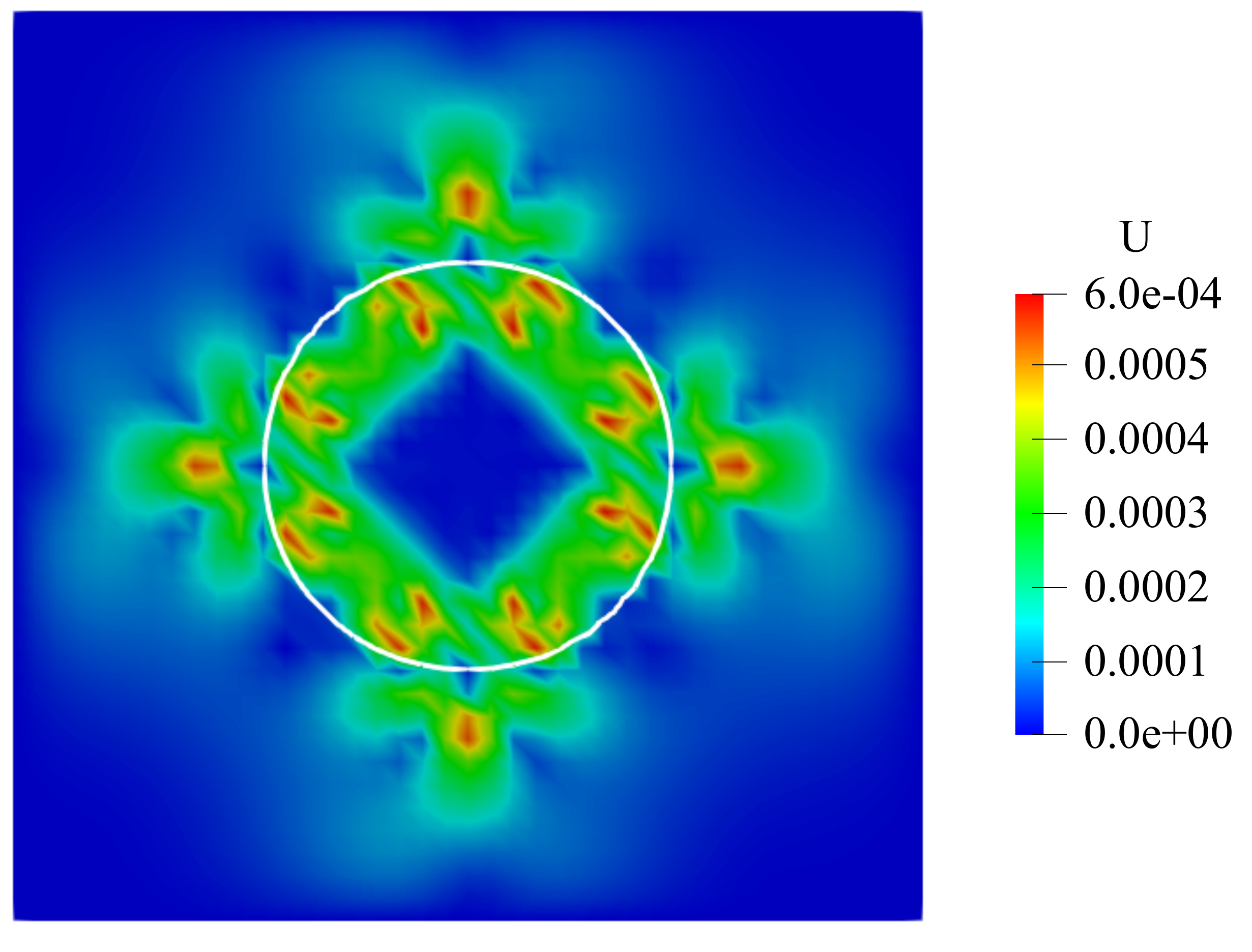}
    \caption{}
    \label{meshRefine1}
  \end{subfigure}
  \hfill
  \begin{subfigure}[h]{0.5\textwidth}
    \includegraphics[width=\textwidth]{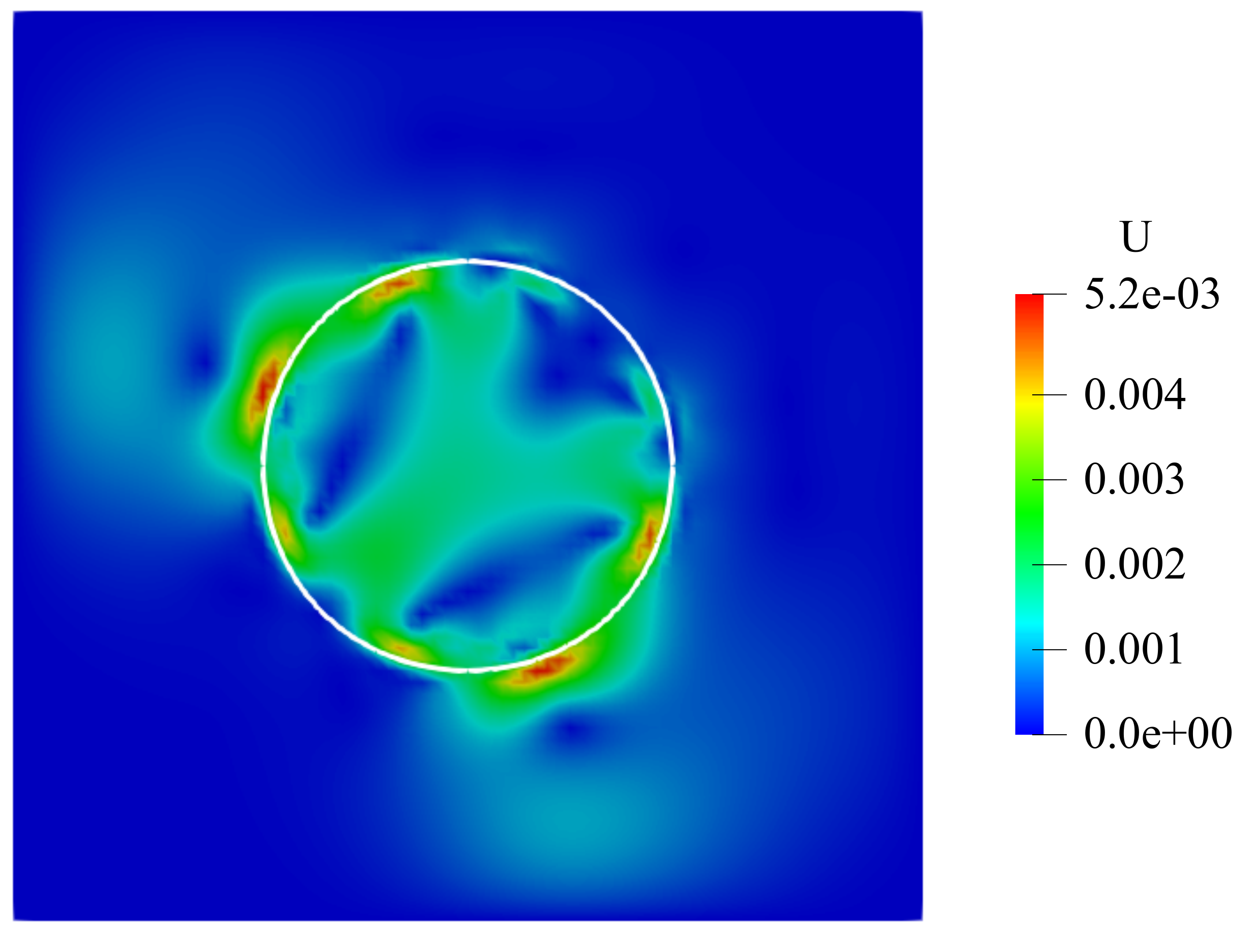}
    \caption{}
    \label{meshRefine2}
  \end{subfigure}
  \caption{Spurious velocities around the interface (represents with the solid white circle) with two different mesh sizes : (a) coarse mesh (cell number: 40$\times$40), (b) fine mesh (cell number: 80$\times$80).}
\label{meshRefine}
\end{figure}

Both the VoF and LS methods suffer from un-physical spurious velocities, which destabilize the numerical simulations. Spurious velocities occur due to inaccurate interface curvature calculations. One method that attempted to improve the curvature calculation was to refine the mesh. However, we found that a finer mesh could not reduce the un-physical velocity and indeed enhanced it (see Fig. \ref{meshRefine}); indeed, the same conclusion can also be found in the literature \cite{guo2015implementation}.
% 2 figures side by side

\subsection{The i-CLSVoF framework}
In this work, we combine the advantages of VoF and LS by using the so-called Coupled LS and VoF (CLSVoF) approach to capture the free surface. The CLSVoF approach improves the mass-conserving issues of LS while also guaranteeing a sharp interface \cite{sussman2000coupled}. However, although it improves the calculation of the interface curvature \cite{albadawi2013influence, yamamoto2017validation}, the conventional CLSVoF, also known as the simple CLSVoF (s-CLSVoF) approach, still suffers from un-physical spurious velocities. The filtering surface tension model based on the VoF approach is reported to suppress un-physical spurious currents, especially for droplets interacting with substrates \cite{raeini2012modelling}. Accordingly, we propose the improved CLSVoF (i-CLSVoF) method to suppress the spurious currents further and improve the numerical stability by  extending the filtering method to filter un-physical spurious velocities further. In contrast to the s-CLSVoF method, our i-CLSVoF framework incorporates an improved surface tension force model to calculate surface-tension forces more accurately and to filter and reduce spurious velocities further by additional filtering steps, which are discussed further in the following part.

\begin{figure}[h]
  \begin{center}
    \includegraphics[width=1.0\textwidth]{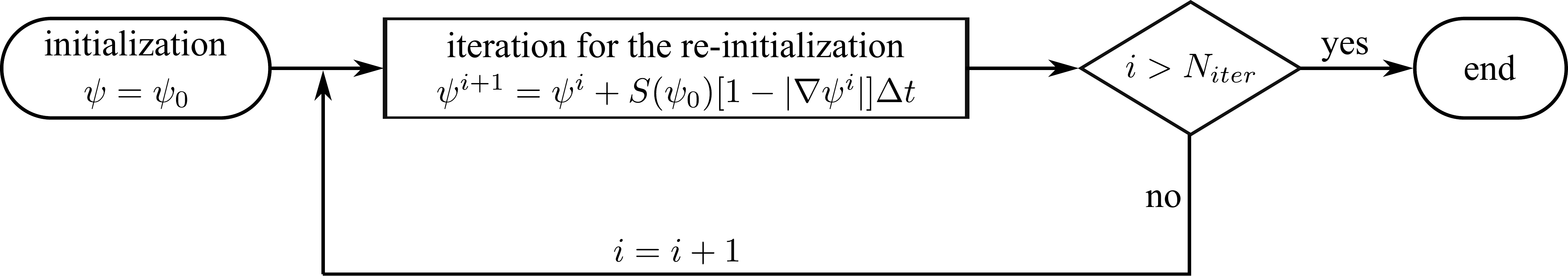}
  \end{center}
  \caption{Flowchart for solving the Hamilton–Jacobi equation.}
  \label{re-initializationScheme}
\end{figure}

The basic idea behind the i-CLSVoF framework is that initializing the initial signed distance function $\psi_0$ from the liquid volume fraction field $\alpha_l$ with the initialization function as
\begin{equation} \label{initializationFunc}
\psi_0 = (2\alpha_l-1)\Gamma,
\end{equation}
with the dimensionless quantity $\Gamma = 0.75 \Delta{x}$, where $\Delta{x}$ is the minimum mesh size around the interface, and is dimensionless as well (as $\Delta{x}$ is artificially divided by a dimensional quantity with a value of $1$ and the dimension of metre). Normally, the LS function gradually loses its property to have a value of zero at the interface and cannot be sharp enough after moving with the convection velocity. Therefore, a re-initialization step is adopted to recover its sharpness. The Hamilton–Jacobi equation 
\begin{equation} \label{HJEqn}
\frac{\partial \psi}{\partial \tau}-S(\psi_0)(1-|\nabla \psi|)=0,
\end{equation}
is used to re-initialize the LS function with the initial condition $\psi = \psi_0$ \cite{sussman1994level}. Here, $\tau$ is an artificial time step, and the smoothed out Sign function $S(\psi_0)$ is defined as
\begin{equation} \label{signFunc}
S(\psi_0)=\frac{\psi_0}{\sqrt {\psi_0^2+ \Delta{x}^2}}.
\end{equation}
Similar to the conclusion in \cite{sussman1998improved}, our benchmarking case study showed that the smoothed out Sign function $S(\psi_0)$ can further reduce the un-physical velocity and guarantee more numerical stability than the conventional Sign function. The Hamilton–Jacobi equation needs to be solved with continuous numerical iterations until $|\nabla \psi|=1$, and the iteration time $N_{iter}$ around $15$ can be enough \cite{yamamoto2017validation}. The re-initialization scheme is outlined in Fig. \ref{re-initializationScheme}.

Before introducing the new surface tension force model implemented in the i-CLSVoF framework, two widely-used surface tension models are also listed below for completeness. The Continuum Surface-tension Force (CSF) model approximates the surface tension with the help of the $\alpha_l$ gradient \citep{popinet2018numerical, brackbill1992continuum}. This surface tension force model is given by
\begin{equation} \label{FstVoF}
\sigma K(\alpha_l) \nabla \alpha_l,
\end{equation}
where $\sigma$ is the surface tension coefficient, and $K(\alpha_l)$ the interface mean curvature defined as
\begin{equation} \label{curvatureAlpha}
K(\alpha_l)=-\nabla \cdot \frac{\nabla \alpha_l}{|\nabla \alpha_l|+\delta_n}.
\end{equation}
Here, $\delta_n$ is a stabilization factor that is used to guarantee a non-zero denominator, and $\delta_n$ can be calculated by
\begin{equation} \label{deltaN}
\delta_n = \frac{10^{-8}}{(\sum\limits_{i=1}^N V_i /N)^{\frac{1}{3}}}, 
\end{equation}
where $N$ is the number of the cells in the computational domain, and $V_i$ is the volume of the $i^{th}$ cell.

As the liquid volume fraction field $\alpha_l$ is not continuous, the calculation of its gradient is another source of numerical error. The LS method uses the signed LS function $\psi$ to calculate the interface curvature, which is more accurate as $\psi$  ensures continuity along with the interface normal \citep{sussman1994level}. The interface curvature with the LS method can be calculated as
\begin{equation} \label{curvaturePsi}
K(\psi)=-\nabla \cdot \frac{\nabla \psi}{|\nabla \psi|+\delta_n}.
\end{equation}
The improved curvature calculation method is incorporated into the conventional CLSVoF approach to improve the surface tension calculations \citep{yamamoto2017validation,albadawi2013influence}. As an alternative to Eqn. \ref{FstVoF}, the surface tension force is then given as
\begin{equation} \label{FstCLSVoF}
\sigma K(\psi) \delta_{\psi} \nabla \psi,
\end{equation}
where the smoothed delta function $\delta_{\psi}$ is given by
\begin{equation} \label{deltaFunc}
\delta_{\psi}=
\begin{cases}
\frac{1}{2\epsilon} \left (1+\text{cos}(\frac{\pi \psi}{\epsilon}) \right) & |\psi| < \epsilon,\\
0 & \text{otherwise,}
\end{cases}
\end{equation}
and $\epsilon$ is the interface thickness; usually taken as $\epsilon = 1.5 \ \Delta{x}$ \cite{wen2020development}. As discussed in the literature, $\epsilon$ can also range from $1.0 \ \Delta{x}$ to $1.5 \ \Delta{x}$ depending on the mesh type \cite{singh2018coupled}. 

\begin{figure}[h]
  \begin{center}
    \includegraphics[width=0.5\textwidth]{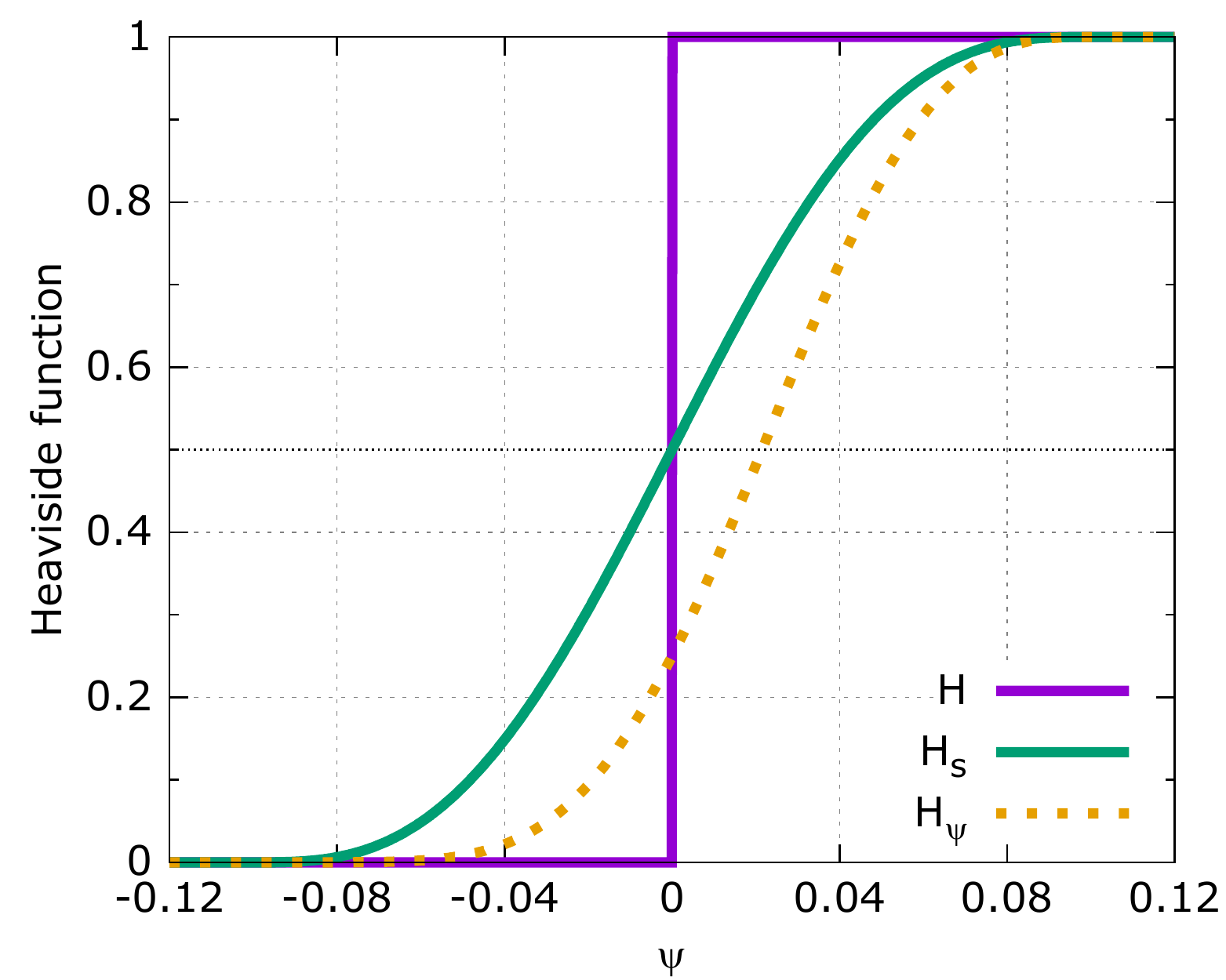}
  \end{center}
  \caption{Three different Heaviside functions ($H$: the Heaviside step function, $H_s$: the symmetrical Heaviside function, $H_{\psi}$: the non-symmetrical Heaviside function).}
\label{HeavisideFuncs}
\end{figure}

The conventional CLSVoF approach is reported to suppress un-physical spurious currents; however, some further work can be done to refine the model and further reduce the un-physical velocities (some improvements are detailed below). Therefore, the non-symmetrical Heaviside function $H_{\psi}$ is incorporated into our new surface tension force model. The reason is that the non-symmetrical Heaviside function is reported to improve the numerical stability compared to the symmetrical Heaviside function $H_s$ \cite{yokoi2014density,yamamoto2017validation}. The symmetrical Heaviside function $H_s$ is obtained by smoothing out the Heaviside step function (purple line in Fig. \ref{HeavisideFuncs}). 
\begin{equation} \label{Hs}
\ H_s =
\begin{cases}
0 & \psi < -\epsilon, \\
\frac{1}{2}[1+\frac{\psi}{\epsilon}+\frac{1}{\pi}\text{sin}(\frac{\pi \psi}{\epsilon})] & |\psi| \leq \epsilon, \\
1& \psi > \epsilon.
\end{cases}
\end{equation}

The non-symmetrical Heaviside function $H_{\psi}$ (yellow dotted line in Fig. \ref{HeavisideFuncs}) is given by
\begin{equation} \label{Hpsi}
\ H_{\psi}=
\begin{cases}
0 & \psi < -\epsilon, \\
\frac{1}{2}[\frac{1}{2}+\frac{\psi}{\epsilon}+\frac{\psi^2}{2 \epsilon^2}-\frac{1}{4 \pi^2} \left(\text{cos}(\frac{2 \pi \psi}{\epsilon})-1 \right)+\frac{\epsilon+\psi}{\epsilon \pi}\text{sin}(\frac{\pi \psi}{\epsilon})] & |\psi| \leq \epsilon,\\
1& \psi > \epsilon.
\end{cases}
\end{equation}

The difference among the three different Heaviside functions is shown in Fig. \ref{HeavisideFuncs}. 

The new and improved surface tension force model developed for our i-CLSVoF framework is given accordingly as
\begin{equation} \label{FstiCLSVoF}
\mathbf{F}_{st}=\sigma K(\psi)\nabla H_{\psi}.
\end{equation}

As inspired by the VoF-based surface tension model, the surface tension related pressure term $p_{st}$ is separated from the total pressure to avoid difficulties in the discretization of the pressure jump \cite{raeini2012modelling}. The pressure equation is given by
\begin{equation} \label{pEqn}
\nabla \cdot \nabla p_{st} = \nabla \cdot \mathbf{F}_{st},
\end{equation}
and the pressure equation can be solved with the prescribed boundary condition
\begin{equation} \label{pEqnBCs}
\frac{\partial p_{st}}{\partial n} = 0.
\end{equation}

To filter spurious currents, the modified indicator function $\alpha_{pc}$ is defined for calculating the new Delta function given as Eqn. \ref{deltaSt}, and $\alpha_{pc}$ is calculated with
\begin{equation} \label{alphaPc}
\alpha_{pc} = \frac{1}{1-C_{pc}} \left[ \text{min} \left( \text{max} (\alpha_l, \frac{C_{pc}}{2}), 1-\frac{C_{pc}}{2} \right) - \frac{C_{pc}}{2}\right],
\end{equation}
where $C_{pc}$ is the sharpening coefficient. $C_{pc}$ equal to $0$ yields the original indicator function $\alpha_l$, which is the liquid volume field and defined by Eqn. \ref{alpha}. Increasing $C_{pc}$ leads to a sharp representation of the interface and can suppress the spurious velocity but also brings numerical instabilities \cite{raeini2012modelling}. 

Finally, employing $p_{st}$ and $\alpha_{pc}$, the filtering surface tension force model is introduced to filter un-physical spurious currents parallel to the free surface and is defined as
\begin{equation} \label{filterFst1}
F_{st,f}^f = F_{st,f} - F_{st,f}^{filt}.
\end{equation}
Here, $F_{st,f}$ is the surface tension force calculated at face centre by $F_{st,f} = \mathbf{F}_{st} \mathbf{n}_f$ with $\mathbf{n}_f$ being the normal vector defined at face centre, and $F_{st,f}^{filt}$ is a time-related term also defined at the face centre. It is calculated from
\begin{equation} \label{filterFst2}
F_{st,f}^{filt} = \frac{\delta_{st}}{|\delta_{st}| + \delta_n} \left( {R_f (F_{st,f}^{filt})}_{i-1} + C_{fc} \bigg \langle \nabla p_{st}- (\nabla p_{st} \cdot \mathbf{n}_s) \mathbf{n}_s \bigg \rangle _f \cdot \mathbf{n}_f \right),
\end{equation}
where $R_f$ is a relaxation factor, and ${(F_{st,f}^{filt})}_{i-1}$ the value of $F_{st,f}^{filt}$ in the previous time step, and $\langle \rangle_f$ denotes the interpolation from cell centre to cell face in OpenFOAM. $C_{fc}$ is the coefficient determining how fast the spurious velocity is filtered, and $\mathbf{n}_s$ is the normal vector defined at cell centre ($\mathbf{n}_s=\nabla \alpha_l/|\nabla \alpha_l|$). $\delta_{st}$ is a newly defined Delta function based on the previously introduced sharpening indicator function $\alpha_{pc}$, and its definition is
\begin{equation} \label{deltaSt}
\delta_{st}  = \nabla_f^{\Gamma} \alpha_{pc}.
\end{equation}
Here, $\nabla_f^{\Gamma}$ denotes the gradient normal to the interface. 

The final step is to define the threshold for filtering the surface-tension flux, also called the capillary flux ($\phi_{cf}= F_{st,f}|S_f|$ with $|S_f|$ being the magnitude of face area). We artificially set the capillary flux as zero when the capillary flux is smaller than the threshold, where the filtering capillary flux is defined as
\begin{equation} \label{capillaryFlux}
\phi_{cf}^{filt} = \phi_{cf} - \text{min}\left(\text{max}(\phi_{cf}, -\phi_{cf}^{thre}), \phi_{cf}^{thre} \right ).
\end{equation}
Here, the $\phi_{cf}^{thre}$ is the threshold value below which the capillary flux is regarded as zero, and it can be calculated by
\begin{equation} \label{capillaryFluxThre}
\phi_{cf}^{thre} = C_{filt} \bar{|F_{st,f}|} |S_f|,
\end{equation}
where $C_{filt}$ is the filtering coefficient. It is normally set as $0.01$, which means that the capillary flux can be regarded as zero when its magnitude is less than $1\%$ of the average capillary flux. $\bar{|F_{st,f}|}$ is the magnitude of the average surface-tension force.

\subsection{The i-CLSVoF framework with evaporation}
Modelling micro-sized droplet evaporation suffers from un-physical velocities, and the i-CLSVoF framework proposed for suppressing un-physical velocities employs an extension to model droplet evaporation. In this section, three different evaporation models are introduced. These models share the same governing equations with the exception of the calculation of the source term per unit area $J$.

We start with the most simple evaporation model, where the only parameter needing to be defined is the constant mass flux per unit area $J$. In contrast to calculating the mass flux $J$ with complex equations, for instance, by temperature difference (Eqn. \ref{massFluxT}), a given constant, e.g., $ J = 1.25\times10^{-2}$ is specified in this work. The constant mass flux evaporation model can be used to validate the implementations of the governing equations before implementing complex approaches to conduct calculations of the source terms.

\begin{figure}[h]
  \begin{center}
    \includegraphics[width=0.5\textwidth]{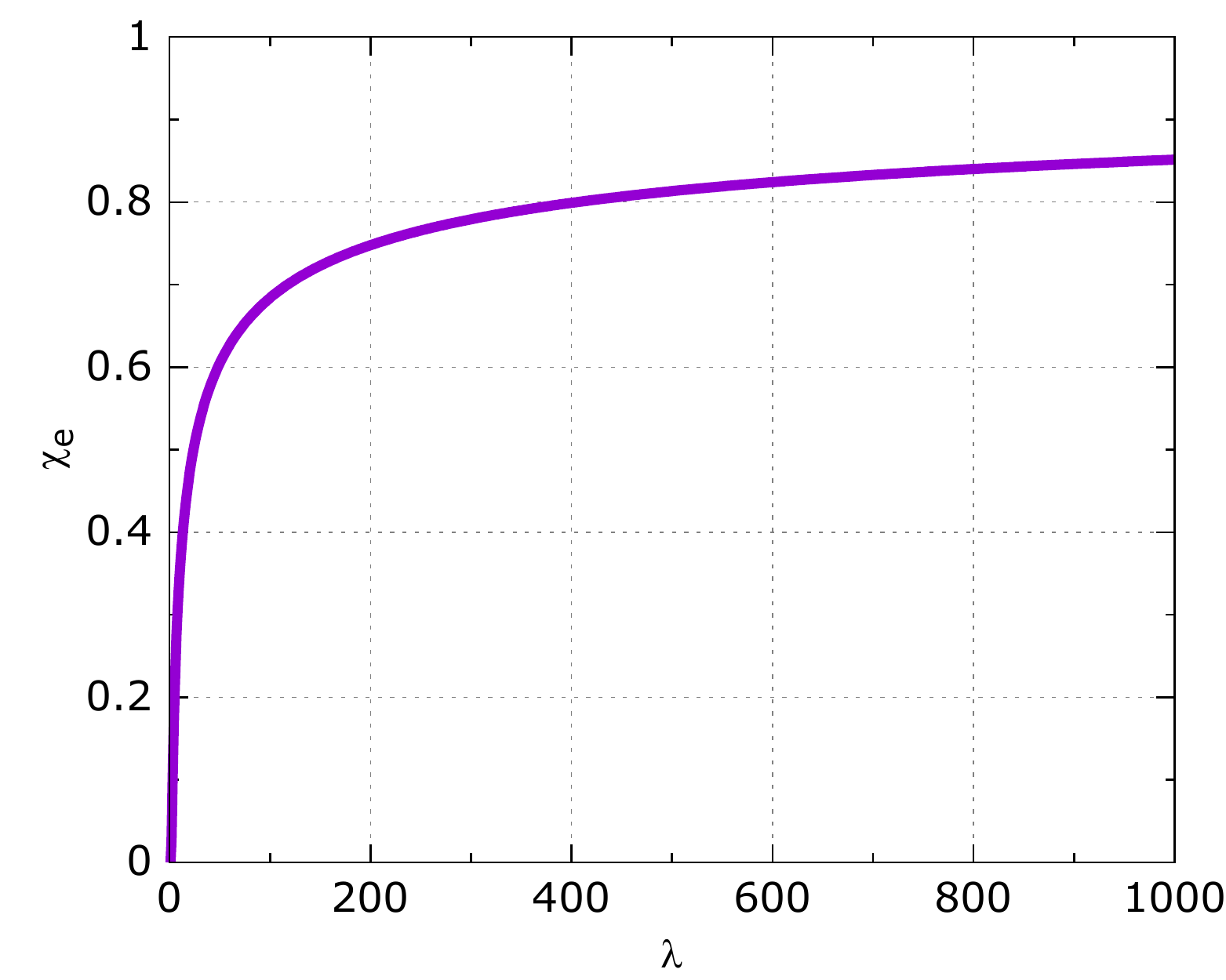}
  \end{center}
  \caption{The evaporation coefficient $\chi_{e}$ calculated by Eqn. 39 versus the density ratio $\lambda$ in the range upto $1000$.}
\label{evapCoeffFig}
\end{figure}

The thermally driven evaporation drives the phase change from liquid to vapour when the temperature around the liquid interface is higher than its saturation temperature $T_{sat}$, and the mass flux depends on the temperature difference at the interface \cite{hardt2008evaporation, montazeri2018microscopic}. The mass flux $J$ is given by
\begin{equation} \label{massFluxT}
J=\frac{T-T_{sat}}{R_{int}h_{ev}},
\end{equation}
where $h_{ev}$ is the latent heat of evaporation and $R_{int}$ the heat resistance of the liquid$–$vapour interface. $R_{int}$ is calculated by
\begin{equation} \label{RInt}
R_{int}=\frac{2-\chi_{e}}{2 \chi_{e}}\frac{\sqrt{2\pi R_{gas}}}{h_{ev}^2}\frac{T_{sat}^{3/2}}{\rho_g},
\end{equation}
where $R_{gas}$ is the gas constant and $\chi_{e}$ the evaporation coefficient which depends on the density ratio $\lambda = \rho_l/\rho_g$ between liquid and gas phase \cite{das2020sharp}. The density ratio dependent evaporation coefficient
\begin{equation} \label{evapCoeff}
\chi_{e}=\left\{1-\left(\frac{1}{\lambda} \right)^\frac{1}{3}\right\} \text{exp}\left(-\frac{1}{2 \lambda^\frac{1}{3}-2}\right),
\end{equation}
is adopted in this work instead of determining $\chi_{e}$ empirically \cite{kharangate2017review}. The evaporation coefficient $\chi_{e}$ increases gradually with the density ratio $\lambda$ and tends to reach a plateau at high density ratios (shown in Fig. \ref{evapCoeffFig}).

The third model implemented in this work is to model droplet evaporation at room temperature. The mass flux per unit area $J$ can be calculated by the mass balance across the interface for a single-component liquid as $J(1-Y ^\Gamma) = \rho_g D_v \nabla_{\Gamma} Y \mathbf{n}^\Gamma$, which leads to the formula 
\begin{equation} \label{massFluxY}
J = \frac{\rho_g D_v \nabla^{\Gamma} Y \mathbf{n}^\Gamma}{1-Y ^\Gamma}
\end{equation}
for the mass flux per unit area, where $\mathbf{n}^\Gamma$ is the unit interface normal and $\nabla^{\Gamma}$ denotes the gradient at the interface. Furthermore, $Y^\Gamma$ is the saturation vapour mass fraction, and it is given by the Clausius–Clapeyron relation
\begin{equation} \label{CCRelation1}
Y^{\Gamma}=\frac{X_v M_l}{(1-X_v)M_g+X_v M_l},
\end{equation}
where $X_v$ is further given by
\begin{equation} \label{CCRelation2}
X_v = \text{exp}[\frac{-h_{ev}M_l}{R_{gas}}(\frac{1}{T}-\frac{1}{T_{sat}})]
\end{equation}
with $M_l$ and $M_g$ being the molar mass of liquid and gas respectively  \cite{palmore2019volume}.

\begin{figure}[h]
  \begin{subfigure}[h]{0.5\textwidth}
    \includegraphics[width=\textwidth]{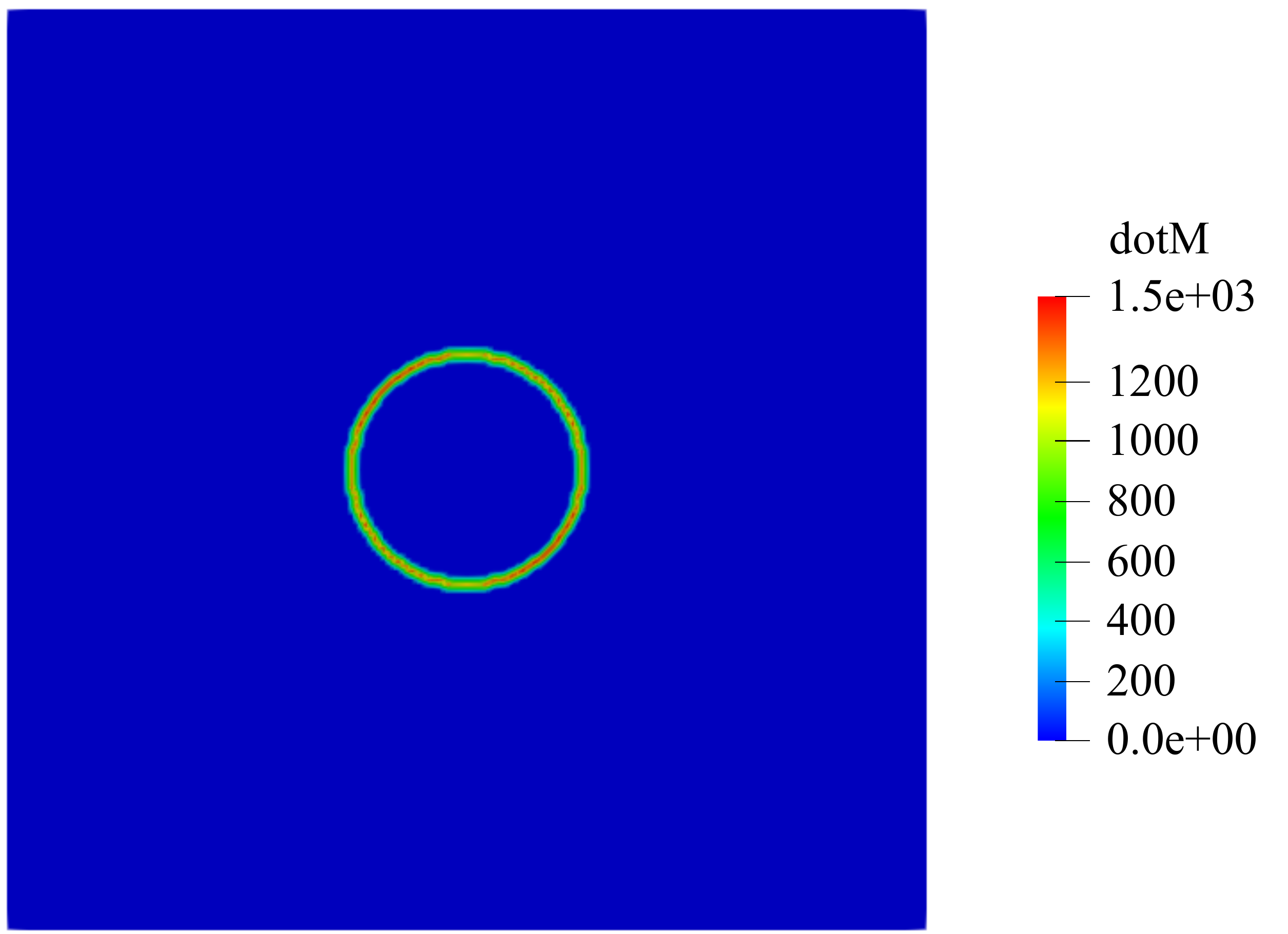}
    \caption{}
    \label{smear1}
  \end{subfigure}
  \hfill
  \begin{subfigure}[h]{0.5\textwidth}
    \includegraphics[width=\textwidth]{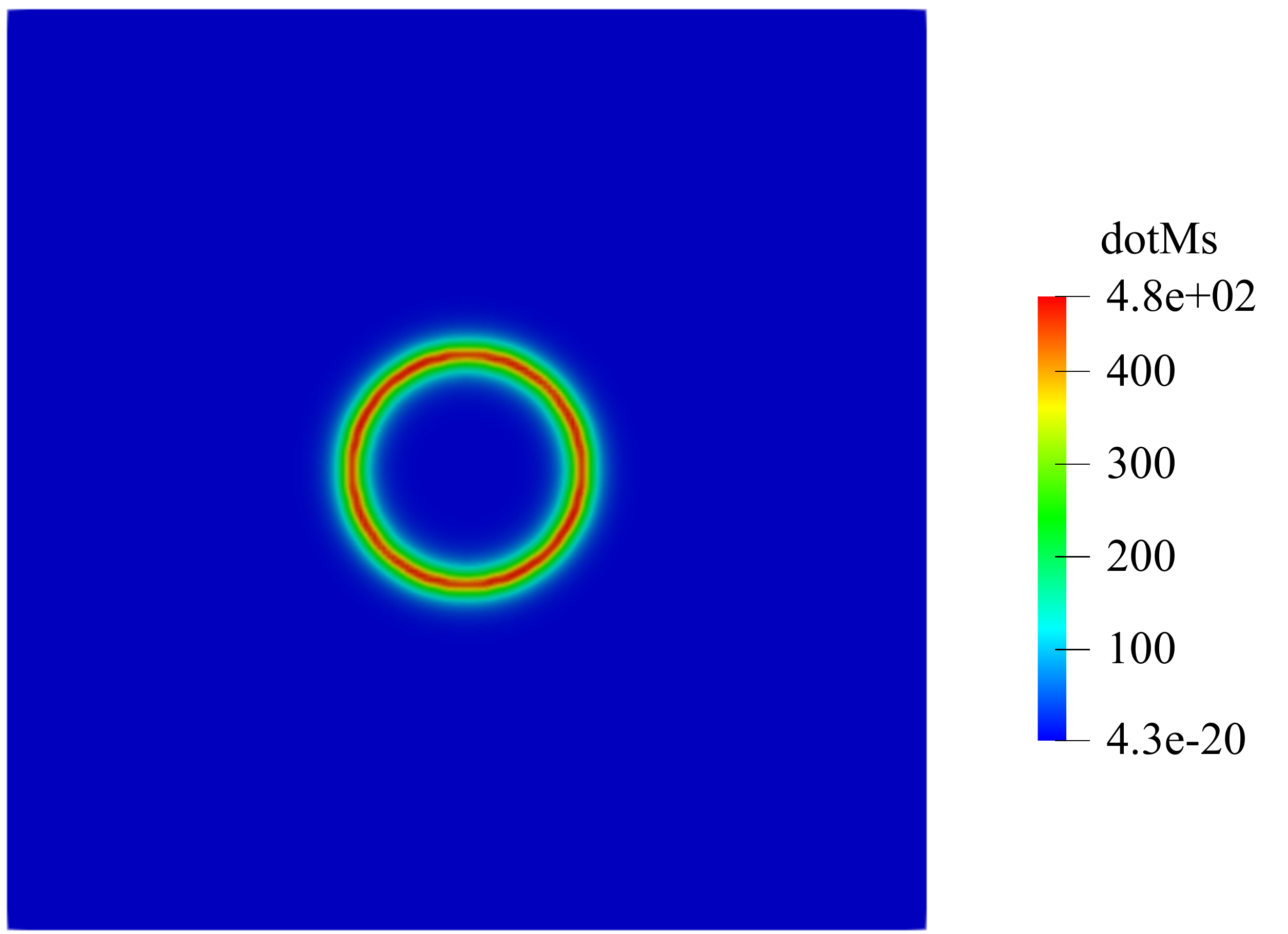}
    \caption{}
    \label{smear2}
  \end{subfigure}
  \caption{The mass source distributions for two different cases: (a) without smearing, (b) with smearing.}
\label{smear}
\end{figure}

The mass source term per unit area $\dot{m}$ can be calculated by
\begin{equation} \label{dotM}
\dot{m} = J \lvert \nabla \alpha_l \rvert
\end{equation}
once the mass source term per unit area $J$ is calculated. Normally, the mass source term $\dot{m}$ is only non-zero at a thin layer around the droplet interface (see Fig. \ref{smear1}), and our preliminary numerical study showed that it leads to numerical instability, especially for evaporation with large mass flux. 
% 2 figures side by side

\begin{figure}[h]
  \begin{center}
    \includegraphics[width=0.5\textwidth]{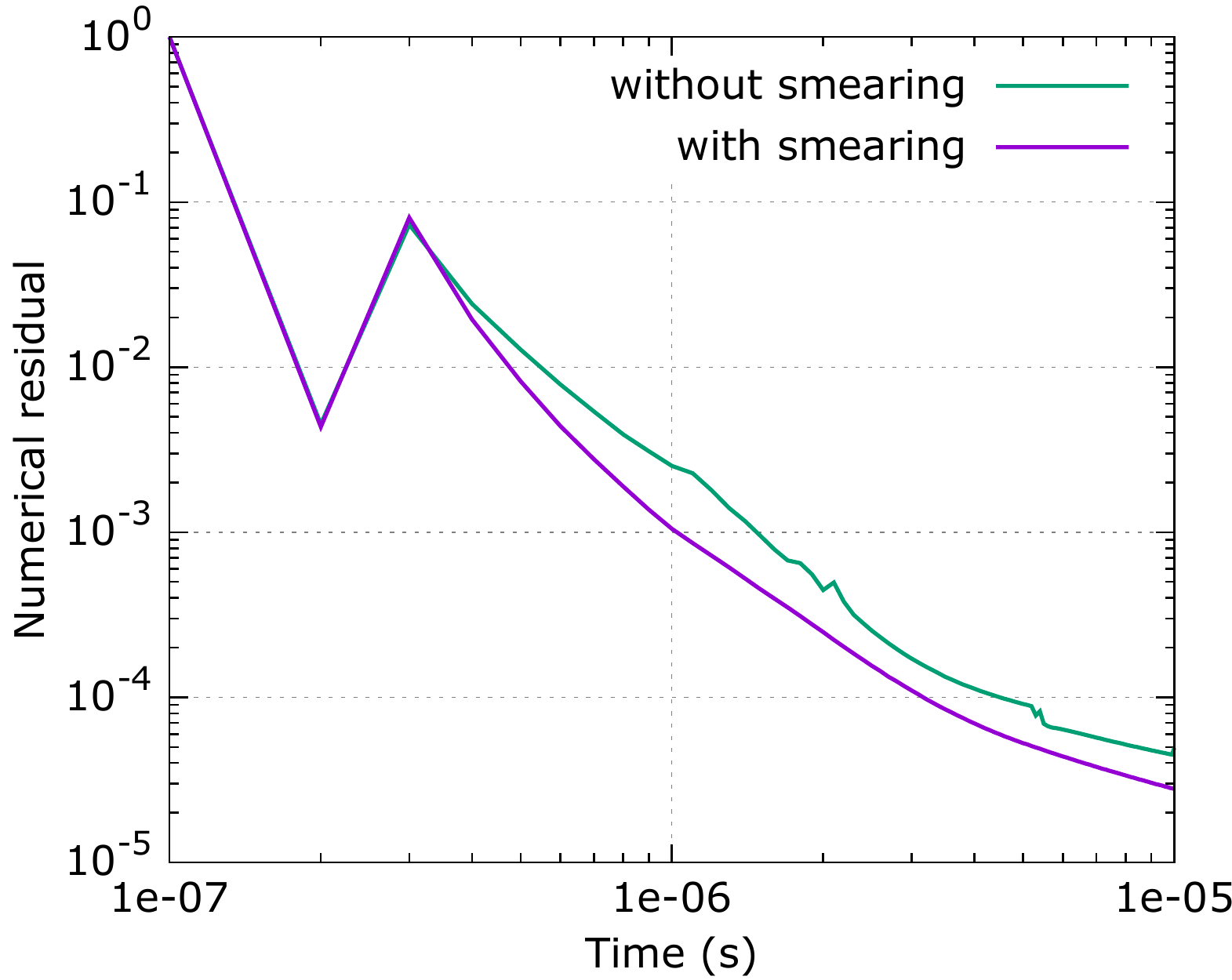}
  \end{center}
  \caption{Evolution of the numerical residual for two different cases with and without smearing the mass source term.}
\label{numResidual}
\end{figure}

Our improved solution in this work is to extend the distribution of $\dot{m}$ to a wide band by smearing $\dot{m}$ over several adjacent cells near the interface (see Fig. \ref{smear2}). The basic idea is to solve the Helmholtz equation 
\begin{equation} \label{HelmholtzEqn}
\dot m_s = \dot m + (\Delta x N)^2 \Delta {\dot m_s}
\end{equation}
with the homogeneous Neumann boundary conditions \cite{hardt2008evaporation}. Here, $\dot m_s$ is the smeared mass source term, $\Delta x$ the minimal mesh size and $N$ is the number of cell over which the mass source term is smeared along the interface normal. Employing smearing of the mass source term, the numerical residual for simple 2D static droplet evaporation cases have been recorded and it turned out that the smeared approach can guarantee a smaller numerical residual (as can be seen in Fig. \ref{numResidual}).

\subsection{The semi-discretized form of the equations}
The momentum equation after numerical discretization and linearization with the FVM in OpenFOAM is outlined in this section. The detailed derivations and numerical issues can be found in the literature \cite{jasak1996error}. The velocity field can be predicted by
\begin{equation} \label{velPredicted}
\mathbf{U} = \mathbf{U}^{\star} - \frac{1}{A_p} \nabla p_{rgh},
\end{equation}
where $p_{rgh}$ is the modified pressure field and $p_{rgh} = p-\rho \mathbf{g} \cdot \mathbf{h}$ with $\mathbf{h}$ being the position vector. $\mathbf{U}^{\star}$ is given as
\begin{equation} \label{velStar}
\mathbf{U}^{\star} = \frac{\mathbf{H}(\mathbf{u})+\mathbf{F}_b}{A_p},
\end{equation}
where $A_p$ is a scalar field corresponding to the linear algebraic equations after discretizing the momentum equation. $\mathbf{H}(\mathbf{u})$ accounts for the residual that is left after extracting the diagonal from the coefficient matrix except pressure gradient and body forces, and $\mathbf{F}_b$ is defined by $\mathbf{F}_b=\rho \mathbf{g}+\mathbf{F}_{st} - \nabla p_{st}$.

The pressure equation is then derived by substituting Eqn. (\ref{velPredicted}) into the continuity equation (\ref{continuityEqnPC}), which yields
\begin{equation} \label{discretizingEqn}
\nabla \cdot \left( \frac{1}{\langle A_p \rangle _f} \nabla p_{rgh} \right ) = \nabla \cdot \mathbf{U}^{\star}_f - \dot{m} (\frac {1}{\rho_g}-\frac {1}{\rho_l}),
\end{equation}
where $\mathbf{U}^{\star}_f$ is the predicted velocity on the cell face. It is defined as
\begin{equation} \label{velStarF}
\mathbf{U}^{\star}_f = \frac{\langle \mathbf{H}(\mathbf{u}) \rangle _f  + \langle \mathbf{F}_b \rangle _f}{\langle A_p \rangle _f},
\end{equation}
where the last term on the RHS of Eqn. \ref{discretizingEqn} accounts for the phase change from liquid to vapour, and this term vanishes for the incompressible flow without phase change. The face flux $\phi$ is then corrected as
\begin{equation} \label{faceFlux}
\phi = \mathbf{U}^{\star}_f \cdot \mathbf{S}_f - \frac{|S_f|}{\langle A_p \rangle _f} \nabla_f^{\Gamma} p_{rgh},
\end{equation}
where $\mathbf{S}_f$ is the surface vector and then the corresponding corrected velocity field is reconstructed from the corrected face flux $\phi$.

\subsection{Time step constraint for stable simulations}
The minimum time step for solving the governing equation and ensuring spurious currents do not enhance over time is estimated from two constraints. The first constraint is 
\begin{equation} \label{deltaT1}
\Delta_{t{\sigma}} < \sqrt \frac{\rho_{\text{avg}} {\Delta x}^3}{2 \pi \sigma},
\end{equation}
where $\rho_{\text{avg}}$ is the average density of the phases. It is proposed for the explicit treatment of the surface tension force term \citep{brackbill1992continuum}. Another more comprehensive time step constraint is given by 
\begin{equation} \label{deltaT2}
\Delta_{tc} < \frac{1}{2} \left(C_2 \tau_{\mu} + \sqrt{(C_2 \tau_{\mu})^2+4C_1 {\tau_{\rho}}^2} \right),
\end{equation}
which involves the density and the viscosity of the multiphase system. $\tau_{\mu} $ and $\tau_{\rho}$ are given as $\mu_{\text{avg}} \Delta x /\sigma$ and $\sqrt{\rho_{\text{avg}} \Delta x^3 / \sigma}$ respectively with $\mu_{\text{avg}}$ being the average dynamic viscosity between phases \cite{galusinski2008stability}. Accordingly, the minimum time step size for stable numerical simulations is given as
\begin{equation} \label{deltaT}
\Delta_t < \text{min}(\Delta_{t\sigma}, \Delta_{tc})C_{\Delta t}
\end{equation}
with $C_{\Delta t}$ being the stabilization factor. A range of $C_{\Delta t}$ between $0.3$ and $0.7$ is recommended for more stable constraints, especially for cases with phase change \cite{scapin2020volume}.

\subsection{The overall solution procedure}
\begin{figure}[h]
  \begin{center}
    \includegraphics[width=0.5\textwidth]{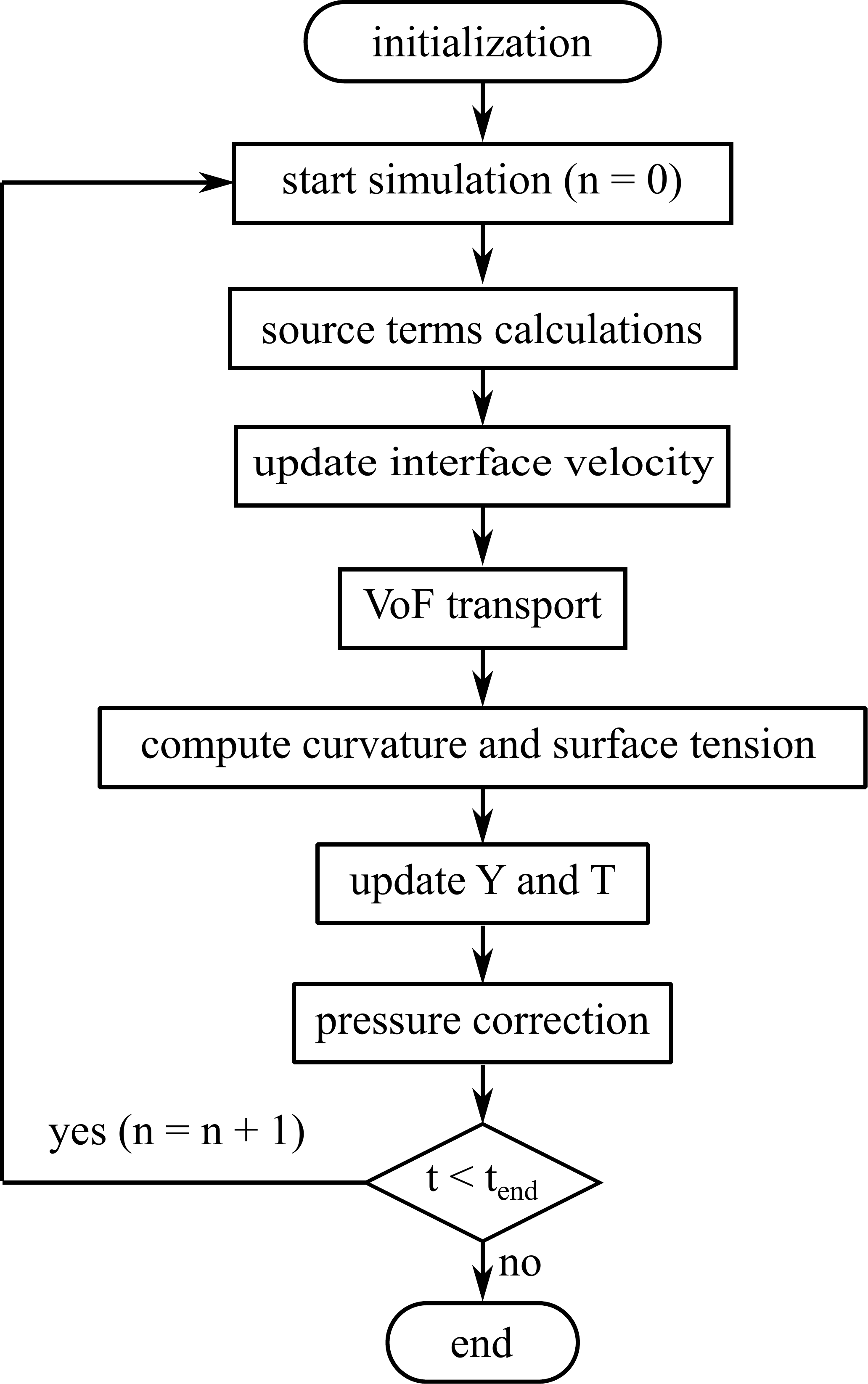}
  \end{center}
  \caption{Flowchart of the i-CLSVoF framework with evaporation.}
\label{solutionProcedure}
\end{figure}

The overall solution procedure is outlined in Fig. \ref{solutionProcedure}.
After initializing the essential fields, such as liquid volume fraction, pressure, velocity, etc., the free surface is advected by solving the VoF equation and then the LS field can also be calculated. Accordingly, the interface curvature and the surface tension can be updated then. To calculate the mass flux, the temperature distribution or the vapour mass fraction needs to be updated by solving the corresponding equations. The velocity and the pressure fields are calculated by solving the continuity and the pressure correction equations. The new divergence-free velocity and the interface velocity are reconstructed then. The reconstructed velocity field is used for solving the VoF equation in the next cycle until the pre-defined total simulation time $t_{end}$ is reached.

\section{Results and discussion}
Representing a sharp interface by the i-CLSVoF framework is demonstrated by the well-known 2D dam-break benchmark case in the Appendix. Several numerical benchmark cases are conducted to validate the i-CLSVoF framework proposed in this work. We first study the 2D droplet cases in static equilibrium to compare the suppression of un-physical spurious velocities with three different surface-tension force models (the VoF based surface tension force (Eqn. \ref{FstVoF}), the CLSVoF based surface tension force (Eqn. \ref{FstCLSVoF}) and the i-CLSVoF based surface tension force (Eqn. \ref{FstiCLSVoF})).

\subsection{Suppression of un-physical spurious velocities}
The simple yet widely-used case to demonstrate the suppression of spurious velocities is to study the surface-tension induced relaxation of a $2$D cubic droplet (density: $1000 \ kg/m^3$, viscosity: $10^{-3} \ Pa \cdot s$) immersed in a base fluid (density: $1000 \ kg/m^3$, viscosity: $10^{-3} \ Pa \cdot s$) \citep{raeini2012modelling}. A constant surface tension coefficient is assumed as $0.07 \ N/m$. The initial configuration is a $2$D square droplet (side length: $40 \ \mu m$) sitting at the centre of a square $2$D computational domain (side length: $100 \ \mu m$). Gravity is absent, and the surface tension force is the only external force acting on the droplet. Accordingly, the surface tension gradually deforms the droplet from its initial square shape to its equilibrium shape, i.e., a $2$D circle. 

The maximum velocity of the system is recorded, and we compare the numerical results of our i-CLSVoF framework to simulation results with the conventional VoF and CLSVoF methods. As an alternative to track the maximum velocity $|\mathbf{U}_{max}|$, the dimensionless Capillary number $Ca = \frac{\mu |\mathbf{U}_{max}|}{\sigma}$ is used to quantify the evolution of spurious velocities in the literature \cite{shams2018numerical}. The convergence of the Capillary number with our i-CSLVoF framework is also promising as the ratio between liquid dynamic viscosity $\mu=10^{-3}$ and surface-tension coefficient $\sigma=0.07$ is smaller than unity in our benchmark cases. Concerning the total simulation time for droplet relaxation, $0.001 \ s$ is enough to guarantee the maximum velocity converges to zero numerically (less than $1.0 \times 10^{-8}$ in our model).
\begin{figure}[h]
  \begin{center}
    \includegraphics[width=0.5\textwidth]{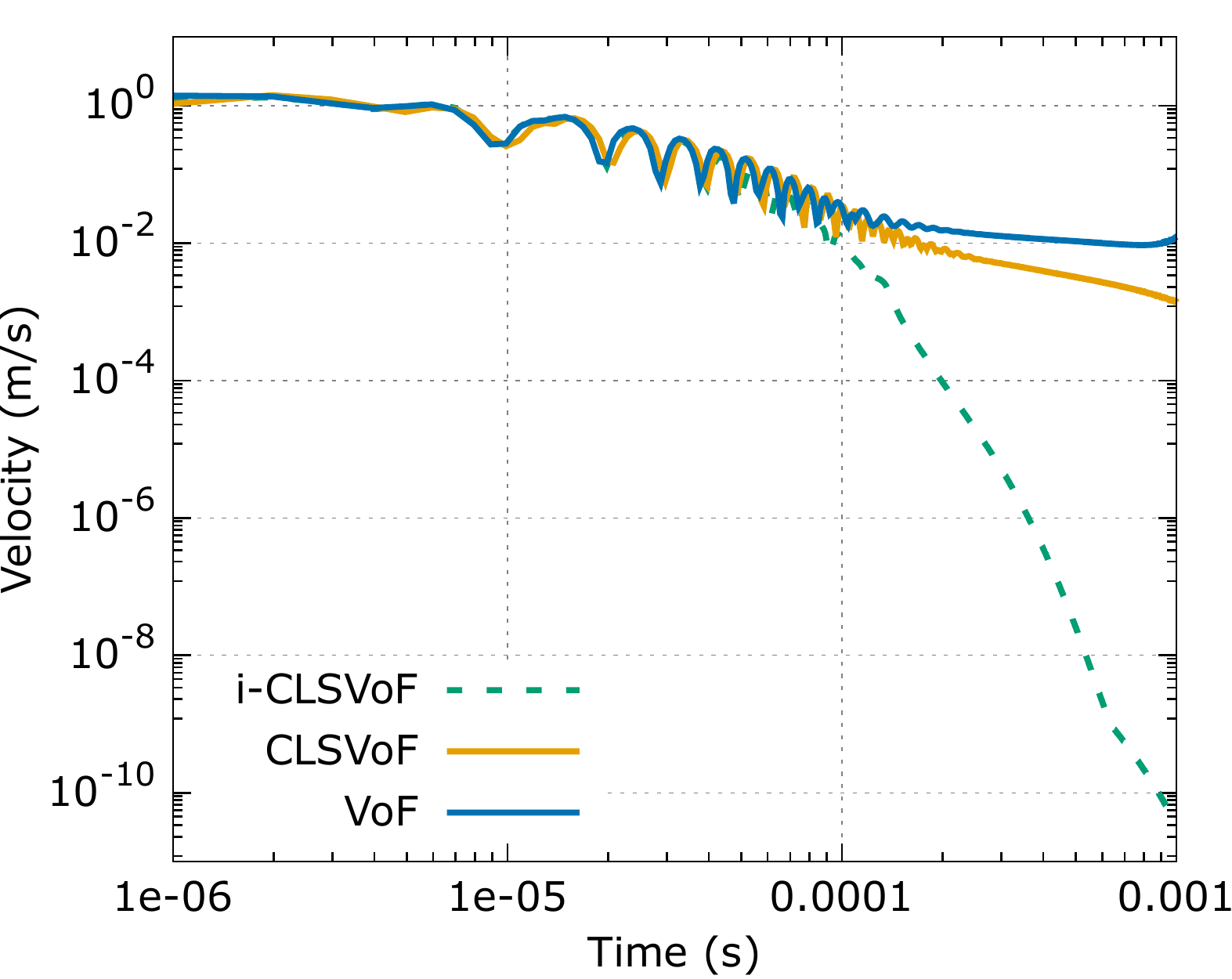}
  \end{center}
  \caption{Evolution of the un-physical velocities with three different surface tension force models.}
\label{velEvolution}
\end{figure}
As shown in Fig. \ref{velEvolution}, the conventional CLSVoF approach can improve the suppression of un-physical spurious velocity better than the VoF approach; however, the un-physical velocities are still too large to guarantee numerical stabilities. The dashed line represents the convergence of the velocity within one millisecond by the i-CLSVoF method, and the velocity converges to $10^{-10} \ m/s$ which is small enough to eliminate the influence of un-physical spurious currents on the numerical stabilities. 

\begin{figure}[h]
  \begin{subfigure}[h]{0.5\textwidth}
    \includegraphics[width=\textwidth]{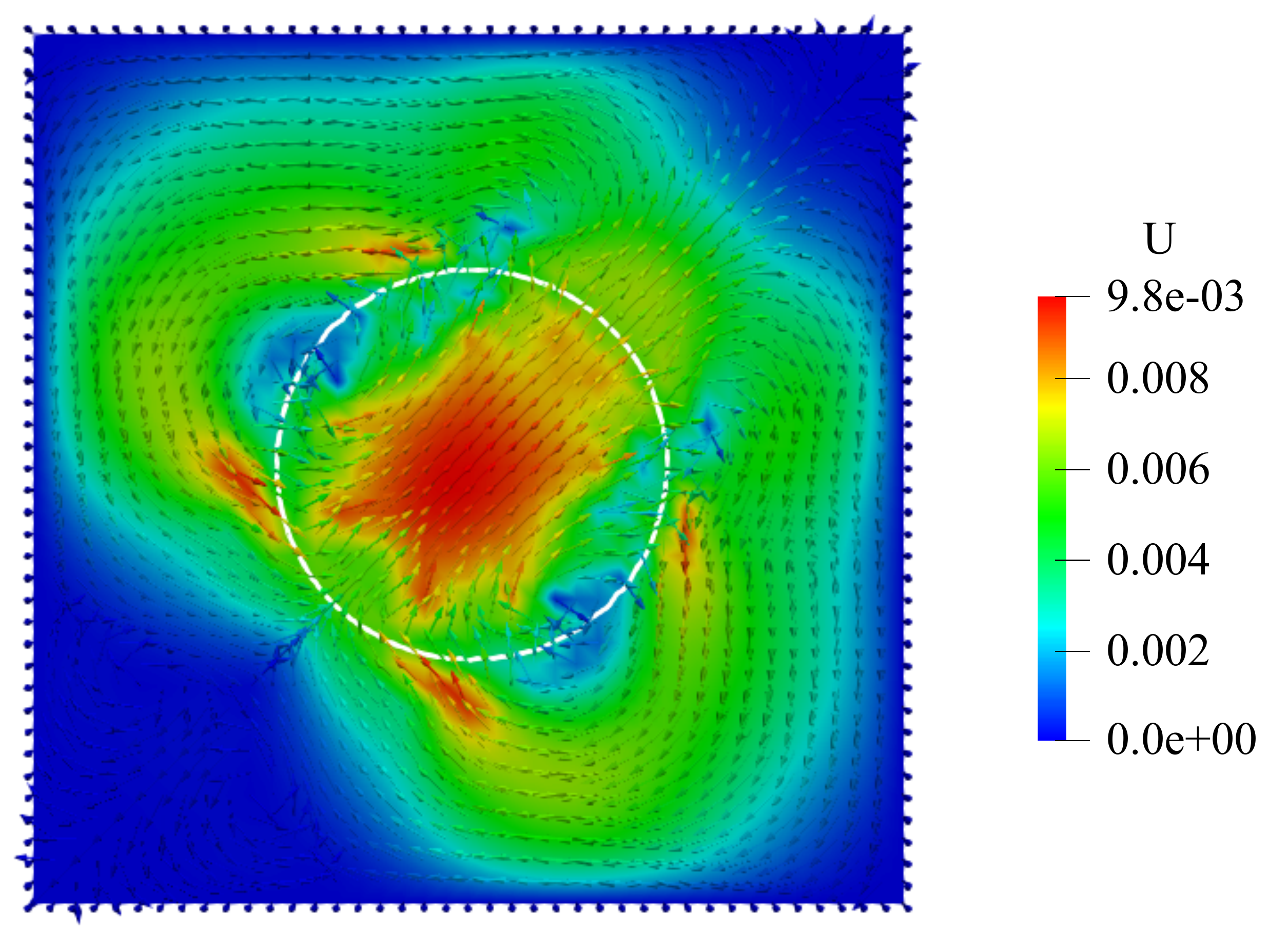}
    \caption{}
    \label{velContour1}
  \end{subfigure}
  \hfill
  \begin{subfigure}[h]{0.5\textwidth}
    \includegraphics[width=\textwidth]{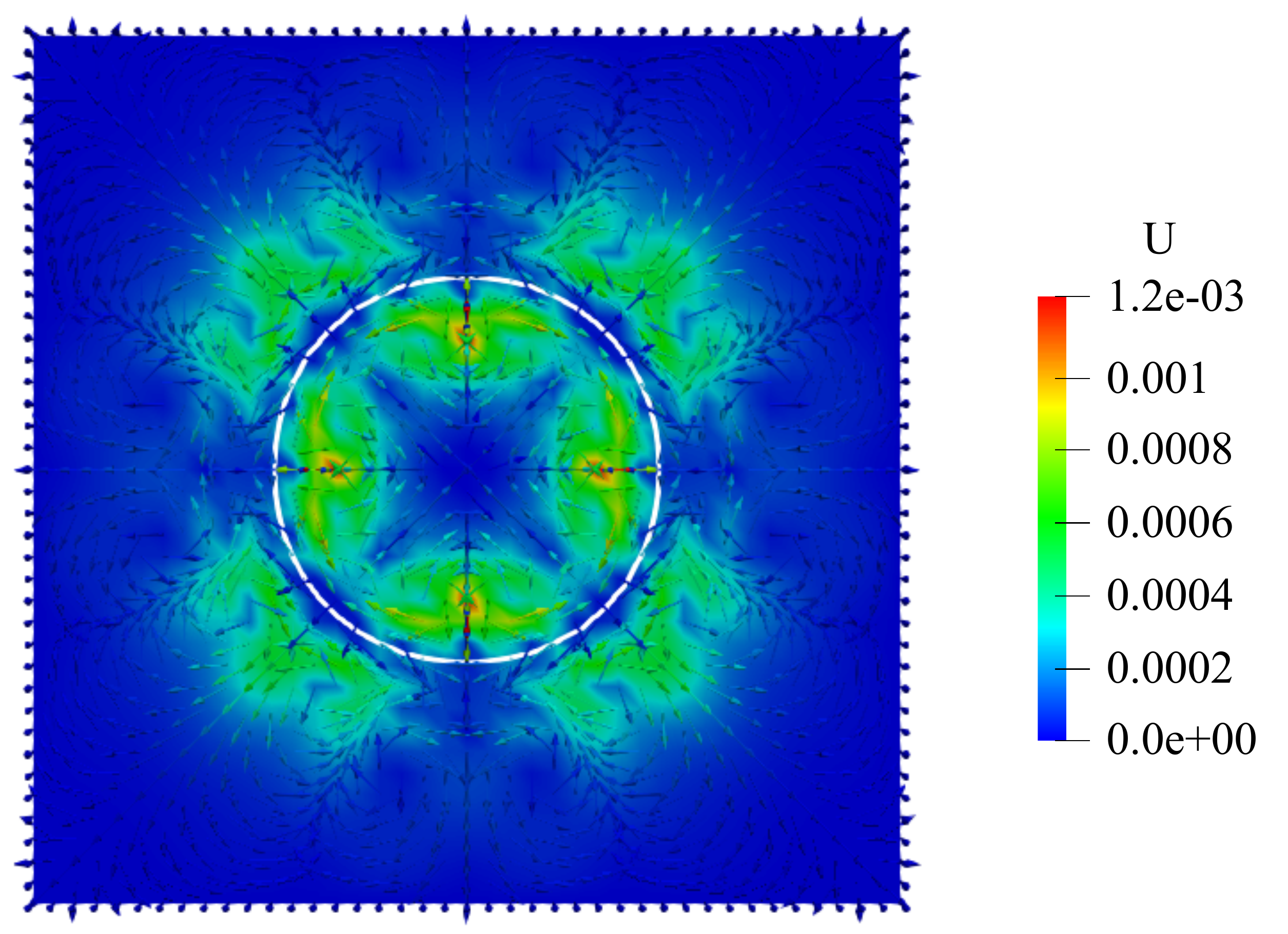}
    \caption{}
    \label{velContour2}
  \end{subfigure}
  \hfill
   \begin{subfigure}[h]{\linewidth}
   \centering
    \includegraphics[width=0.5\textwidth]{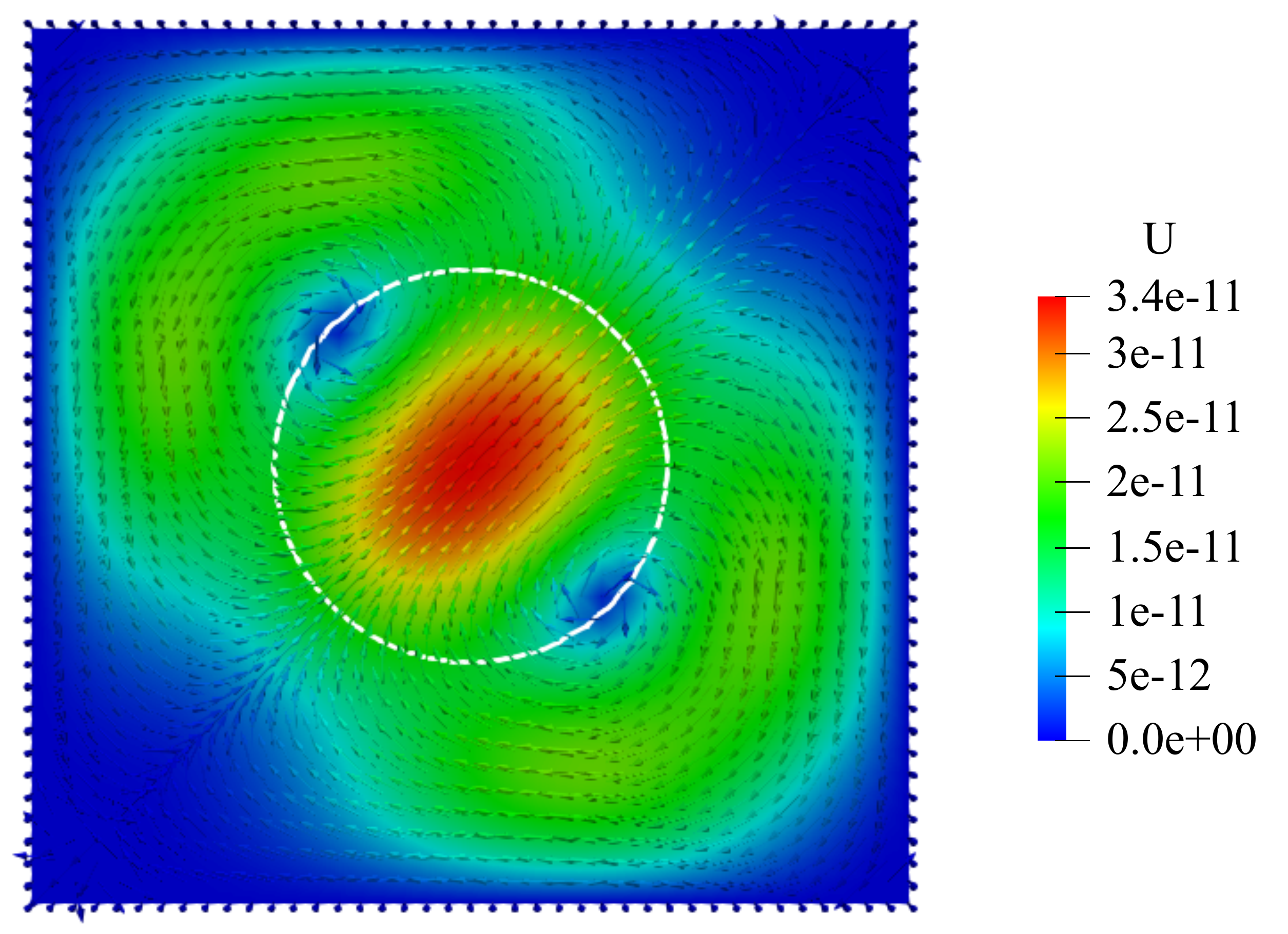}
    \caption{}
    \label{velContour3}
  \end{subfigure}
  \caption{Velocity vector contours at 0.001 s with three different surface tension models (white circle represents the $0.5$ iso-surface for $\alpha_l$ field): (a) VoF, (b) CLSVoF, (c) i-CLSVoF.}
\label{velContour}
\end{figure}

The corresponding velocity vector contours at $0.001 \ s$ with three different surface tension models (the VoF based surface tension force (Eqn. \ref{FstVoF}), the CLSVoF based surface tension force (Eqn. \ref{FstCLSVoF}) and the i-CLSVoF based surface tension force (Eqn. \ref{FstiCLSVoF}) are given in Fig. \ref{velContour}, and the white circles represent the $0.5$ iso-surface for the liquid volume fraction field $\alpha_l$. The distributions of spurious vortices for the three different surface-tension models are different. For the normal VoF approach, four large spurious vortices appear around the free surface and point toward four different directions, which lead to the strong spurious velocities that deform the free surface of the droplet and then move the droplet randomly away from its centre. Concerning the spurious vortices of the CLSVoF approach, we can also see four main vortices pointing inward but the overall distribution is symmetrical along with the horizontal and vertical directions. For the simulation with the i-CLSVoF method, the maximum velocity is located inside the droplet, however, no large spurious vertices are found around the free surface. 
% 3 figures side by side
The velocity distribution is symmetrical along with the diagonal of the computational domain, and the magnitude of the maximum velocity is tiny enough to avoid the influence of spurious currents on the numerical stability.

\begin{figure}[h]
  \begin{subfigure}[h]{0.495\textwidth}
    \includegraphics[width=\textwidth]{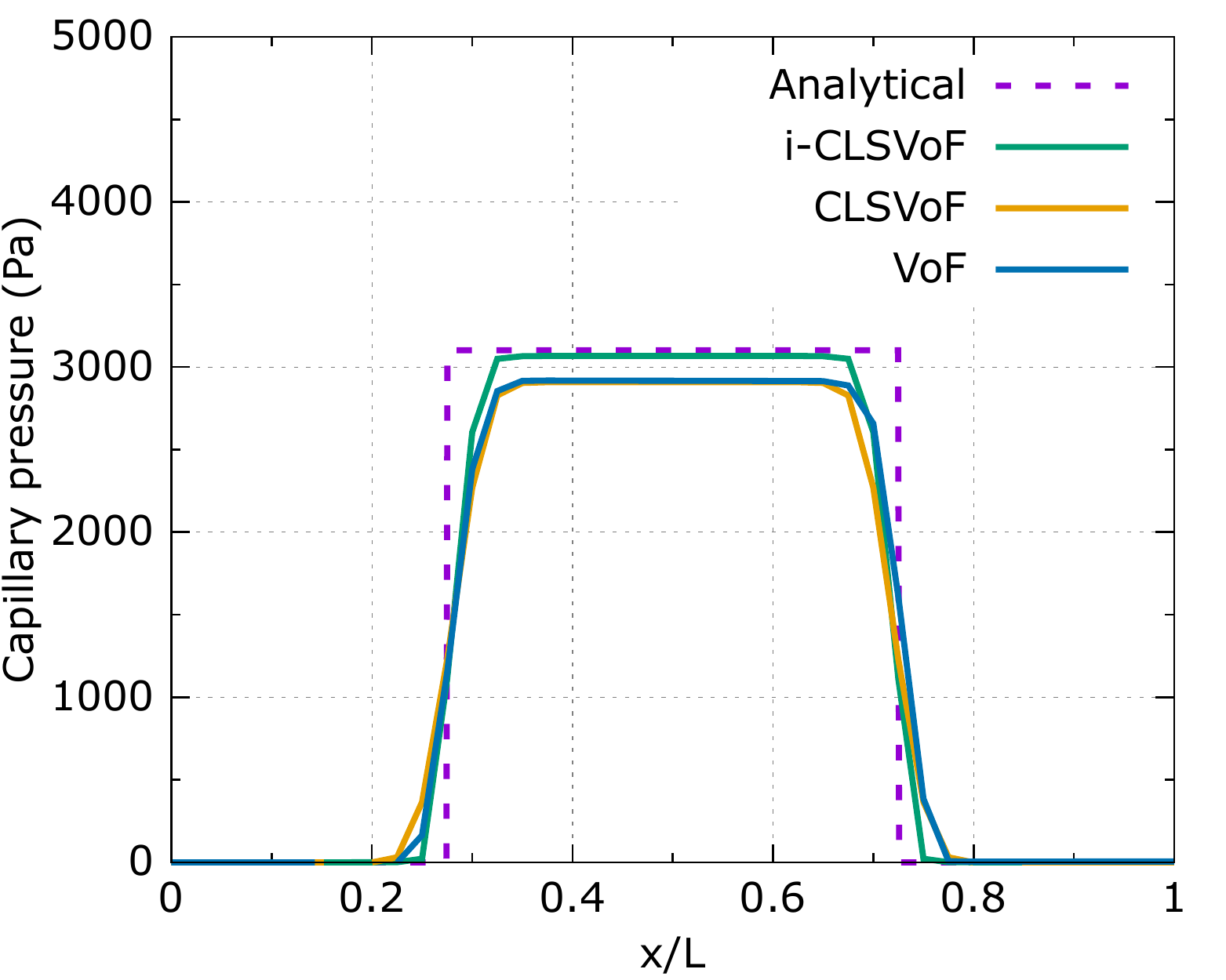}
    \caption{}
    \label{pcJump1}
  \end{subfigure}
  \hfill
  \begin{subfigure}[h]{0.495\textwidth}
    \includegraphics[width=\textwidth]{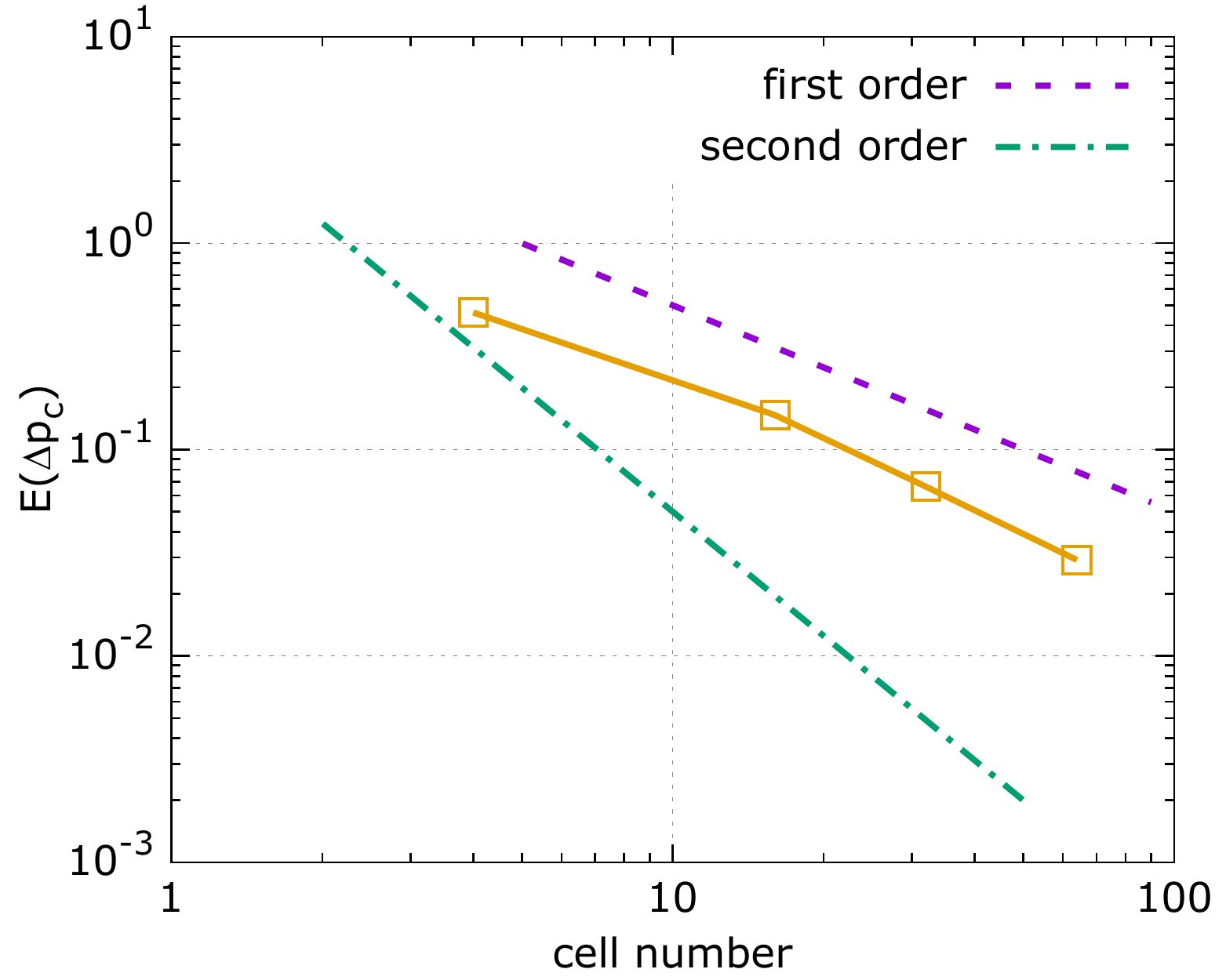}
    \caption{}
    \label{pcJump2}
  \end{subfigure}
  \caption{(a) Capillary pressure fields of droplets with three different surface tension force models (dotted purple line represents the analytical solution), (b) Relative errors for capillary pressure jump with different mesh resolutions.}
\label{pcJump}
\end{figure}

The analytical solution of the capillary pressure jump across droplets is given by the Young–Laplace equation, and $\Delta p_c=\frac{\sigma}{R}=\frac{0.07}{40/\sqrt{\pi}}=3101.8 \ Pa$ is the theoretical solution for 2D droplets (where $R$ is the droplet radius) \cite{shams2018numerical}. The relative error for capillary pressure jump is given by
\begin{equation} \label{relError}
E(\Delta p_c) = \frac{|p_c^n-p_c^a|}{p_c^a},
\end{equation}
where $p_c^a$ and $p_c^n$ are the analytical and numerical capillary pressure jump, respectively. The comparison among three different surface-tension force models in predicting capillary pressure is shown in Fig. \ref{pcJump1}. The i-CLSVoF developed in this work demonstrates the best agreement between numerical prediction and the analytical solution. The quantitative study on the relative error in predicting capillary pressure jump with the i-CLSVoF framework is shown in Fig. \ref{pcJump2}, and first-order convergence is found with our model.

The sharpening coefficient $C_{pc}$ is a key parameter governing the suppression of un-physical spurious currents. An additional parameter study demonstrates the effect of the sharpening coefficient on the suppression of the spurious currents. 
\begin{figure}[h]
  \begin{center}
    \includegraphics[width=0.5\textwidth]{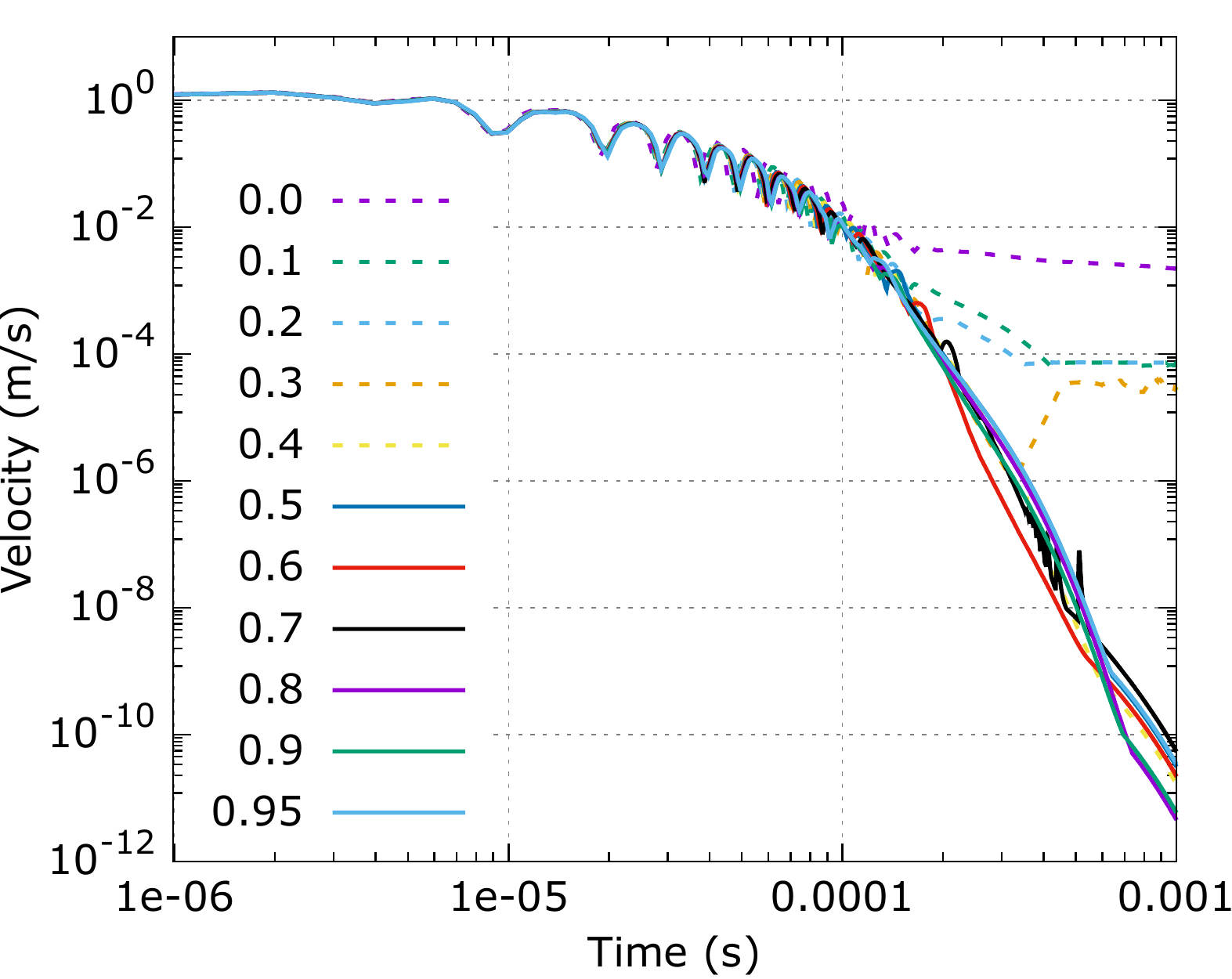}
  \end{center}
  \caption{Effect of the sharpening coefficient $C_{pc}$ on the suppression of spurious velocity.}
\label{sharpeningCoeff}
\end{figure}
As shown in Fig. \ref{sharpeningCoeff}, the suppression of un-physical velocities is improved with increasing sharpening coefficient $C_{pc}$, especially for relatively smaller sharpening coefficient values ($0.1$ - $0.4$). However, for a relatively large sharpening coefficient ($0.5$ - $0.95$), the corresponding results give no major improvement. Note that too large sharpening coefficient may lead to numerical instability. 

\subsection{Numerical validation of the i-CLSVoF framework with evaporation}
\begin{figure}[h]
  \begin{center}
    \includegraphics[width=0.5\textwidth]{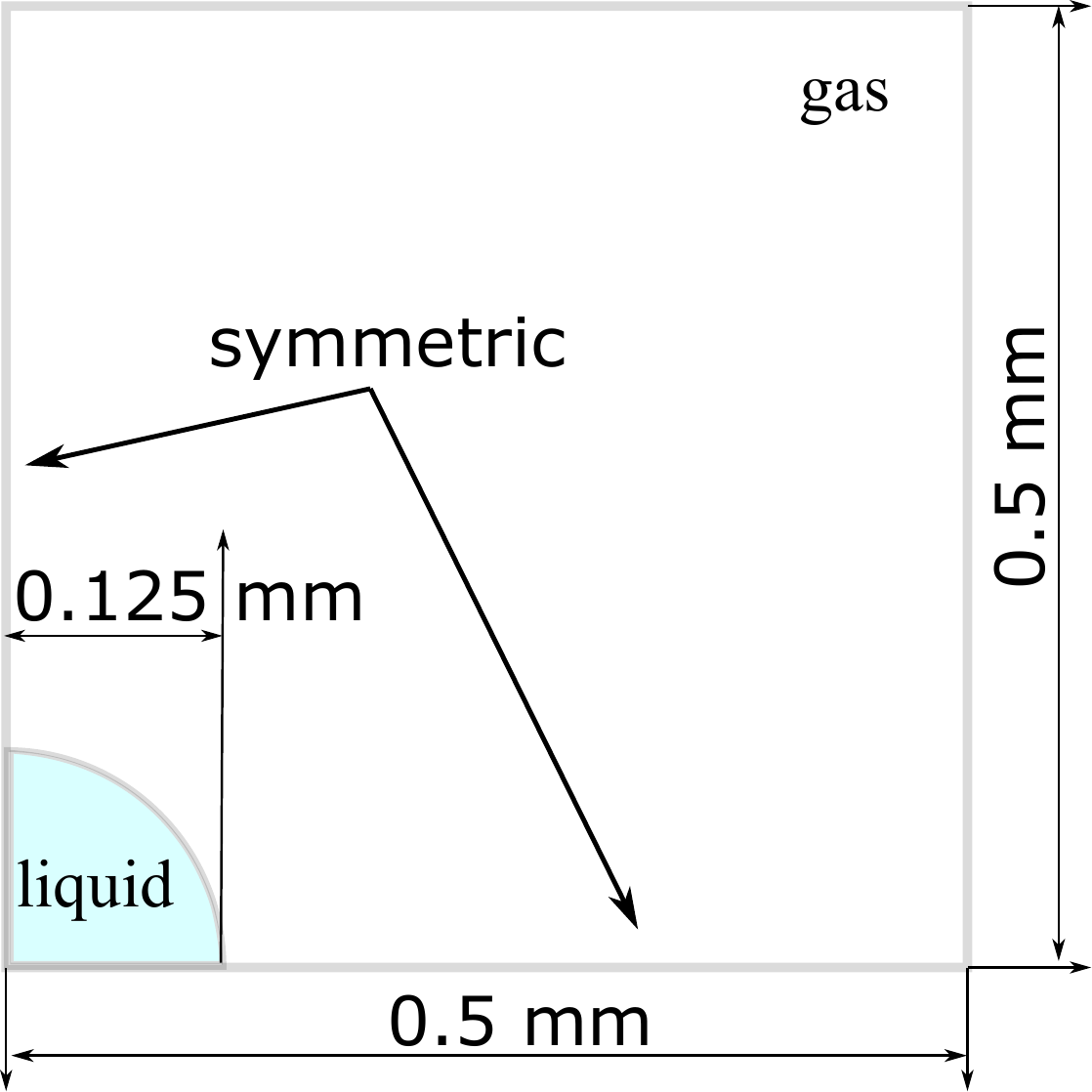}
  \end{center}
  \caption{The numerical setup for $2D$ droplet evaporation.}
\label{numSetUp}
\end{figure}

Several benchmark cases are conducted to validate the i-CLSVoF framework with evaporation. For saving computational cost, symmetrical model is adopted in the current study. As shown in Fig. \ref{numSetUp}, only a quarter of a $2$D droplet with an initial radius of $125 \ \mu m$ is investigated, and symmetry boundary conditions on the left and bottom sides are applied. Outflow boundary conditions (a Dirichlet boundary condition for the pressure field and a Newmann boundary condition for velocity field) are applied on the rest two sides so as to let the newly generated vapour from the liquid surface leave the domain freely.

The parameters used in the evaporation cases are listed in Table \ref{paras}. In order to save the computational cost, the scaled liquid density $10$ is used in the current study, and this density scaling approach was also used in several previous studies \cite{ledesma2014lattice, malan2021geometric, scapin2020volume, irfan2017front}. A constant surface-tension coefficient is used here, meaning that the effect of temperature on the surface-tension coefficient is not considered in this work.

\begin{table}[h]
\normalsize
\centering
\caption{Physical properties for liquid and gas phases.}
\label{paras}
\begin{tabular}{lccll}
\cline{1-4}
Property                    & Liquid      & Gas    & Units      &  \\ \cline{1-4}
Density $\rho$              & 10 & 1   & {[}$kg \cdot m^{-3}${]} &  \\
Dynamic viscosity $\mu$     & $1\times10^{-3}$  & $1\times10^{-5}$ &{[}$Pa \cdot s${]}&  \\
Thermal conductivity $k$    & 0.1         & 0.01   &{[}$W \cdot m^{-1} \cdot K^{-1}${]}&  \\
Specific heat capacity $c_p$      & 4181        & 1900   &{[}$J \cdot kg^{-1} \cdot K^{-1}${]}&  \\
Molar mass $M$              & 0.018 & 0.029   & {[}$kg/mol${]} &  \\
Entropy of evaporation $h_{ev}$ & $1\times10^{6}$  & - &{[}$J \cdot kg^{-1}${]}&  \\
Surface-tension coefficient $\sigma$ & 0.072       & -   &{[}$N \cdot m^{-1}${]}&  \\ 
Vapour diffusivity $D_v$ & -  & $1\times10^{-5}$ &{[}$m^2/s${]}&  \\ \cline{1-4}
\end{tabular}
\end{table}

\subsubsection{Sessile droplet evaporation with the constant mass flux evaporation model}
First, we study the evaporation of $2$D sessile droplets with the pre-defined constant mass flux $J$, as we can validate the implementation of the governing equations in simple manner without taking the calculation of source terms into account. The analytical solution for $2$D sessile droplet evaporation with constant mass flux is derived in this section. Let $R$ and $R_0$ be the shrinking and the initial droplet radius, respectively. The droplet shrinks when the evaporation moves the interface inwards with the interface velocity $\mathbf{U}_{\Gamma}$. For the $2$D sessile droplet cases, the interface velocity $\mathbf{U}_{\Gamma}$ equals to the interface recession velocity $\mathbf{U}_{re}$, leading to
\begin{equation} \label{dropR}
R=R_0-|\mathbf{U}_{\Gamma}|t=R_0-\frac{J} {\rho_l}t.
\end{equation}
Let $D$ and $D_0$ denote the shrinking and the initial droplet diameter, respectively,  and $t^*$ the total evaporation time. The dimensionless droplet diameter changes with the dimensionless time during the evaporation process which is given as
\begin{equation} \label{dropD}
\frac{D}{D_0}=1-\frac{2J}{\rho_l D_0}t = 1- \frac{2Jt^*}{\rho_l D_0} \frac{t}{t^*}.
\end{equation}
This formula is the analytical solution for $2$D static droplet evaporation subject to the constant mass flux and is also valid for $3$D static evaporation cases.

\begin{figure}[h]
  \begin{subfigure}[h]{0.495\textwidth}
    \includegraphics[width=1.0\textwidth]{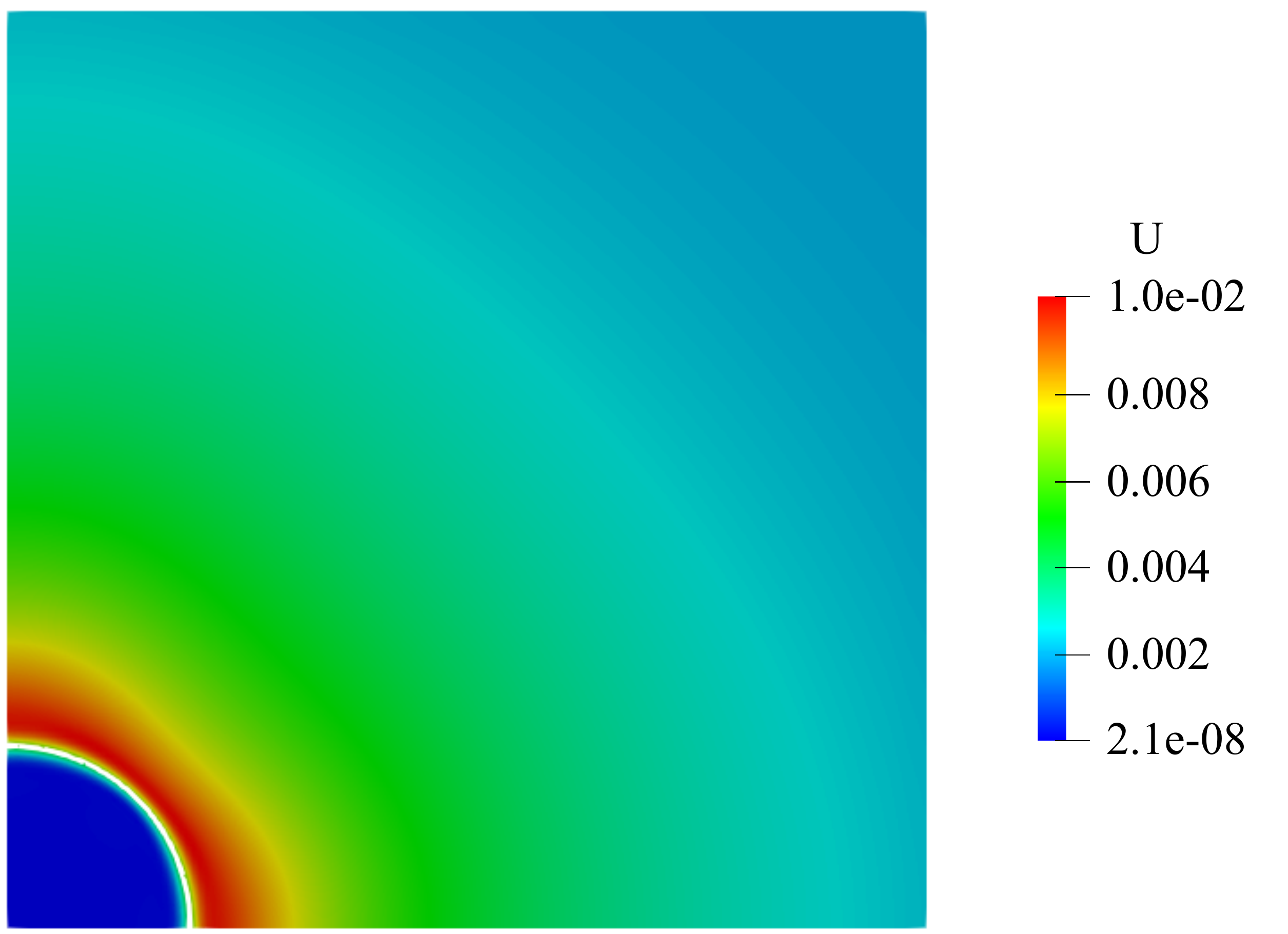}
    \caption{}
    \label{velField1}
  \end{subfigure}
  \hfill
  \begin{subfigure}[h]{0.495\textwidth}
    \includegraphics[width=1.0\textwidth]{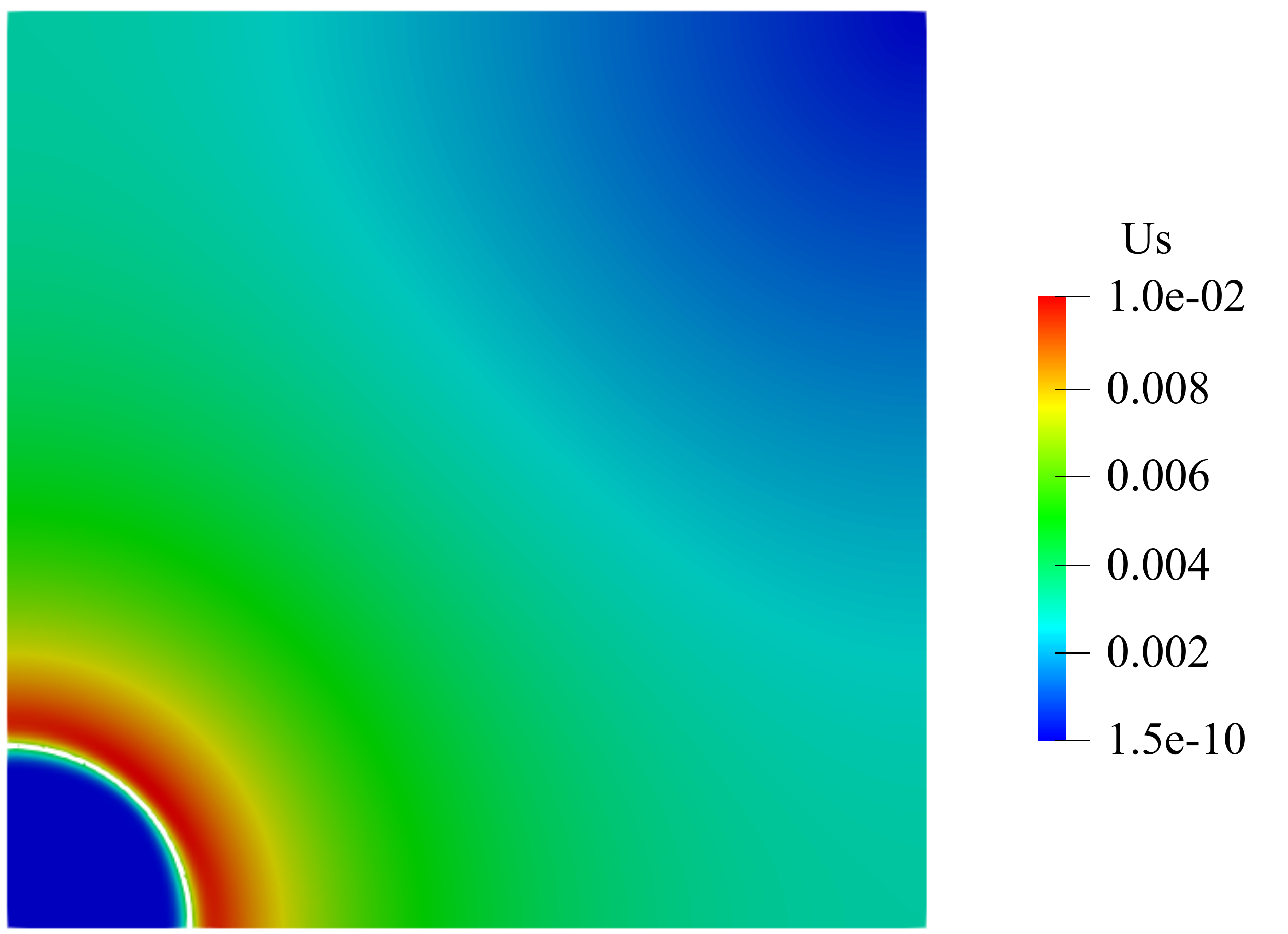}
    \caption{}
    \label{velField2}
  \end{subfigure}
   \hfill
  \begin{subfigure}[h]{0.495\textwidth}
    \includegraphics[width=1.0\textwidth]{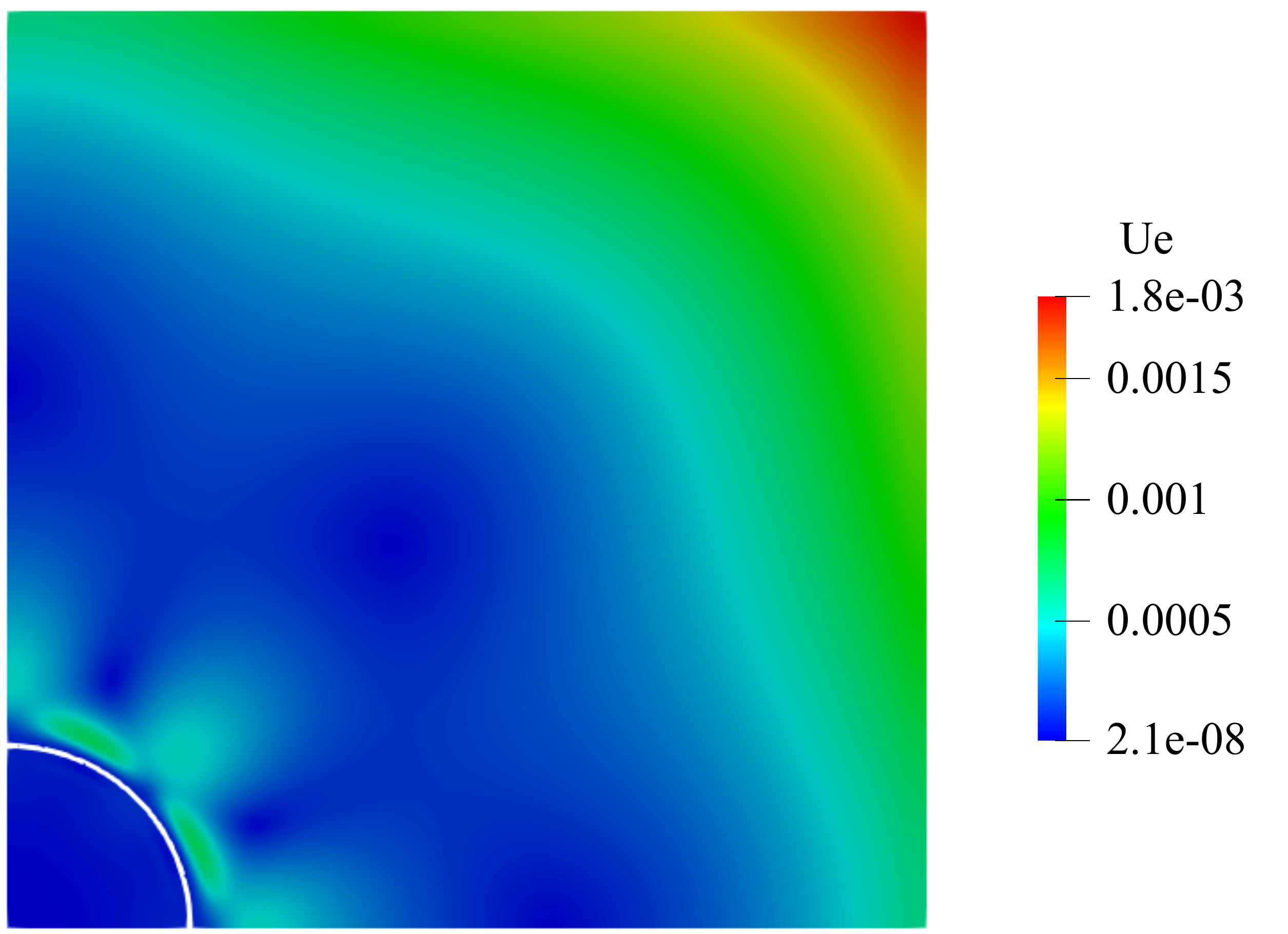}
    \caption{}
    \label{velField3}
  \end{subfigure}
   \hfill
  \begin{subfigure}[h]{0.495\textwidth}
    \includegraphics[width=1.0\textwidth]{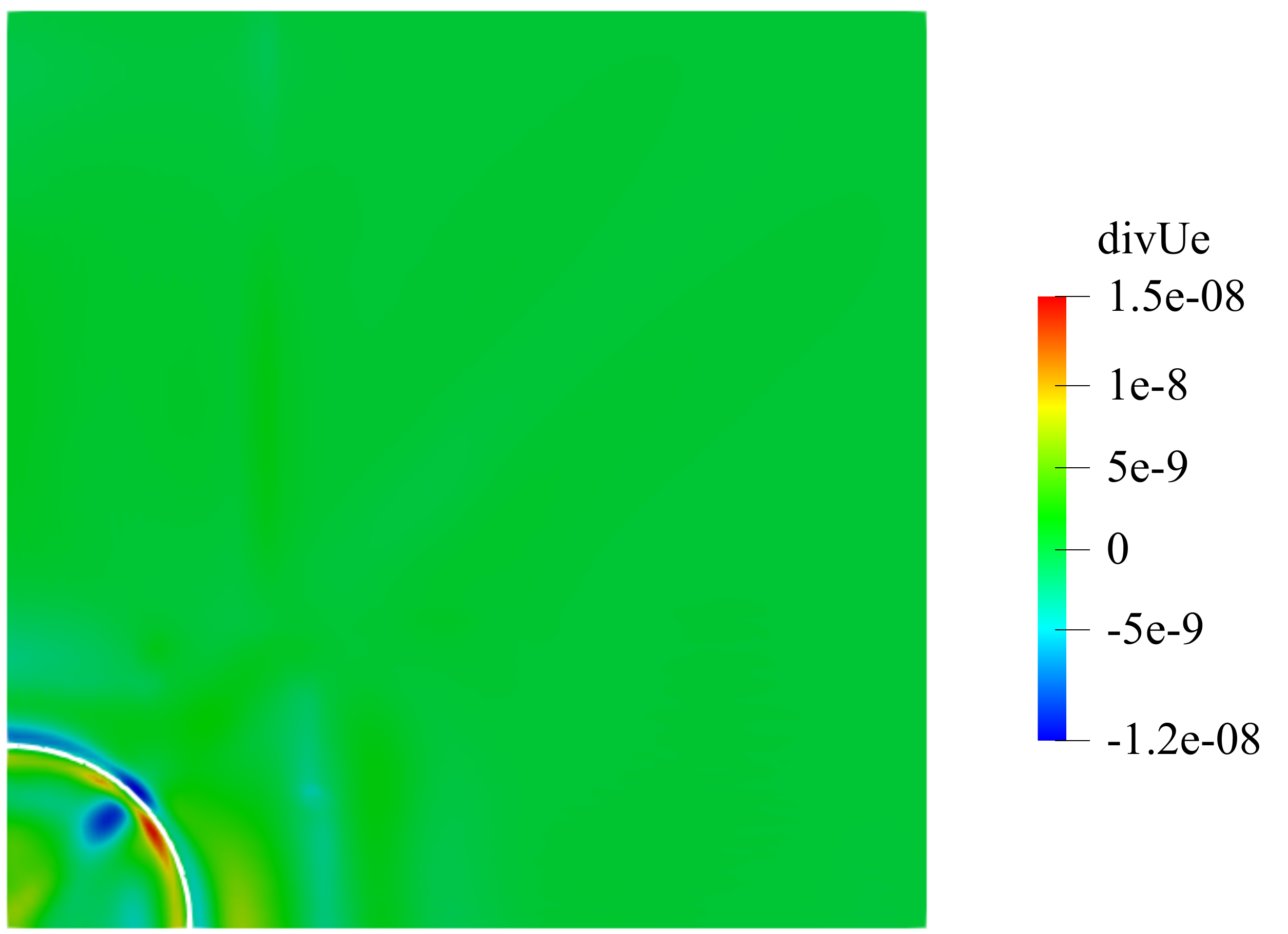}
    \caption{}
    \label{velField4}
  \end{subfigure}
  \caption{Different velocities fields of an evaporating droplet (white curve represents the interface): (a) the one-field velocity field $\mathbf{U}$, (b) the evaporation-induced Stefan flow velocity field $\mathbf{U_s}$, (c) the divergence-free velocity field $\mathbf{U_e}$, (d) the divergence of $\mathbf{U_e}$.}
\label{velField}
\end{figure}

The crucial aspect of modelling droplet evaporation with the i-CLSVoF framework lies in reconstructing the new divergence-free velocity field $\mathbf{U_e}$. After implementing the velocity-potential approach (refer to section 2.2, Eqn. \ref{velPotEqn} and Eqn. \ref{Ue}) proposed in this work for the reconstruction of the new velocity field, the three different velocity fields are obtained by the simulations: the one-field velocity $\mathbf{U}$, the evaporation-induced Stefan flow velocity $\mathbf{U_s}$ and the newly reconstructed velocity $\mathbf{U_e}$ for a $2$D droplet subject to the constant mass flux are shown in Fig. \ref{velField}. 
% 4 figures (2*2)
The maximum magnitudes of $\mathbf{U}$ and $\mathbf{U_s}$ are the same. We can also see the contour of the divergence of velocity field $\mathbf{U_e}$ (as shown in Fig. \ref{velField4}), demonstrating that its magnitude is around $10^{-8}$ which can be regarded as numerically zero. This means that the divergence-free velocity field $\mathbf{U_e}$ is successfully reconstructed. The divergence-free velocity field $\mathbf{U_e}$ for the sessile droplet evaporation case should be zero, but some velocity vertices can be seen around the interface (Fig. \ref{velField3}). The reason lies in both evaporation and surface tension deforming the interface during the evaporation process, where the interplay between evaporation and surface tension produces additional spurious velocities. These additional spurious currents are challenging to reduce, especially with the algebraic VoF approach of OpenFOAM on which our i-CLSVoF framework is based. Furthermore, the distribution of the spurious velocities is symmetrical along the diagonal of the computational domain (as shown in Fig. \ref{velField3}). A symmetrical distribution of spurious velocities around an evaporating droplet is more stable than the case with random distribution. 

\begin{figure}[h]
  \begin{center}
    \includegraphics[width=0.5\textheight]{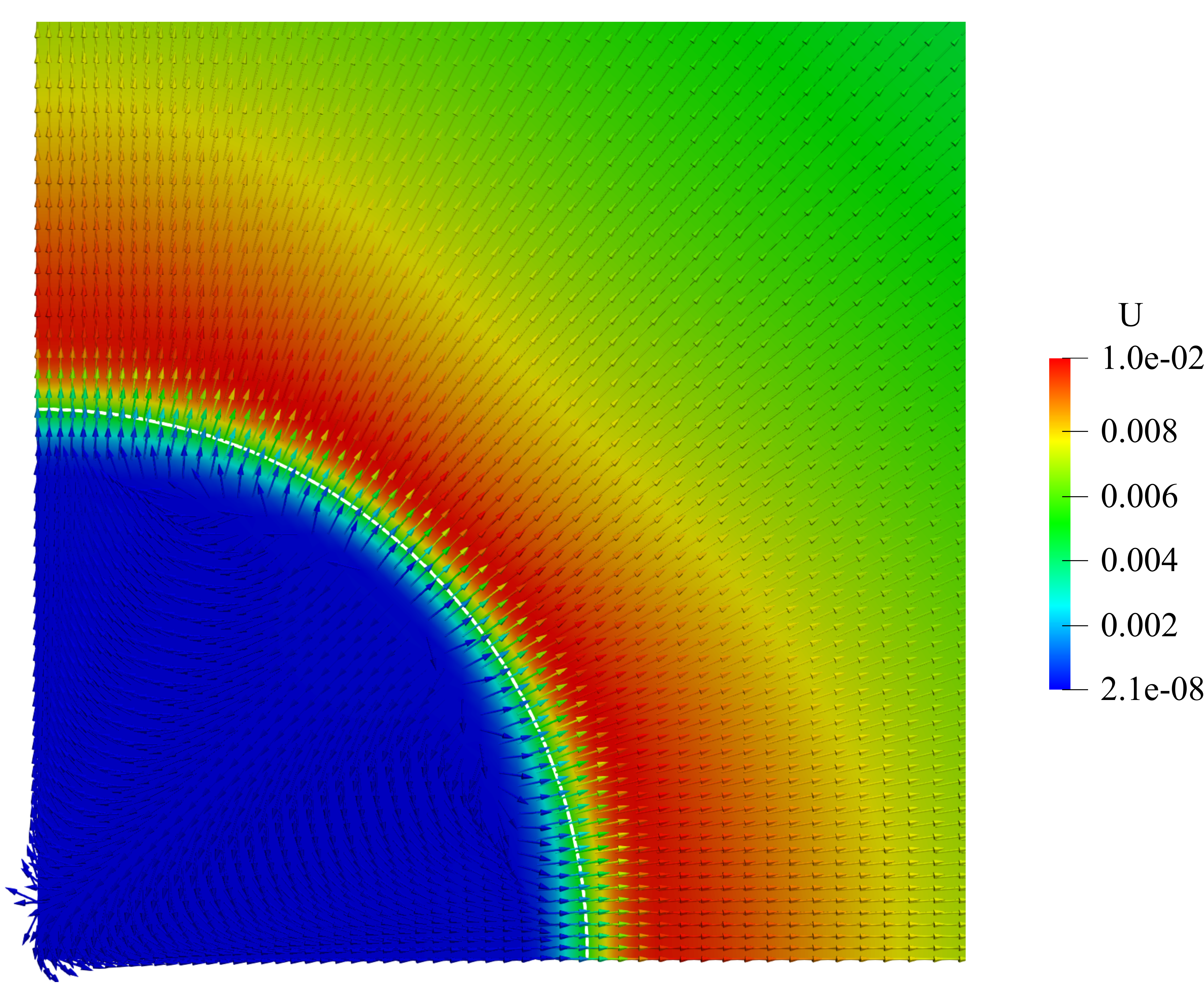}
  \end{center}
  \caption{Vector contour of the one-field velocity field $\mathbf{U}$ around an evaporating sessile droplet (the white solid curve represents the interface).}
\label{velJump}
\end{figure}

\begin{figure}[h]
  \begin{subfigure}[h]{0.5\textwidth}
    \includegraphics[width=\textwidth]{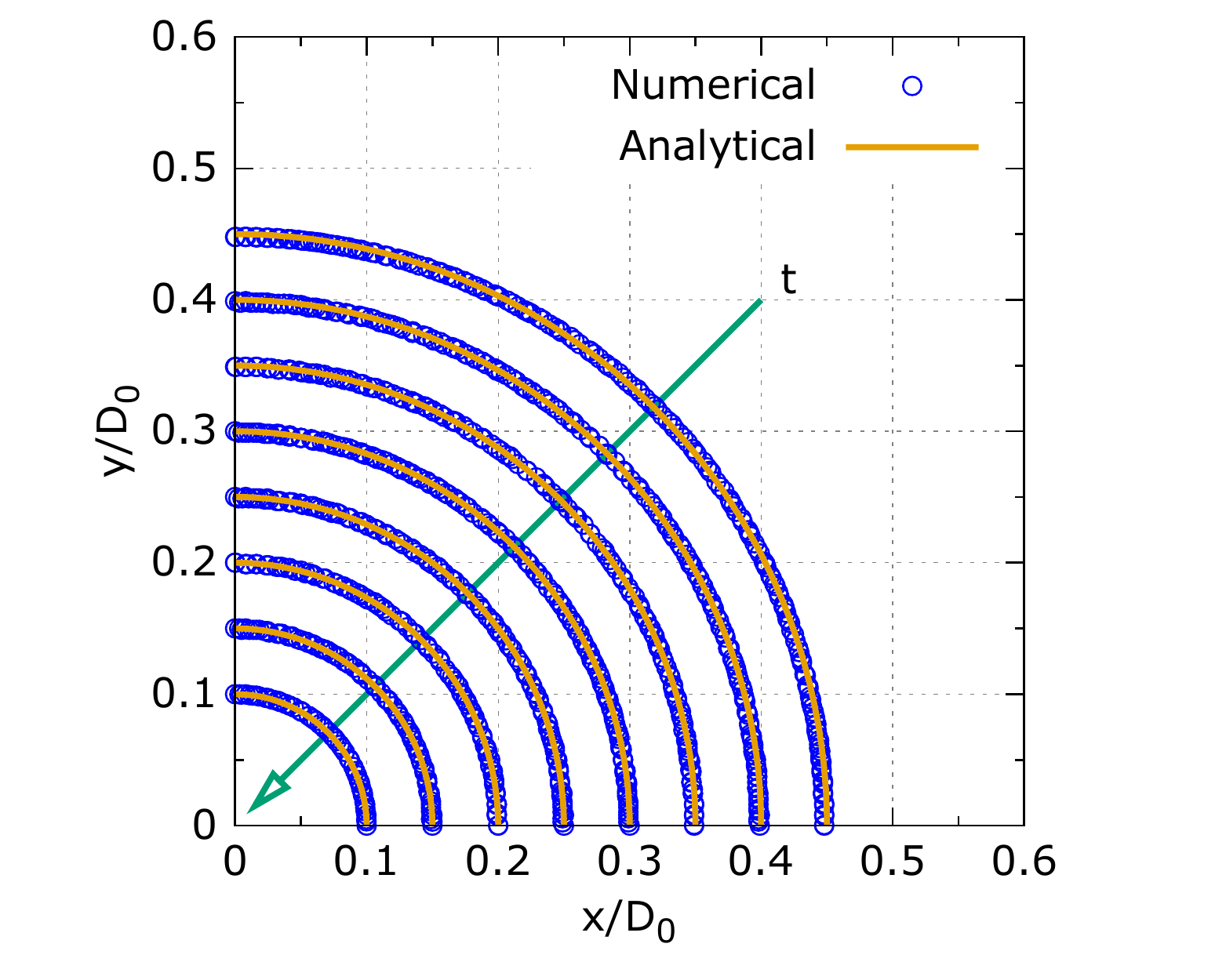}
    \caption{}
    \label{intCap1}
  \end{subfigure}
  \hfill
  \begin{subfigure}[h]{0.5\textwidth}
    \includegraphics[width=\textwidth]{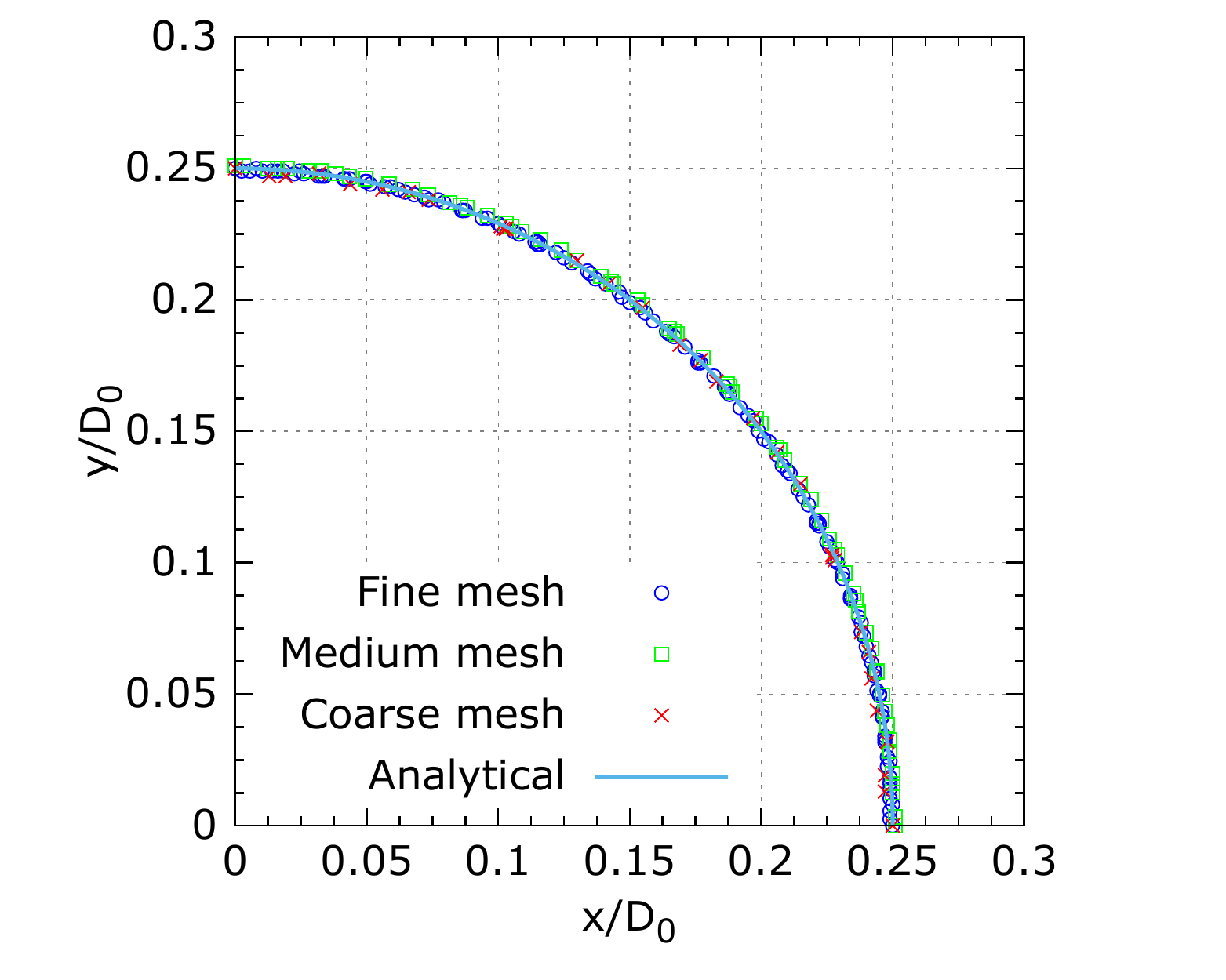}
    \caption{}
    \label{intCap2}
  \end{subfigure}
  \caption{Numerical validations of the interface capturing during evaporation of a sessile droplet: (a) the global plot, (b) the mesh convergence study.}
\label{intCap}
\end{figure}

In Fig. \ref{velJump}, we present the velocity contour of the one-field velocity field $\mathbf{U}$. All the vectors are perpendicular to the interface (represented by the white solid curve) and point from the liquid phase to the vapour phase. Additionally, a velocity jump can also be found around the interface. This demonstrates that the influence of spurious currents on the droplet evaporation is negligible. 
Our further numerical validations (as shown in Fig. \ref{intCap1}) also confirm that these symmetrical spurious velocities never deform the interface in an un-physical way and perfect interface shapes are predicted with our i-CLSVoF framework.

\begin{figure}[h]
  \begin{subfigure}[h]{0.5\textwidth}
    \includegraphics[width=\textwidth]{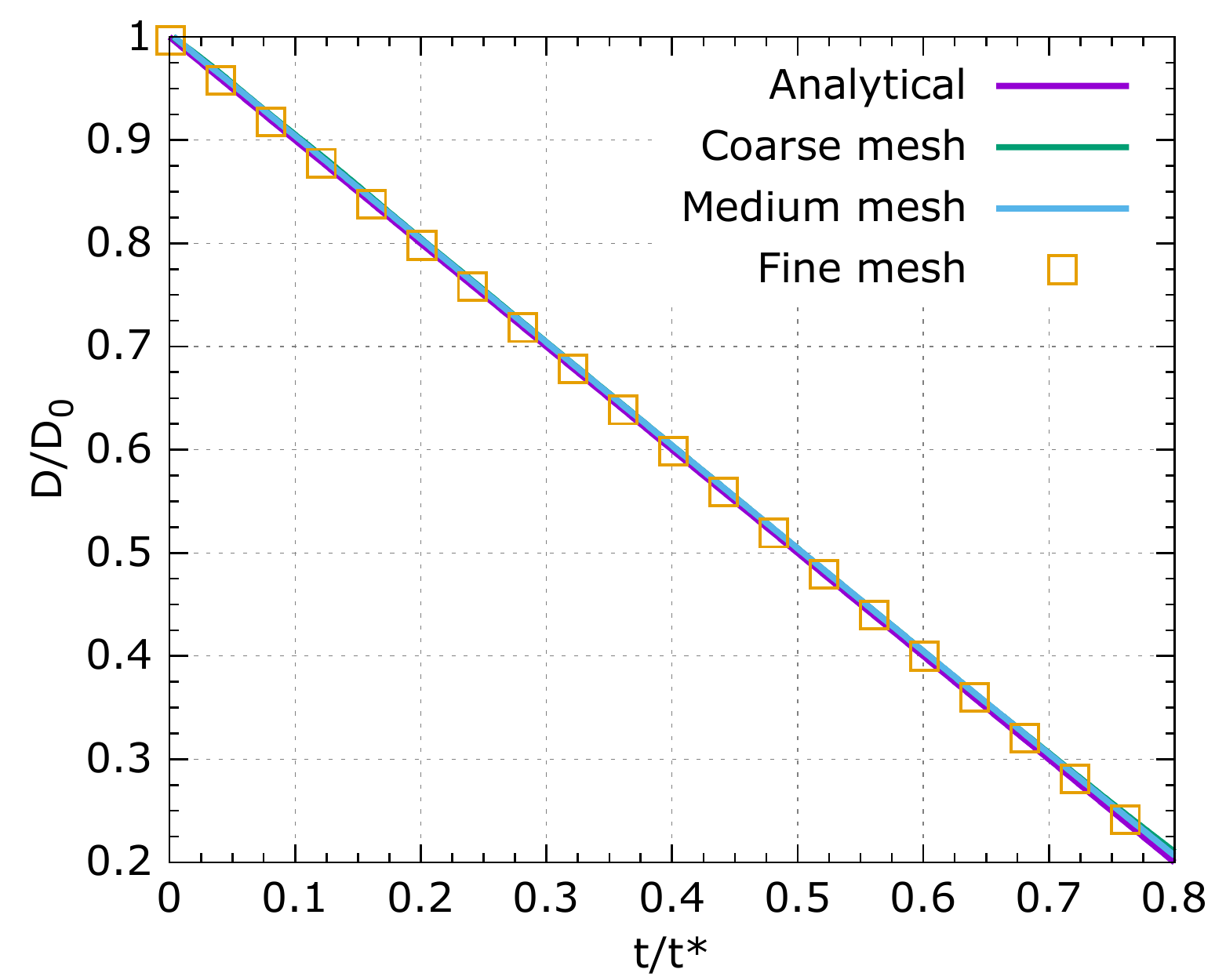}
    \caption{}
    \label{meshConvergence1}
  \end{subfigure}
  \hfill
  \begin{subfigure}[h]{0.5\textwidth}
    \includegraphics[width=\textwidth]{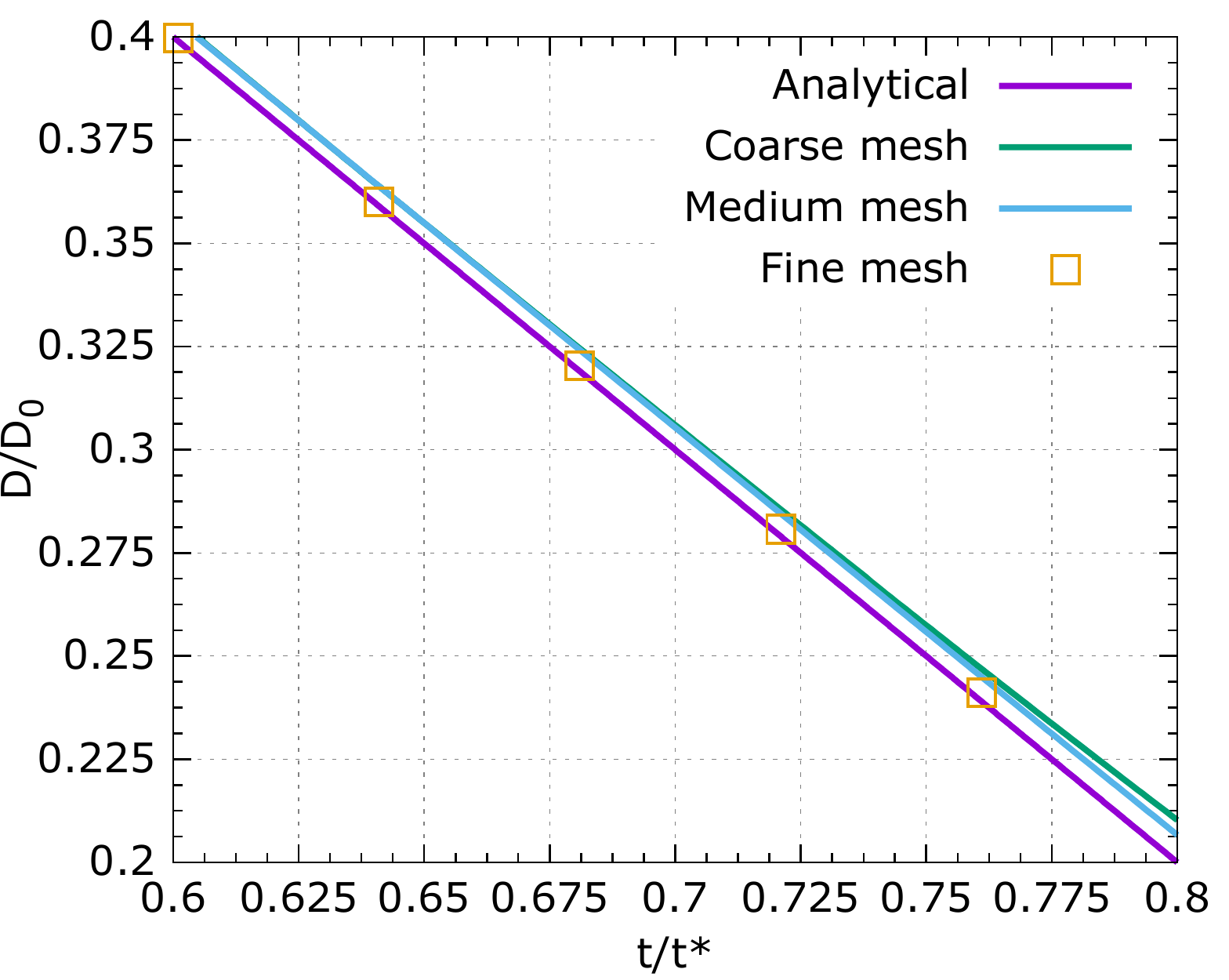}
    \caption{}
    \label{meshConvergence2}
  \end{subfigure}
  \caption{The mesh convergence study: (a) the global plot, (b) the local magnified plot.}
\label{meshConvergence}
\end{figure}

Three different mesh sizes are conducted to study the effect of mesh resolution on the numerical results. As shown in Fig. \ref{meshConvergence}, 
% 2 figures side by side
the time evolution of dimensionless droplet diameter with the dimensionless time until 80\% of the total evaporation time is presented. The agreement between the numerical solution and the corresponding analytical solution is getting better with finer mesh. For the fine mesh, the analytical curve perfectly goes through all the numerical data points (as shown in the locally magnified plot in Fig. \ref{meshConvergence2})).

\begin{figure}[h]
  \begin{center}
    \includegraphics[width=0.5\textheight]{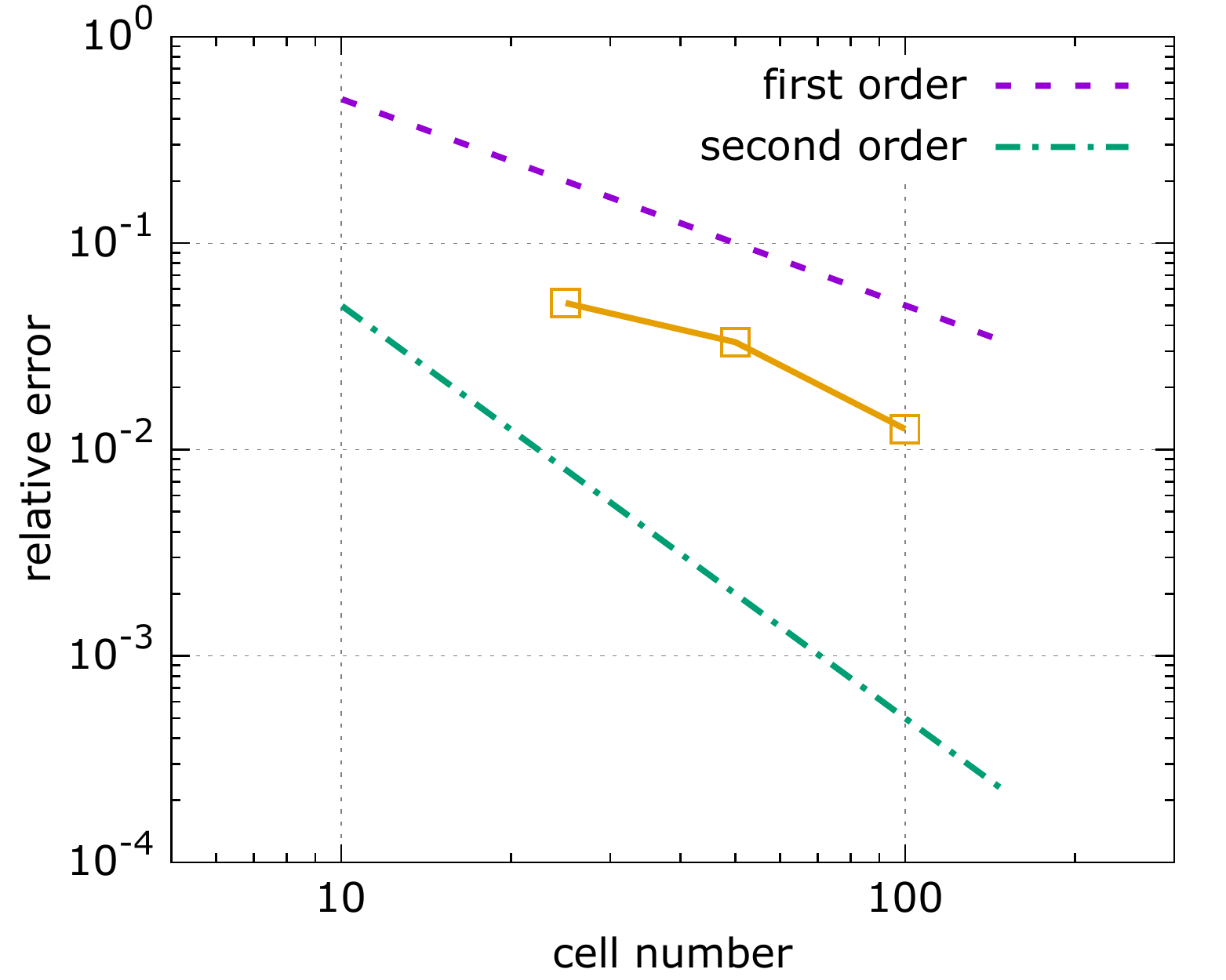}
  \end{center}
  \caption{The numerical error in predicting dimensionless droplet diameter for different mesh resolutions (larger cell number corresponds to finer mesh).}
\label{relErrorD}
\end{figure}

To validate the evaporation model quantitatively, the relative error of the predictions for the shrinking droplet diameter calculated with different mesh resolutions are compared in Fig. \ref{relErrorD} (when $t/t^* = 0.8$).
The relative error is lower than $0.01$ when the number of cells in one direction for $2$D cases is larger than $100$, and first-order convergence is obtained with the evaporation model developed in this work. Additionally, the numerical error for predicting the shrinking droplet diameter is still acceptable (around $5\%$) even for a coarse mesh with our improved numerical model.
% 2 figures side by side

Interface capturing is another crucial aspect in modelling droplet evaporation, and the i-CLSVoF approach developed in this work can capture the free surface in a sharp manner. The corresponding validation is also conducted. The solid yellow curves are the analytical solution in different evaporation stages (as can be seen in Fig. \ref{intCap}), while the blue dots are the numerical data collected on the $0.5$ iso-surface for the eight stages. We can see from Fig. \ref{intCap1} that the agreement between the numerical data and the corresponding analytical solution is perfect. It is also promising to see that the circular droplet shape is maintained, meaning that the influence of spurious currents on destroying the droplet shape is suppressed throughout the whole evaporation process. For a given stage, the mesh convergence study indicates no major difference when refining the mesh (as shown in Fig. \ref{intCap2}), and the reason is that the numerical data are only collected at the $0.5$ iso-surface, which does not depend on mesh resolution. 

\subsubsection{Sessile droplet evaporation with the thermally driven evaporation model}
After validating the sessile droplet evaporation with constant mass flux model, the i-CLSVoF framework is extended to incorporate evaporation with more complex evaporation mass flux calculations. The mass flux of the thermally driven evaporation model depends on the interfacial temperature difference, which drives the phase change from liquid to vapour. The analytical solution for the shrinking droplet diameter square $D^2$ during the evaporation process was detailed derived with the interface energy balance in the literature \cite{stephen2000turns}. The analytical solution is given as the ordinary differential equation
\begin{equation} \label{D2Thermal}
\frac{d D^2}{dt}=-\frac{8k_g}{\rho_l {c_p}_g} \ln(1+B_q),
\end{equation}
where $B_q$ is the Spalding mass transfer number defined by
\begin{equation} \label{Bq}
B_q = \frac{{c_p}_g (T^{\infty}-T_{sat})}{h_{ev}},
\end{equation}
and $T^{\infty}$ here is the temperature on the boundary. This is known as the $D^2$ law, and it is derived for the droplet evaporation in an infinite domain. The modified $D^2$ law takes into account the effect of computational domain size on the evolution of droplet diameter, and this model is more suitable for validating droplet evaporation in a finite computational domain. The modified analytical solution is 
\begin{equation} \label{D2ThermalC}
\frac{d D^2}{dt}=-\frac{8k_g}{\rho_l {c_p}_g} \frac{\ln(1+B_q)} {\ln (D_s/\sqrt{D^2})},
\end{equation}
where $D_s$ is the diameter of the inscribed circle for the computational domain \cite{irfan2017front}. However, the unknown $D^2$ appears in the denominator due to the correction, and consequently we can only solve the Eqn. \ref{D2ThermalC} numerically.

\begin{figure}[h]
  \begin{subfigure}[h]{0.5\textwidth}
    \includegraphics[width=\textwidth]{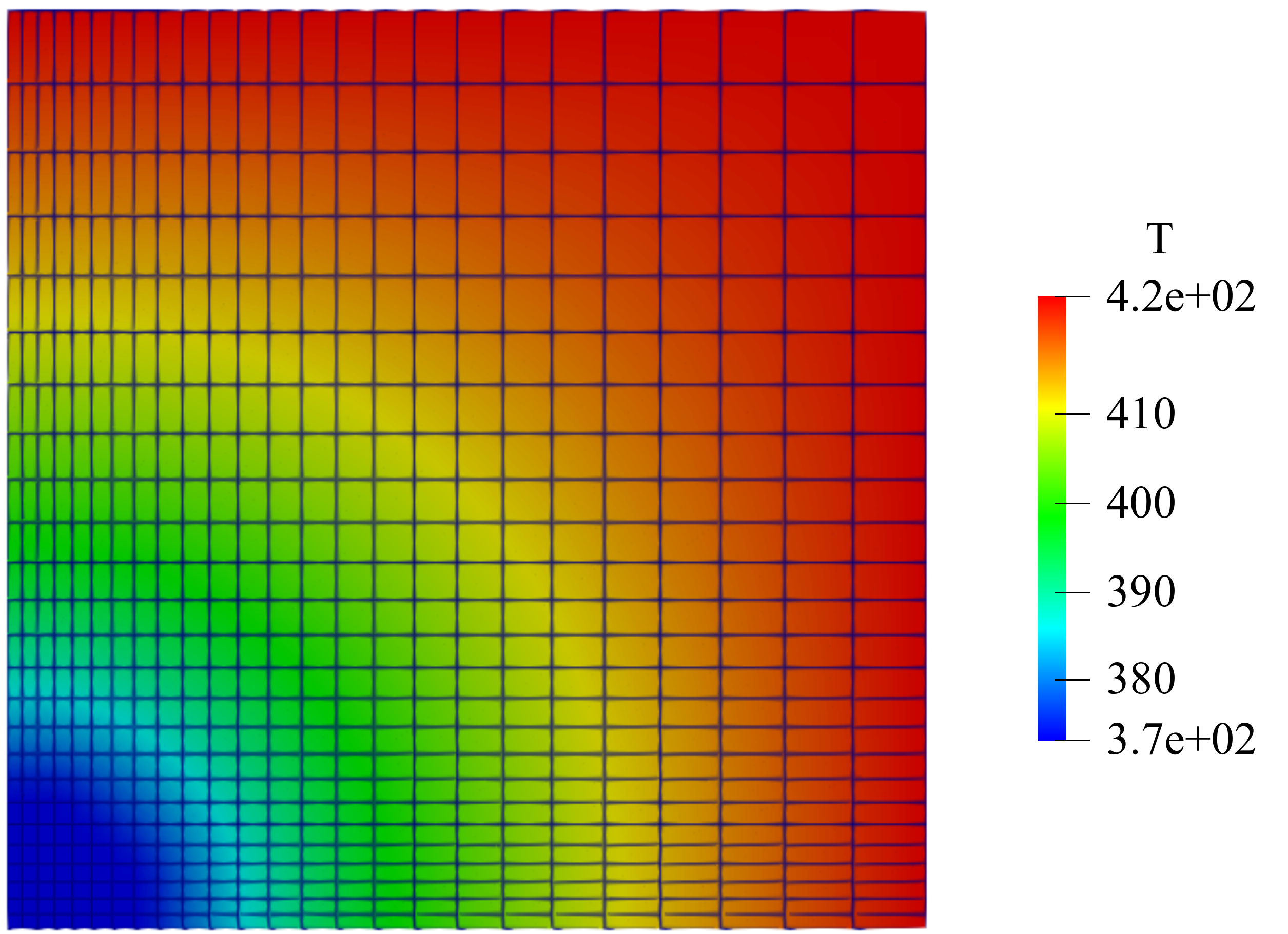}
    \caption{}
    \label{thermal1}
  \end{subfigure}
  \hfill
  \begin{subfigure}[h]{0.5\textwidth}
    \includegraphics[width=\textwidth]{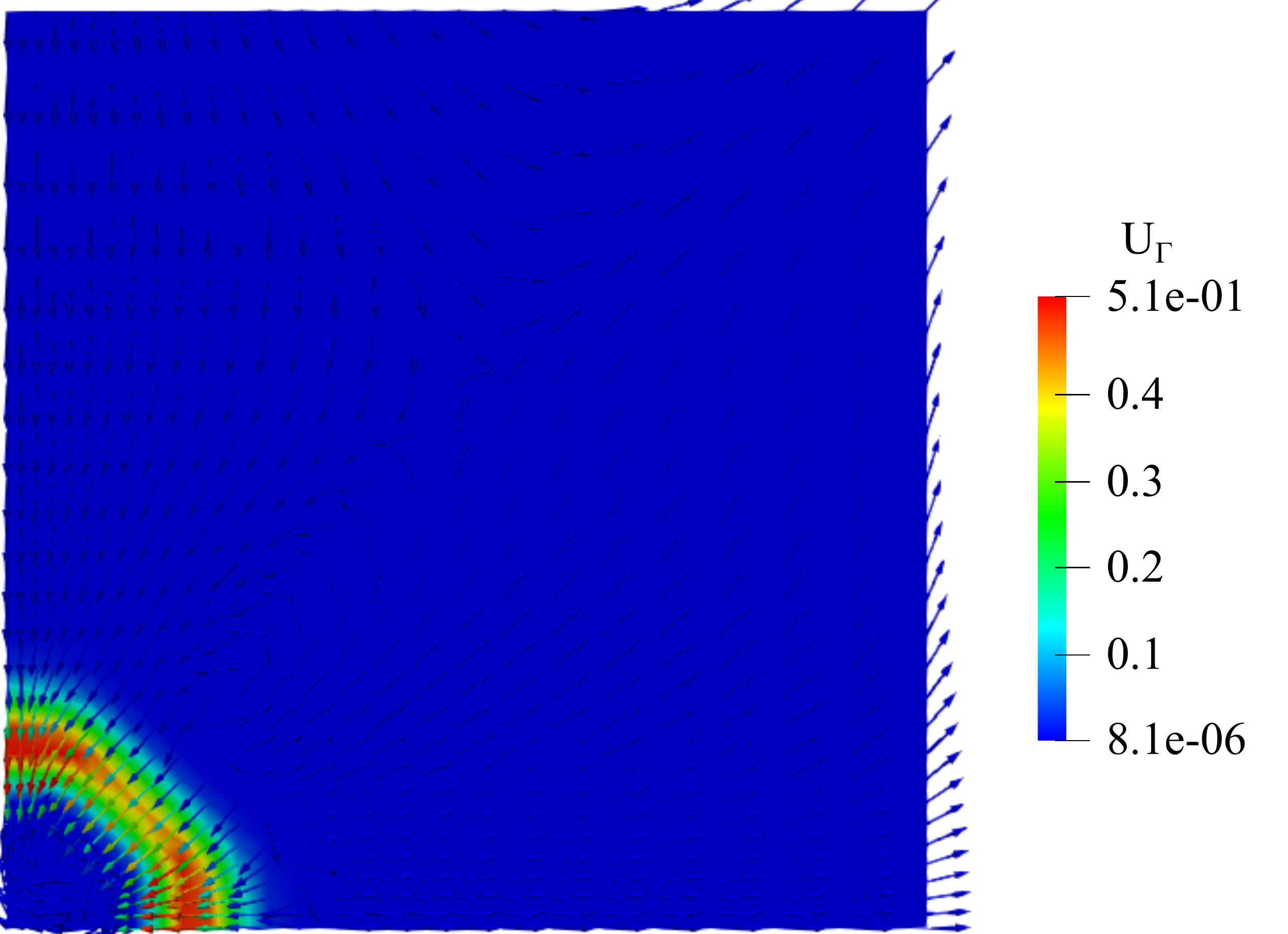}
    \caption{}
    \label{thermal2}
  \end{subfigure}
  \caption{(a) The temperature distribution around the droplet, (b) the interface velocity field $\mathbf{U}_{\Gamma}$.}
\label{thermal}
\end{figure}

The numerical set-up for the $2$D simulation of the thermally driven evaporation model is similar to the set-up as shown in Fig. \ref{numSetUp}. The only difference is that in the initial configuration the temperature for the droplet is equal to its saturation temperature $T_{sat}$ while the temperature for the rest of the domain is higher than the saturation temperature. Additionally, a Dirichlet boundary condition for the temperature field on the boundaries except the symmetry boundaries must be applied. The numerical validation in this part starts with the temperature difference of $50 \ K$, and the corresponding temperature distribution around the evaporating droplet is shown in Fig. \ref{thermal1}. 
% 2 figures side by side
The temperature gradient around the droplet drives the phase change from liquid to vapour and the temperature inside the droplet stays constant and equals the saturation temperature. As shown in Fig. \ref{thermal2}, the interface velocity $\mathbf{U}_{\Gamma}$ points towards the droplet centre, which is related to the evaporation-induced droplet shrinking. Additionally, the interface velocity $\mathbf{U}_{\Gamma}$ is dominant only at the droplet interface, which demonstrates that the divergence-free velocity field is successfully reconstructed also for the thermally driven evaporation model.

\begin{figure}[h]
  \begin{subfigure}[h]{0.49\textwidth}
    \includegraphics[width=\textwidth]{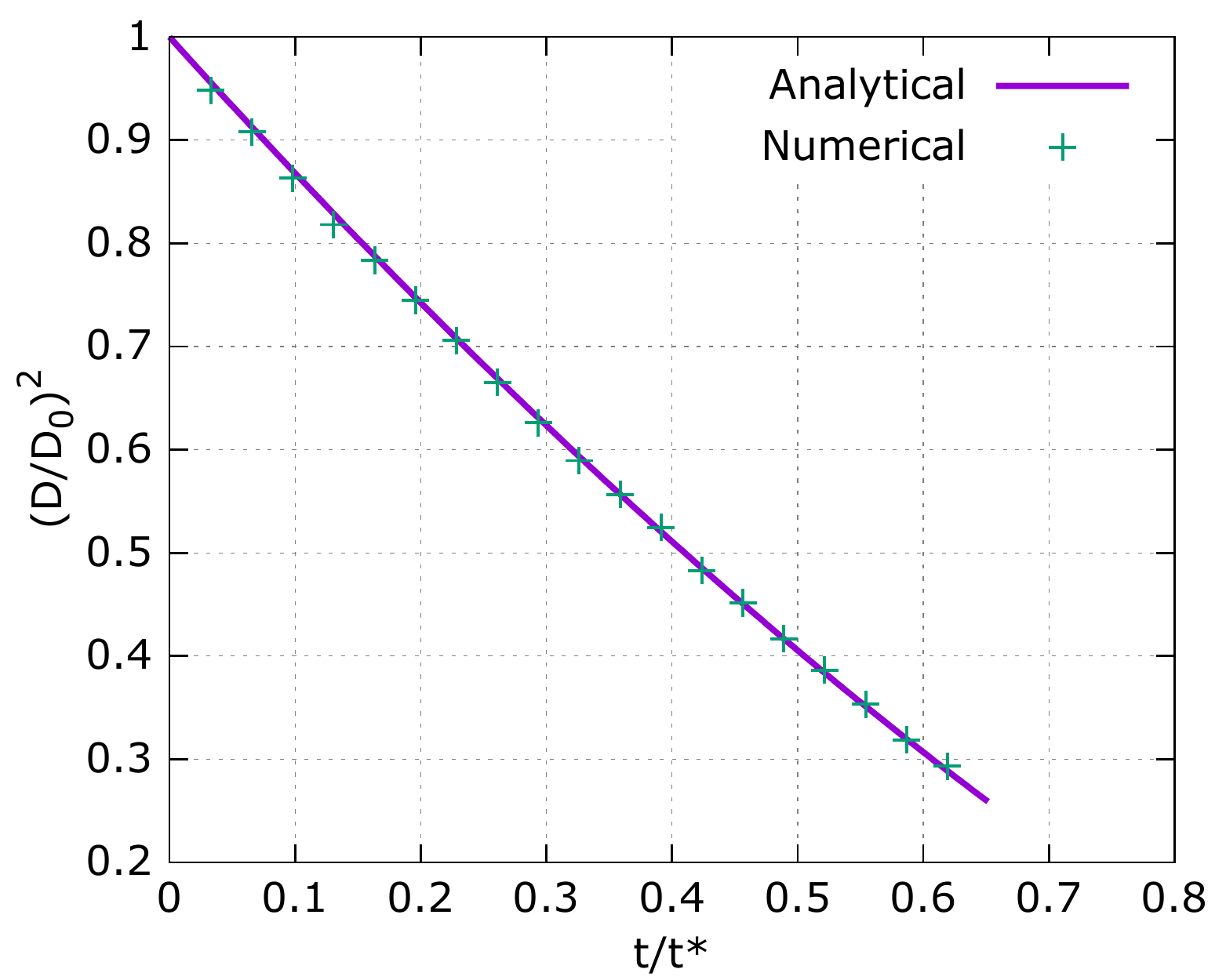}
    \caption{}
    \label{thermalValidation1}
  \end{subfigure}
  \hfill
  \begin{subfigure}[h]{0.51\textwidth}
    \includegraphics[width=\textwidth]{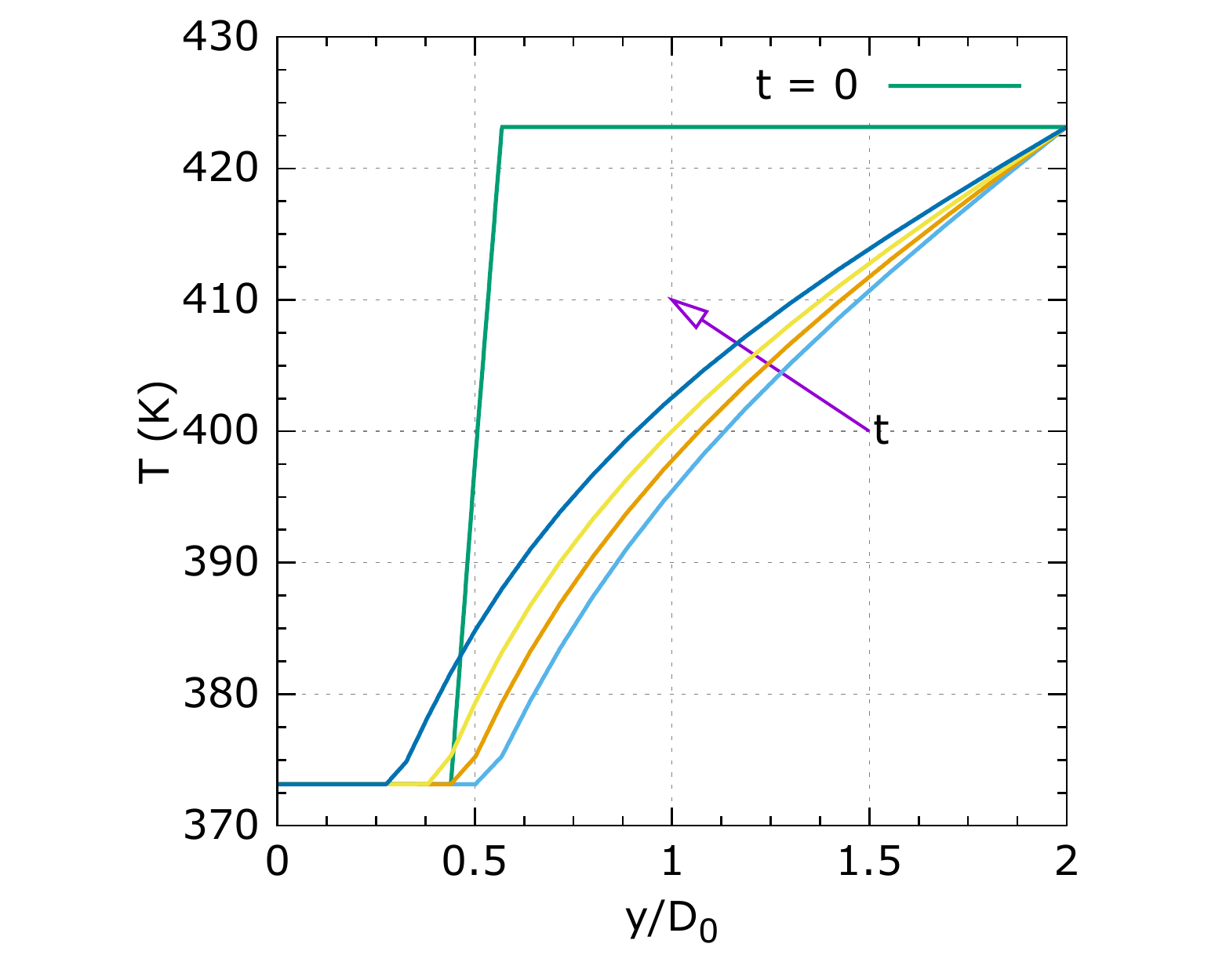}
    \caption{}
    \label{thermalValidation2}
  \end{subfigure}
  \caption{(a) The numerical validation for thermally driven evaporation model, (b) the temperature evolution during the evaporation process ($t=0$ corresponds to the initial temperature field).}
\label{thermalValidation}
\end{figure}

The quantitative study of the shrinking droplet diameter is compared against the corresponding analytical model according to Eqn. \ref{D2ThermalC}. The dimensionless droplet diameter and the dimensionless time are adopted. It can be seen from Fig. \ref{thermalValidation1}, that an accurate solution is achieved for around $65\%$ of the total evaporation time $t^{\star}$ with the graded mesh shown in Fig. \ref{thermal1}. 
% 2 figures side by side
The corresponding temperature evolution collected from the bottom-left corner to the bottom-right corner can be seen in Fig. \ref{thermalValidation2}. The solid green line represents the initial configuration for the temperature field, where the transition band from the saturation temperature to the temperature value corresponding to the boundaries in the initial temperature field is relatively large. Further mesh refinement can shorten this transition band. The other curves in Fig. \ref{thermalValidation2} show the evolution of temperature (the purple arrow indicates time going on) during the evaporation process. The saturation temperature inside the droplet and the temperature at the domain boundary are strictly maintained constant during the evaporation process.

Additionally, droplet evaporation with different Stefan numbers is further studied to validate our model extensively. The Stefan number is defined as
\begin{equation} \label{SetfanNum}
St = \frac{{c_p}_g \Delta T}{h_{ev}},
\end{equation}
where $\Delta T = T_{\infty} - T_{sat}$ is the temperature difference between the saturation temperature $T_{sat}$ inside the droplet and the temperature $T_{\infty}$ at the boundaries, and ${c_p}_g$ the specific heat capacity of the gas/vapour phase. 
\begin{figure}[h]
  \begin{center}
    \includegraphics[width=0.5\textheight]{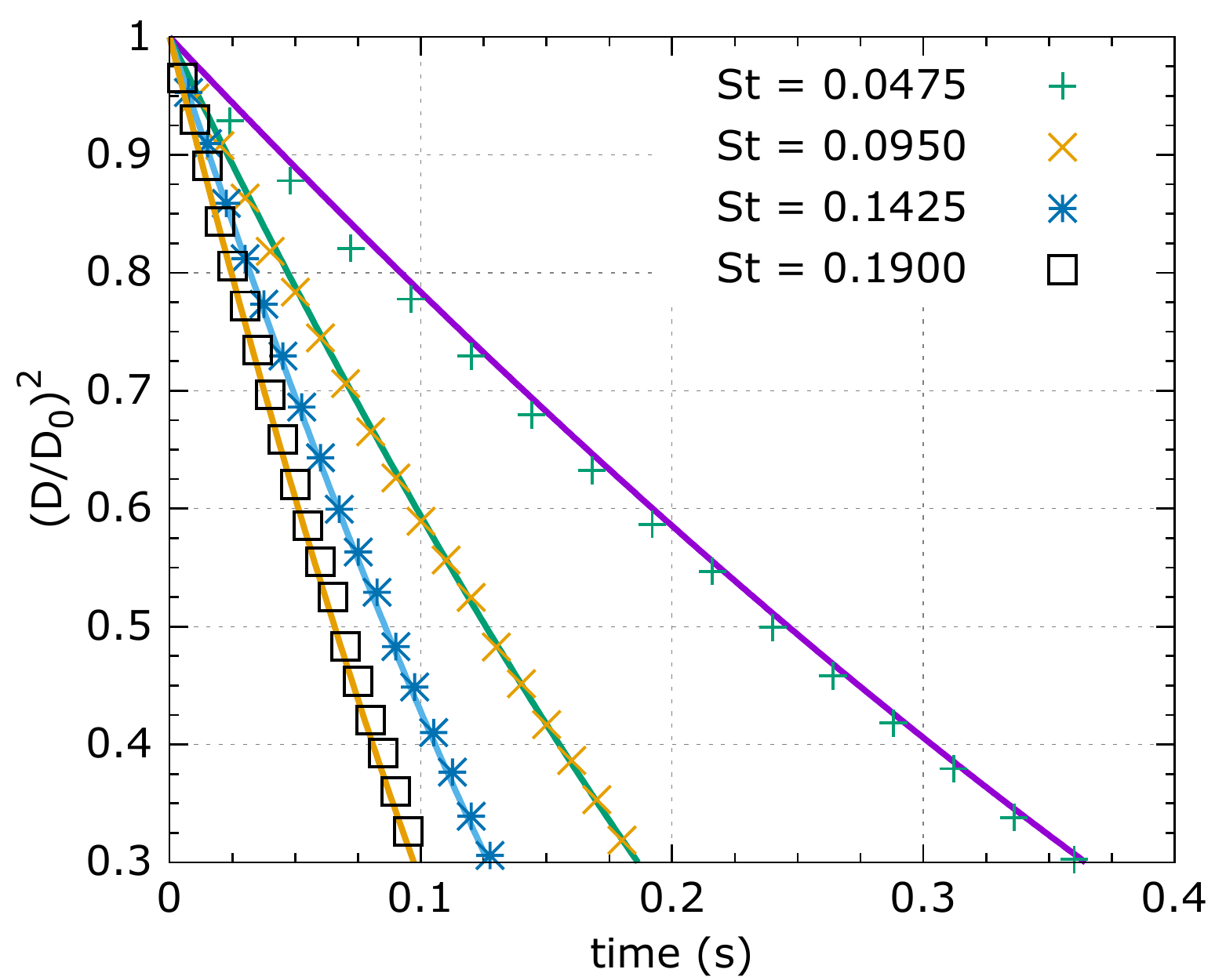}
  \end{center}
  \caption{Validations for four cases with different Stefan number $St$ (solid curves represent the analytical solutions, and points represent the corresponding numerical data).}
\label{effectStefanNum}
\end{figure}
As can be seen from Fig. \ref{effectStefanNum}, four evaporation cases with different Stefan numbers are presented. Cases with a large Stefan number experience faster evaporation, and a good agreement between numerical and the corresponding solution of the analytical model is observed for all cases.

\subsubsection{Sessile droplet evaporation at room temperature}
The numerical set-up for modelling droplet evaporation at room temperature is the same as the two aforementioned evaporation models (refer to Fig. \ref{numSetUp}). In addition to prescribe the outflow boundary conditions, a Dirichlet boundary condition for the vapour mass fraction field is applied on the boundaries except for the symmetry boundaries.
  
\begin{figure}[h]
  \begin{subfigure}[h]{0.49\textwidth}
    \includegraphics[width=\textwidth]{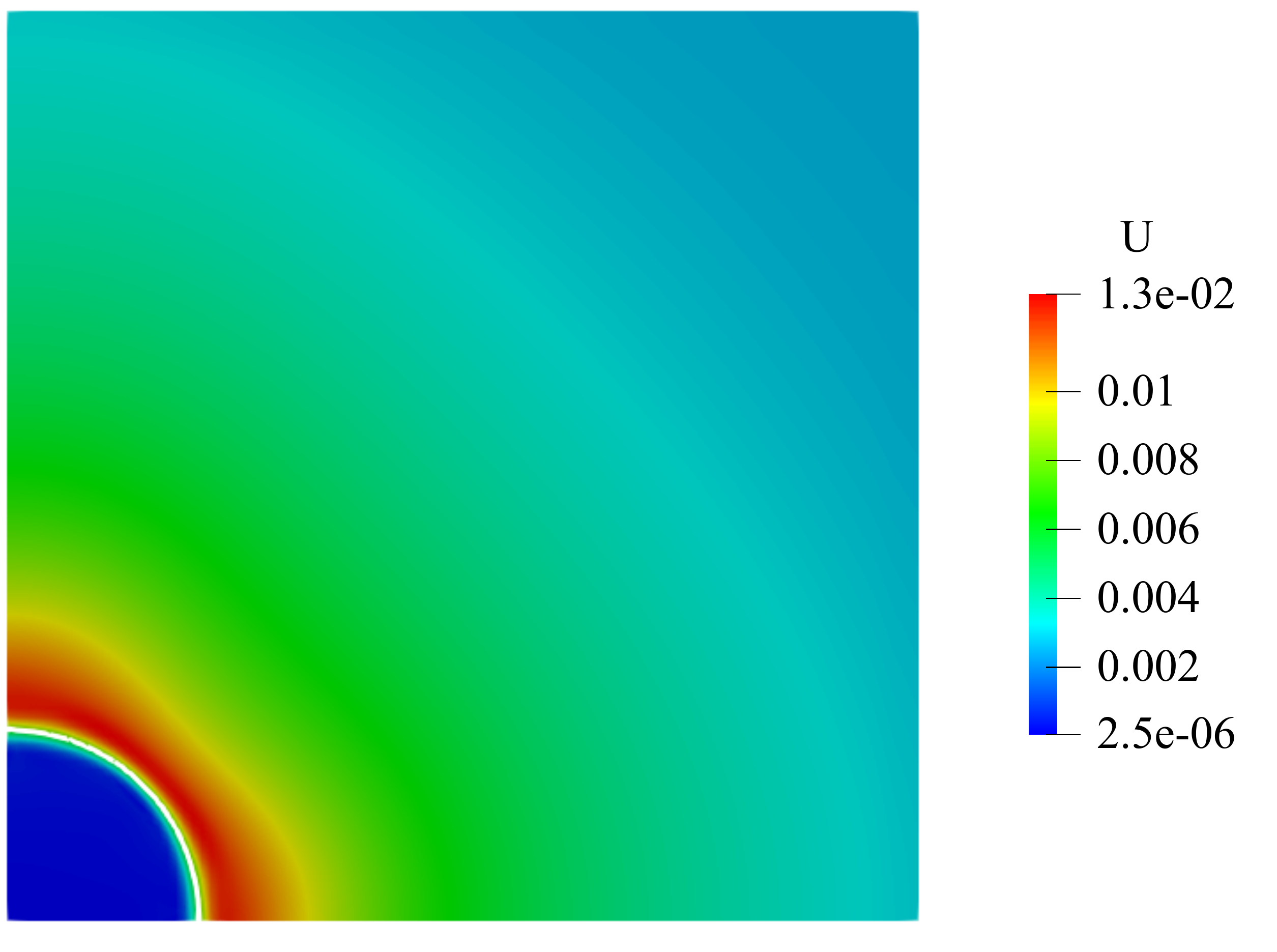}
    \caption{}
    \label{vapourUY1}
  \end{subfigure}
  \hfill
  \begin{subfigure}[h]{0.51\textwidth}
    \includegraphics[width=\textwidth]{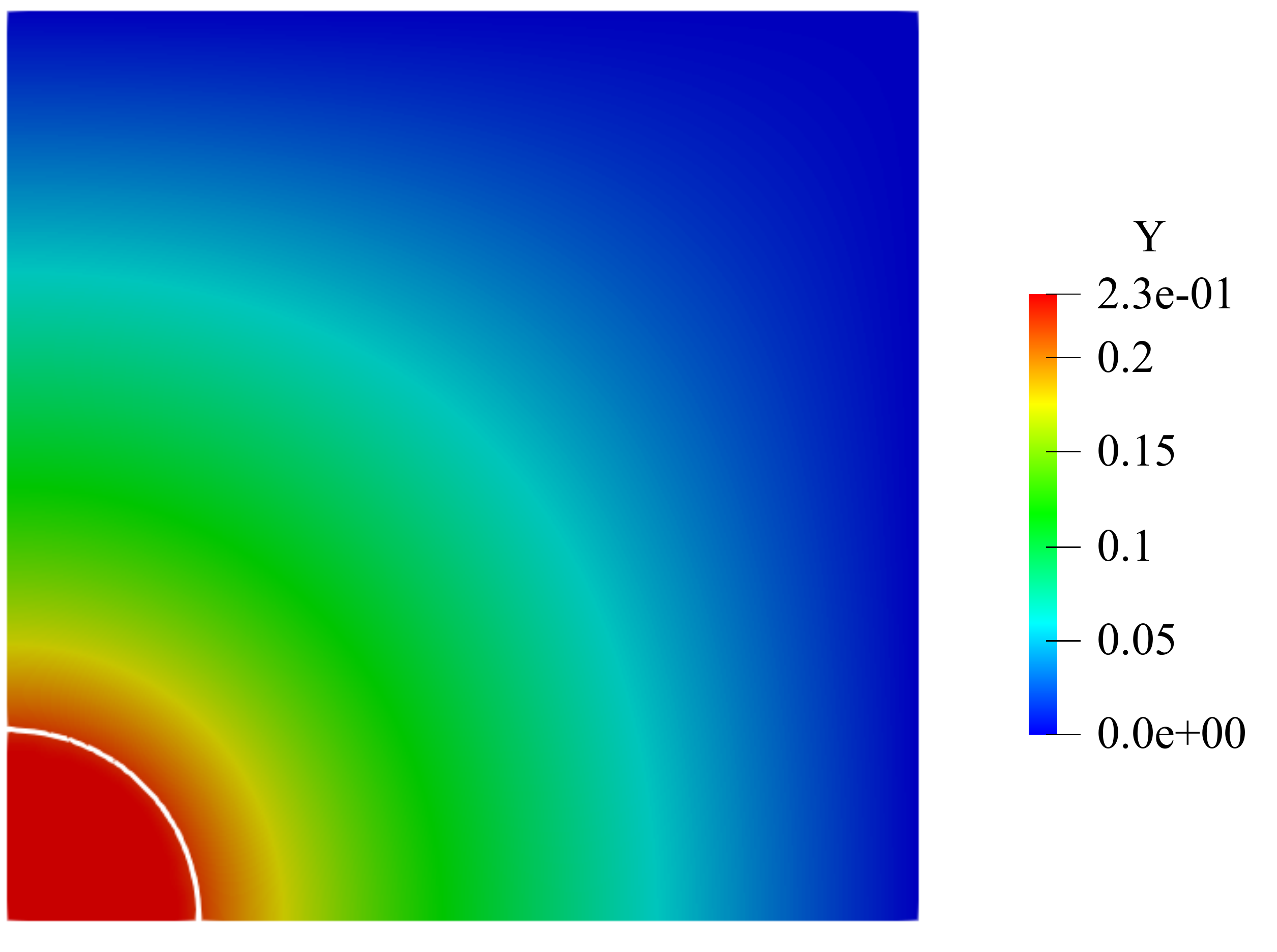}
    \caption{}
    \label{vapourUY2}
  \end{subfigure}
  \caption{(a) The one-field velocity field $\mathbf{U}$ of droplet evaporation at room temperature, (b) The vapour mass fraction field of droplet evaporation at room temperature (white curve represents the interface).}
\label{vapourUY}
\end{figure}

Some essential parameters used in the simulations are listed in Table \ref{paras}. The one-field velocity field during the evaporation process is shown in Fig. \ref{vapourUY1} where a velocity jump can be seen around the interface region (the white curve represents the interface). The corresponding vapour mass fraction field is shown in Fig. \ref{vapourUY2} with the vapour mass faction at the top and right boundaries given as constants prescribed by the Dirichlet boundaries conditions. The vapour mass fraction gradient at the interface drives the droplet evaporation.

The shrinking droplet diameter $D$ is recorded to quantitatively validate the evaporation model by comparing the diameter predicted by the numerical simulations to the analytical solution given by the so-called $D^2$ law \cite{stephen2000turns}. However, for the droplet evaporation in a finite domain, the classical $D^2$ law should be corrected to take the computational domain size into account, and a modified analytical model is adopted in this work \cite{irfan2017front}. This modified $D^2$ law is given by
\begin{equation} \label{D2Y}
\frac{d D^2}{dt}=-\frac{8\rho_g D_v}{\rho_l} \frac{\ln(1+B_y)}{\ln (D_s/\sqrt{D^2})},
\end{equation}
where $D_s$ is the diameter of the inscribed circle for the computational domain, and $B_y$ the Spalding mass transfer number defined as
\begin{equation} \label{By}
B_y = \frac{Y^{\Gamma}-Y^{\infty}}{1-Y^{\Gamma}}
\end{equation}
with $Y^{\infty}$ being the vapour mass fraction far way from droplets \cite{stephen2000turns}.
% 2 figures side by side

\begin{figure}[h]
  \begin{center}
    \includegraphics[width=0.5\textheight]{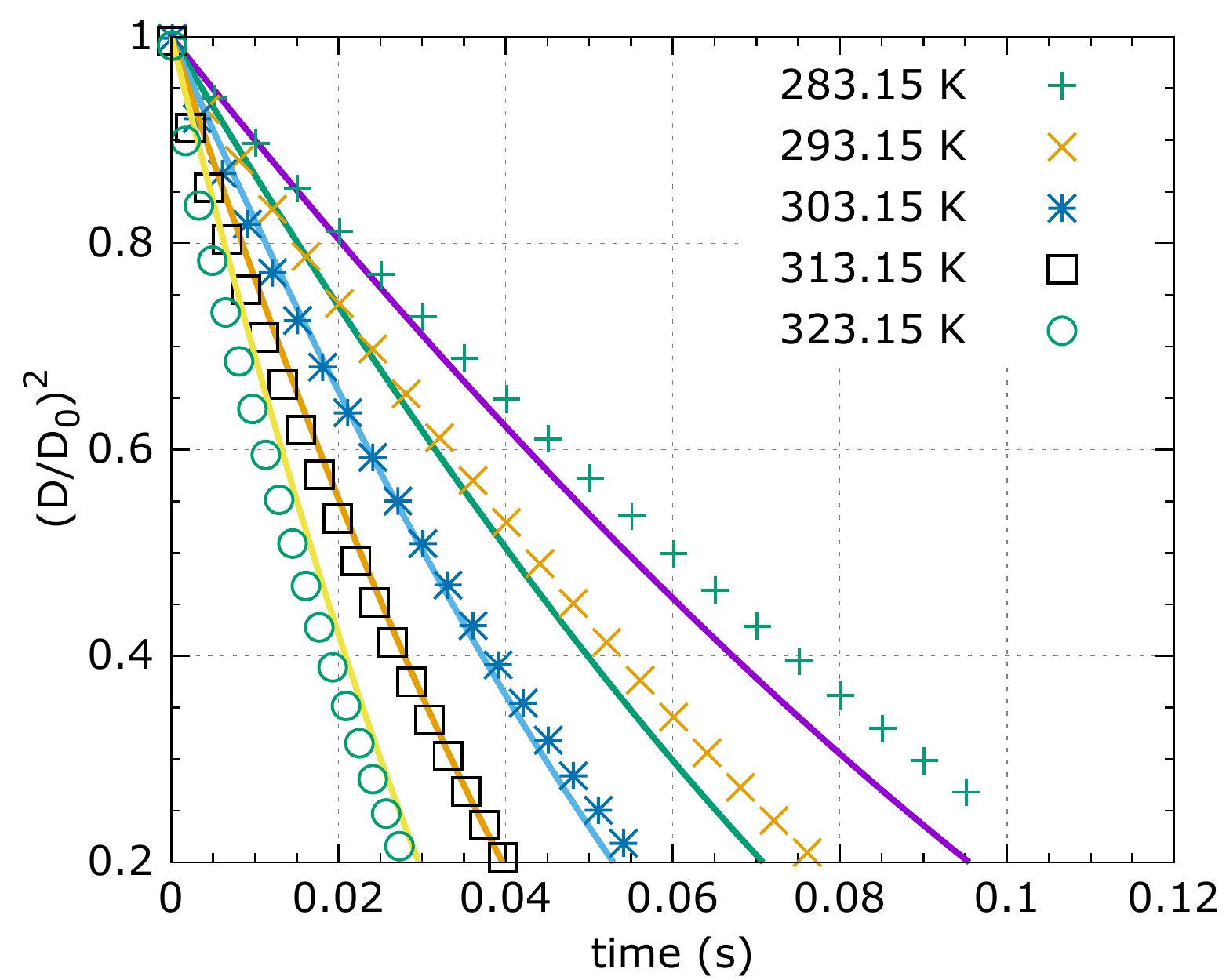}
  \end{center}
  \caption{Validations for five cases with different evaporation temperatures (solid curves represent the analytical solutions, and points represent the corresponding numerical data).}
\label{validationY}
\end{figure}

A parameter study on a wide range of evaporation temperatures was conducted to validate the evaporation model. As shown in Fig. \ref{validationY}, the solid curves represent corresponding analytical solutions given by Eqn. \ref{D2Y} and the points represent numerical solutions predicted by the evaporation model. Good agreement between numerical and analytical solutions are found for evaporation at relatively higher temperatures (303.15 K, 313.15 K and 323.15 K). Some minor discrepancies between numerical and analytical solutions are getting larger, especially, in the late stage of evaporation are found for evaporation at relatively lower temperatures (283.15 K and 293.15 K).

\section{Conclusions}
In this work, the i-CLSVoF framework which combines the sharp-interface LS method and the mass-conserving algebraic VoF method with filtering steps to suppress un-physical velocities, has been developed and implemented in OpenFOAM. The numerical framework was specially developed for modelling surface-tension-dominant flow with and without phase change in a simple yet accurate way.

The governing equation was solved based on the one-field formulation within the FVM framework. An improved surface tension force model and the additional filtering approach were incorporated into the i-CLSVoF framework to further reduce un-physical spurious currents. The superiority of the improved numerical framework over the conventional VoF and the CLSVoF approaches was demonstrated by several benchmark cases. The sharp interface was captured, and interface diffusion was suppressed. A promising amount of suppression of un-physical spurious currents was achieved with our improved numerical framework. 

The i-CLSVoF framework was further extended to model sessile droplet evaporation, especially in micro-scale. A simple approach was proposed and implemented to reconstruct a divergence-free velocity by removing the evaporation-induced irrotational velocity from the one-field velocity field to predict the evaporation rate accurately. The constant mass flux evaporation model was first incorporated into the i-CLSVoF framework to check the implementation of the basic equations. The numerical validations proved the successful construction of the divergence-free velocity field. Besides, the model accurately predicted the droplet evaporation as shown by comparing the shrinking dimensionless droplet diameter to an analytical solution derived in this work. The interface capturing of the i-CLSVoF was also demonstrated to be accurate enough by comparing the numerically captured free surface to the analytical solution; crucially, no spurious velocity induced interface deformation was found during the evaporation process.  The thermally driven evaporation model was implemented to account for the temperature gradient induced phase change. A novel density-ratio dependent evaporation coefficient was used to calculate the mass flux at the interface rather than determining the evaporation coefficient empirically. The modified $D^2$ law was used to validate the numerical model, and an encouraging agreement between the numerical solution and the analytical solution was achieved. Additionally, extensive parameter studies were conducted to demonstrate the accuracy of the thermally driven evaporation model for a wide range of Stefan numbers. The third evaporation model incorporated in this work describes the sessile droplet evaporation at room temperature. This model can predict the evaporation rate accurately for different evaporation temperatures while some minor discrepancy between numerical and analytical solutions is found for relatively small evaporation temperature, which we consider still acceptable. Droplet evaporation with a fixed triple contact line (also known as evaporation with Constant Contact Radius (CCR) mode) is ubiquitous in case of drying droplets on rough substrates. In the future work, the i-CLSVoF framework will be extended to model the evaporation of micro-sized droplets under CCR conditions.

\section * {Acknowledgements}
Huihuang Xia sincerely acknowledged the funding from the China Scholarship Council (CSC) for the financial support (No. CSC201808350108). Some numerical simulations were done using the computational source of the BwUniCluster 2.0. Many insightful suggestions and comments from Prof. Dominic Vella are highly appreciated. A three-month research stay at the University of Oxford was funded by the Research Travel Grant from the Karlsruhe House of Young Scientists (KHYS).

\section * {Appendix}
The current numerical demonstration is to compare capturing the free-surface by the i-CLSVoF model developed in this work to the conventional VoF model (the interDyMFoam (OpenFOAM 5.x) solver is used for comparison). 
\begin{figure}[h]
  \begin{subfigure}[h]{0.495\textwidth}
    \includegraphics[width=\textwidth]{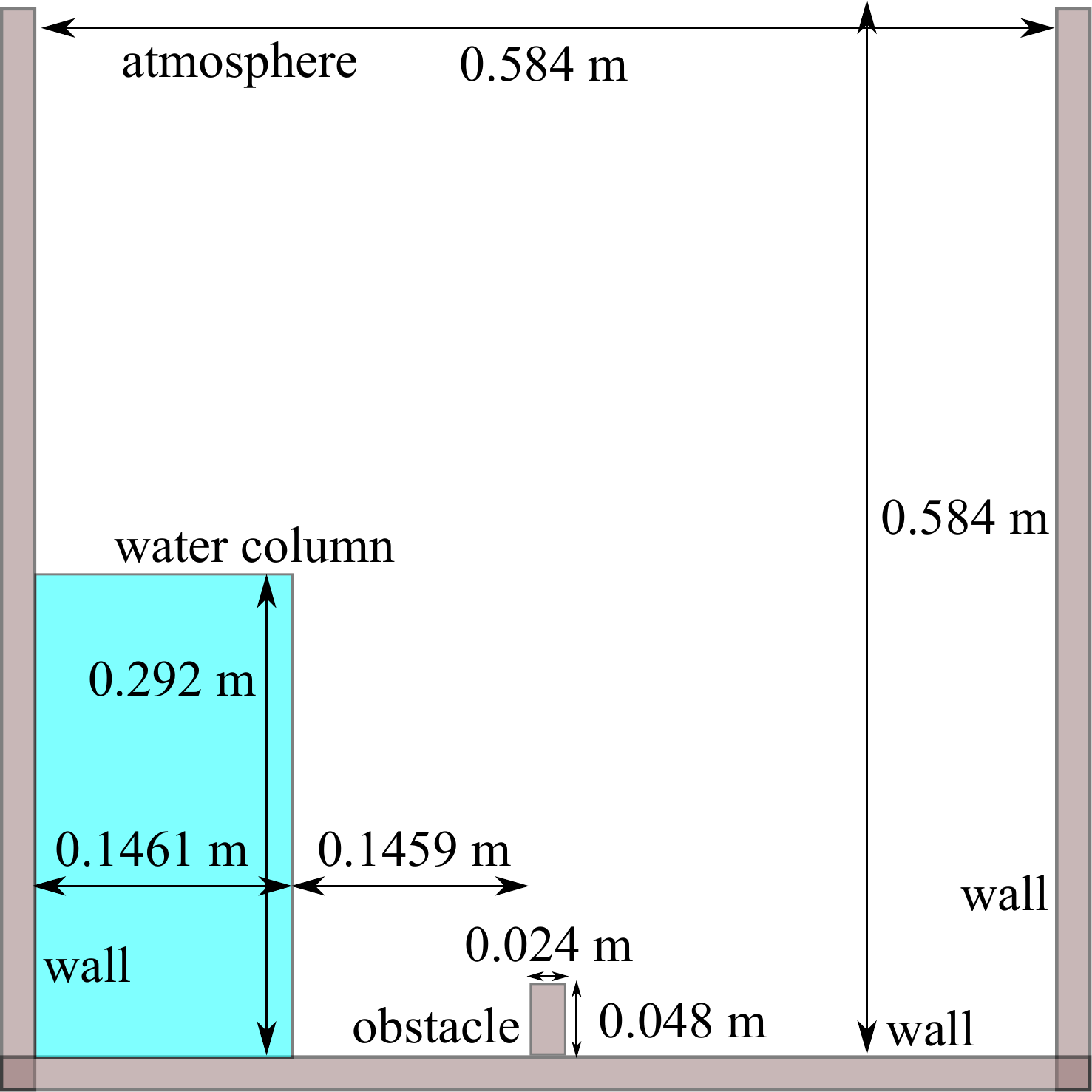}
    \caption{}
    \label{damBreakFig1}
  \end{subfigure}
  \hfill
  \begin{subfigure}[h]{0.495\textwidth}
    \includegraphics[width=\textwidth]{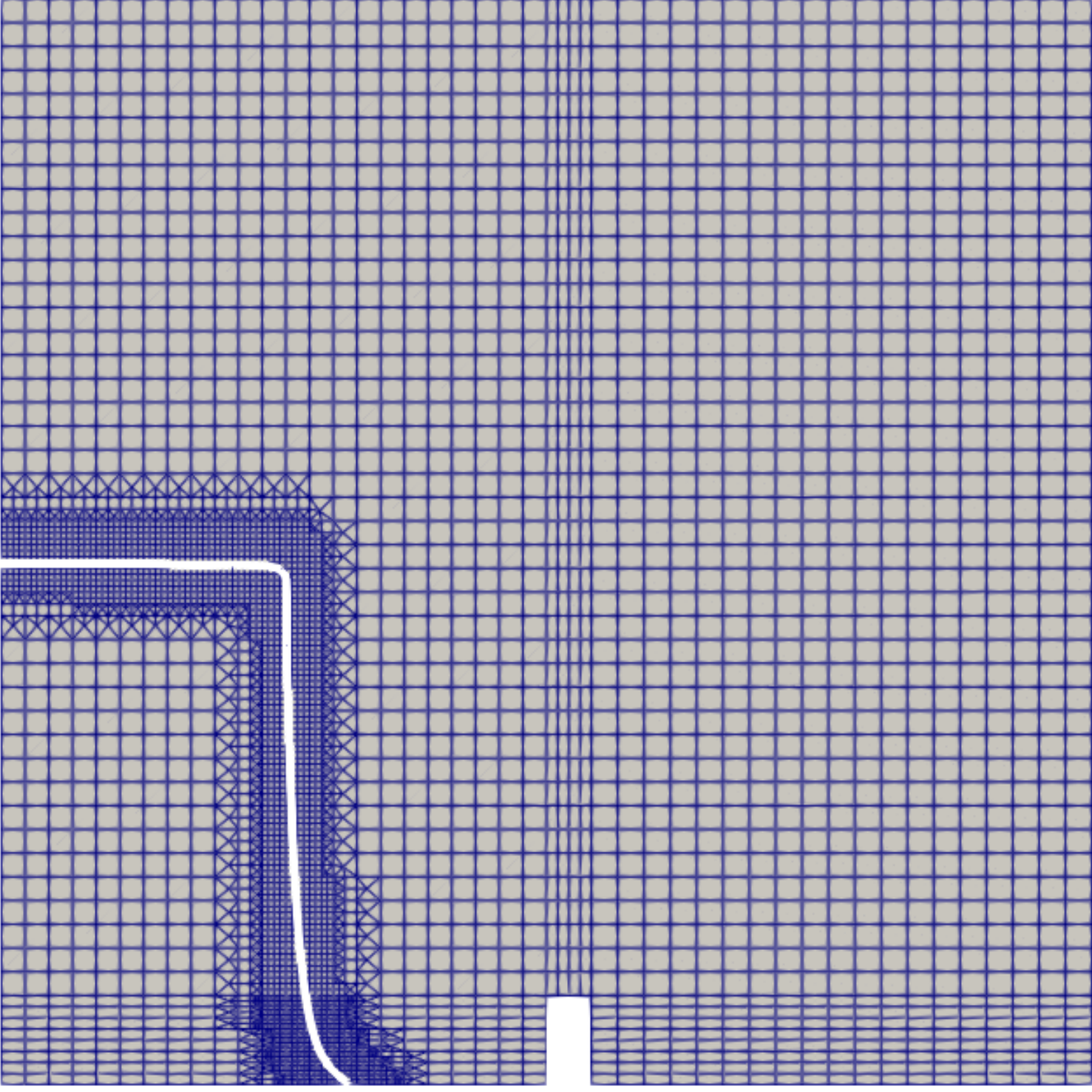}
    \caption{}
    \label{damBreakFig2}
  \end{subfigure}
  \caption{(a) Geometry and boundaries of the 2D dam break simulation, (b) Numerical simulations with 2D AMR (the white curve represents the free-surface).}
\label{damBreakFig}
\end{figure}

In the current case, the surface-tension force is not dominant other than in the droplet relaxation cases. The numerical set-up is that a water column is initially situated at the bottom left of the 2D computational domain with the top as an opening, and the other boundaries are regarded as walls (as shown in Fig. \ref{damBreakFig1}). Some essential parameters for the simulations are outline in Table \ref{damBreak}. 

\begin{table}[h]
\centering
\caption{Parameters for 2D dam break simulations}
\label{damBreak}
\begin{tabular}{ccccc}
\hline
phase & density & dynamic viscosity & surface-tension & gravitational acceleration \\ \hline
water & 1000    & $1\times 10^{-3}$               & 0.07            & -9.81   \\
air   & 1       & $1.48\times 10^{-5}$            & -               & -       \\ \hline
\end{tabular}
\end{table}

% 4 figures (2*2)
\begin{figure}[h]
  \begin{subfigure}[h]{0.495\textwidth}
    \includegraphics[width=\textwidth]{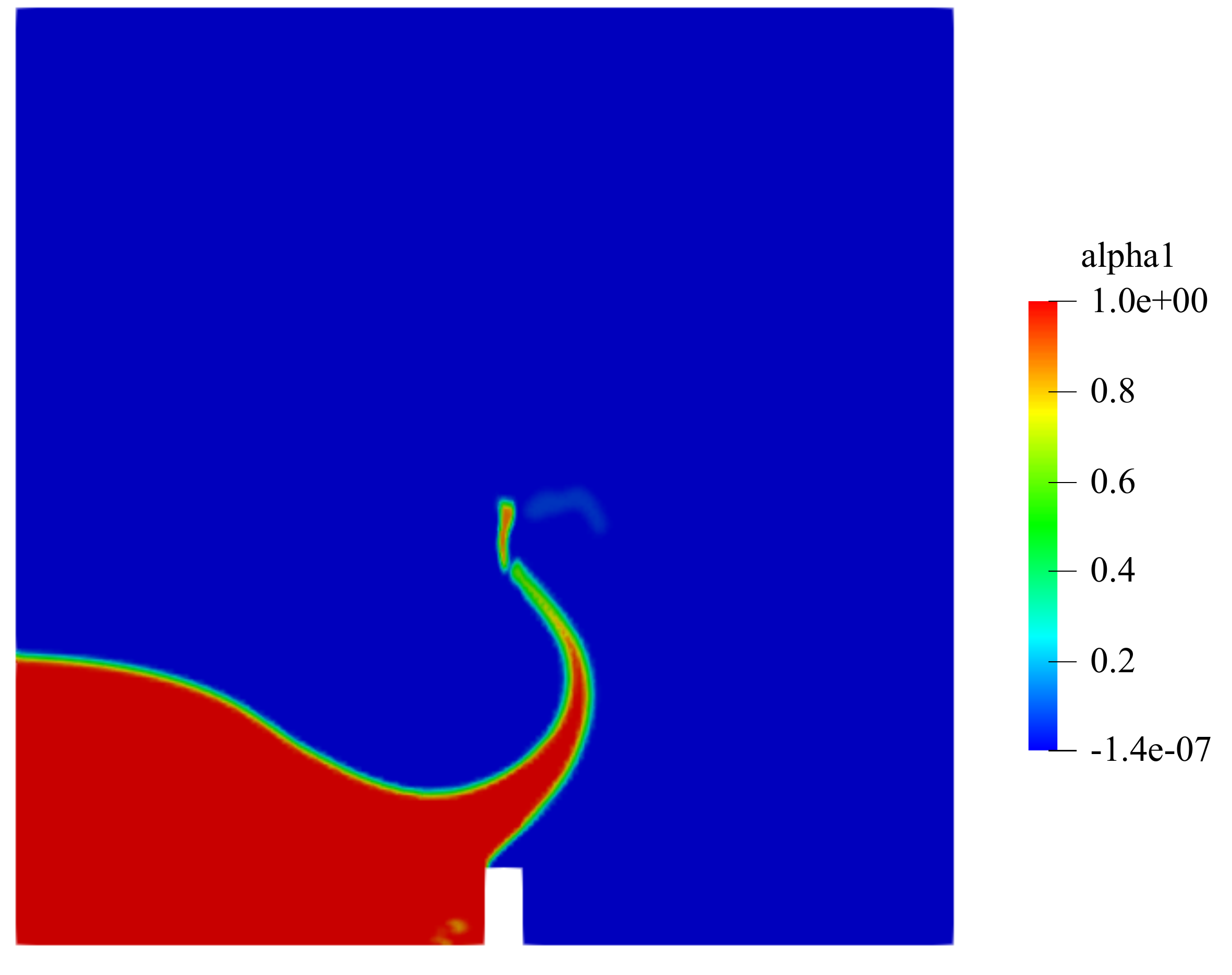}
    \caption{}
    \label{damBreakFreeSurf1}
  \end{subfigure}
  \hfill
  \begin{subfigure}[h]{0.495\textwidth}
    \includegraphics[width=\textwidth]{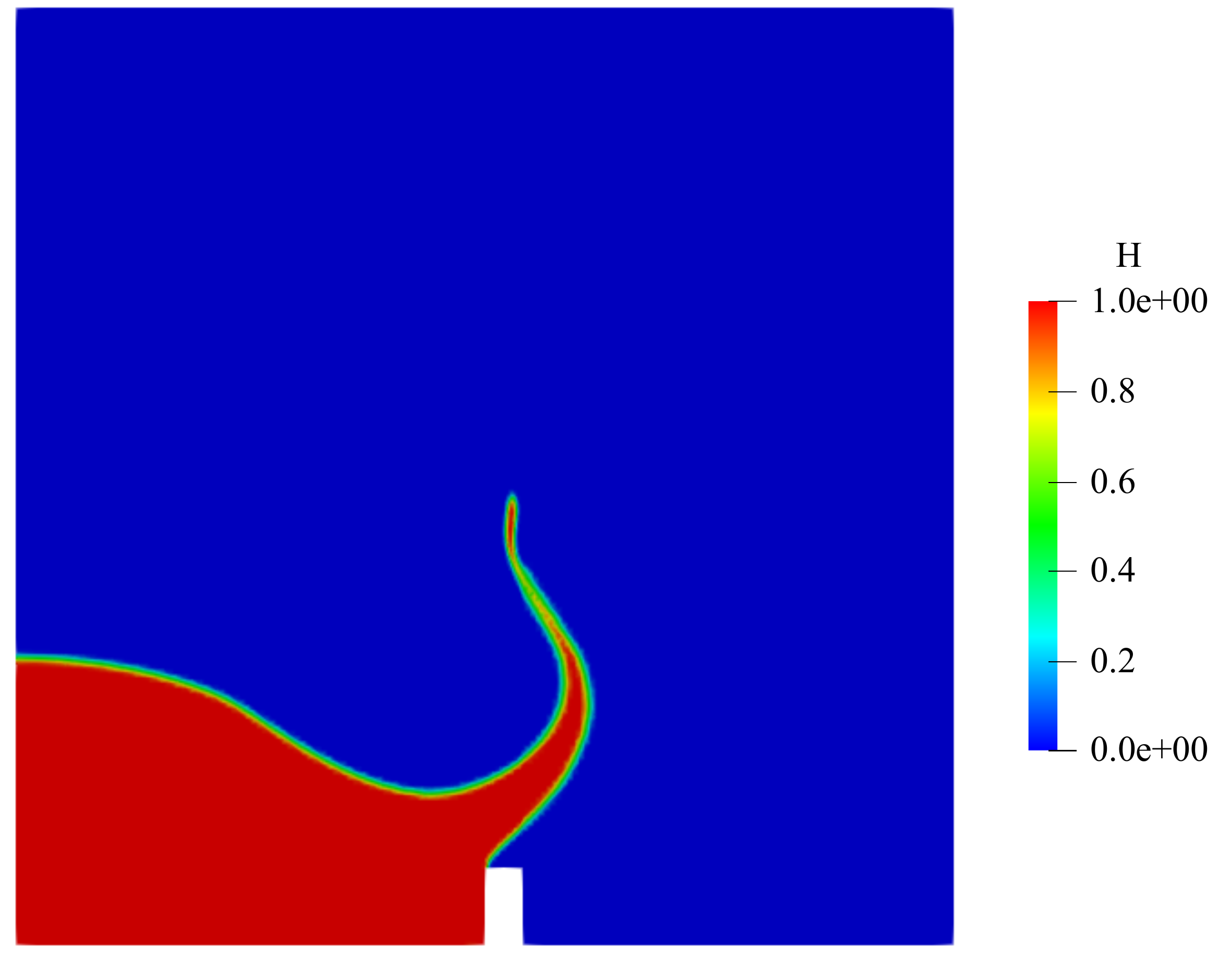}
    \caption{}
    \label{damBreakFreeSurf2}
  \end{subfigure}
    \hfill
  \begin{subfigure}[h]{0.495\textwidth}
    \includegraphics[width=\textwidth]{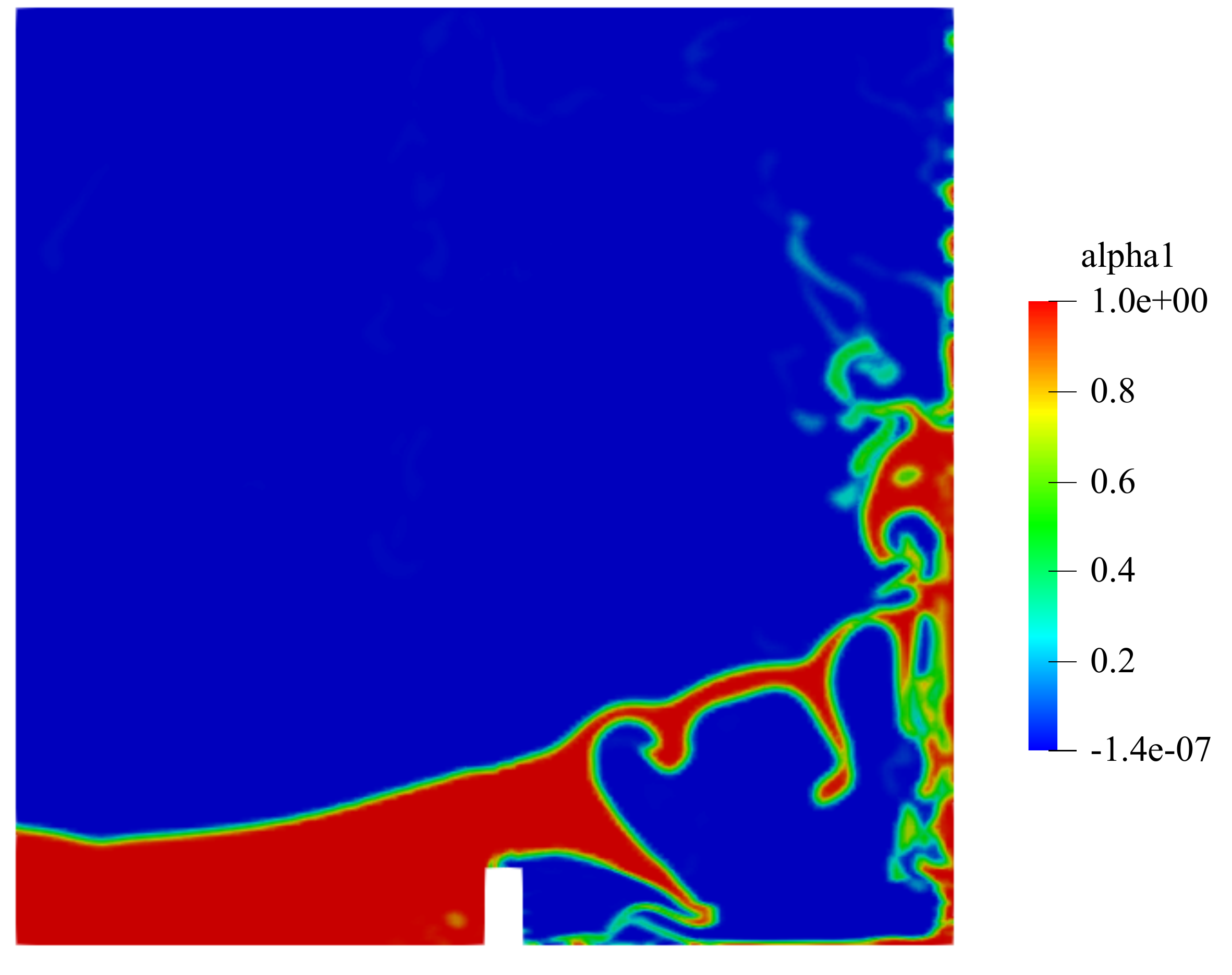}
    \caption{}
    \label{damBreakFreeSurf3}
  \end{subfigure}
    \hfill
  \begin{subfigure}[h]{0.495\textwidth}
    \includegraphics[width=\textwidth]{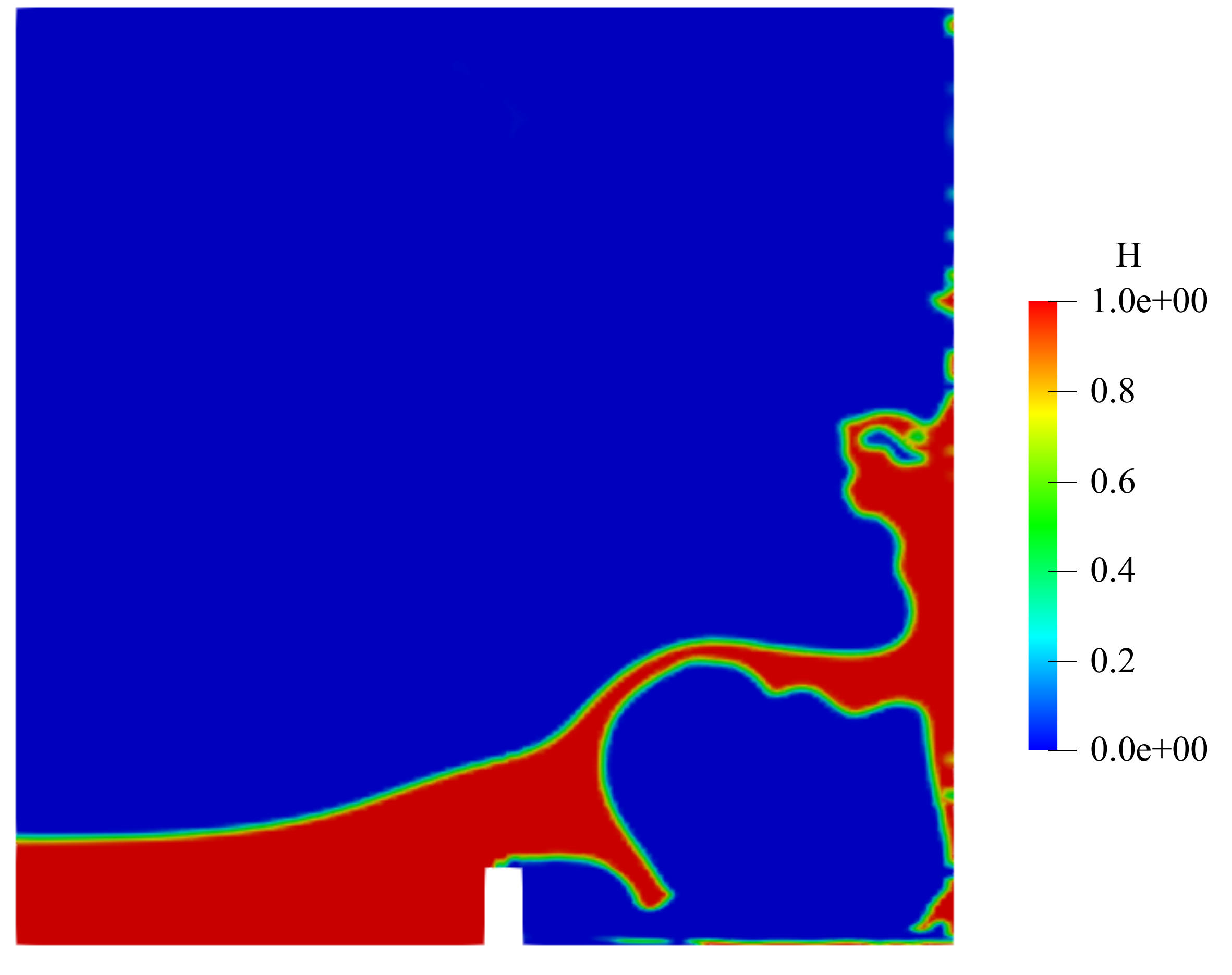}
    \caption{}
    \label{damBreakFreeSurf4}
  \end{subfigure}
  \caption{Evolutions of the free-surface in the dam break simulations: (a) with the VoF method (t = 0.2 s), (b) with the i-CLSVoF method (t = 0.2 s), (c) with the VoF method (t = 0.5 s), (d) with the i-CLSVoF method (t = 0.5 s).}
\label{damBreakFreeSurf}
\end{figure}

The Adaptive Mesh Refinement (AMR) is used to track a sharp interface with a finer mesh around the interface region and a coarser mesh elsewhere to save computational costs (as shown in Fig. \ref{damBreakFig2}).

The total simulation time is $0.5 \,s$ which is enough for water to reach both the bottom obstacle and the right wall. As shown in Fig. \ref{damBreakFreeSurf}, Fig. \ref{damBreakFreeSurf1} and Fig. \ref{damBreakFreeSurf3} are the simulation results with the VoF approach, and Fig. \ref{damBreakFreeSurf2} and Fig. \ref{damBreakFreeSurf4} are with the i-CLSVoF model. When water hits the obstacle and forms a wave over the obstacle, the results with the VoF approach show some interface diffusion (see Fig. \ref{damBreakFreeSurf1}) while the i-CLSVoF captures the sharp interface accurately (see Fig. \ref{damBreakFreeSurf2}). As the simulation progresses, water hits on the right wall and some water is bounced back to the container. Fig. \ref{damBreakFreeSurf3} shows large interface diffusion like rising gas around centre of the right wall for VoF, while Fig. \ref{damBreakFreeSurf4} presents a sharp interface captured by the i-CLSVoF without interface diffusion.

\bibliographystyle{elsarticle-num}
\bibliography{Refs}

\end{spacing}

\end{document}